\documentclass[aps,twocolumn,superscriptaddress,floatfix,longbibliography]{revtex4-1}
\UseRawInputEncoding

\usepackage{lipsum}
\usepackage{graphicx}
\usepackage{amsmath,amssymb}
\usepackage{braket}
\usepackage{mathtools}
\usepackage{hyperref}
\usepackage{multirow}
\usepackage{color}
\usepackage[normalem]{ulem}
\usepackage{amsfonts}
\usepackage{float}
\usepackage{pdfpages} 
\makeatletter
\usepackage{subfigure}
\usepackage{soul, xcolor}
\usepackage[english]{babel}

\usepackage{tikz}

\usepackage{pdfpages} 
\makeatletter
\AtBeginDocument{\let\LS@rot\@undefined}
\makeatother

\usepackage{url}
\usepackage{xr-hyper}
\usepackage{hyperref}

\begin{document}

\title{Universal monitored dynamics in multimode bosonic systems}

 \author{Shivam Patel}
 \thanks{co-first author}
  \affiliation{Department of Physics and Astronomy, 
Rutgers University, Piscataway, NJ 08854, USA}
\affiliation{Center for Materials Theory,  
Rutgers University, Piscataway, NJ 08854, USA}
\author{Catherine McCarthy}
\thanks{co-first author}
\affiliation{Department of Theoretical Physics, University of Geneva, 24 quai Ernest-Ansermet, 1211 Gen\`eve, Switzerland}
 \author{Ahana Chakraborty}
 \affiliation{Department of Physics and Astronomy, Louisiana State University, Baton Rouge, Louisiana 70803, USA}
  \affiliation{Department of Physics and Astronomy, 
Rutgers University, Piscataway, NJ 08854, USA}
\affiliation{Center for Materials Theory,  
Rutgers University, Piscataway, NJ 08854, USA}
\author{Jordan Huang}
  \affiliation{Department of Physics and Astronomy, 
Rutgers University, Piscataway, NJ 08854, USA}
\author{Thomas J. DiNapoli}
  \affiliation{Department of Physics and Astronomy, 
Rutgers University, Piscataway, NJ 08854, USA}
\author{Romain Vasseur}
\affiliation{Department of Theoretical Physics, University of Geneva, 24 quai Ernest-Ansermet, 1211 Gen\`eve, Switzerland}
\author{J. H. Pixley}\email{jed.pixley@physics.rutgers.edu}
  \affiliation{Department of Physics and Astronomy, 
Rutgers University, Piscataway, NJ 08854, USA}
\affiliation{Center for Materials Theory,  
Rutgers University, Piscataway, NJ 08854, USA}
\affiliation{Center for Computational Quantum Physics, Flatiron Institute, 162 5th Avenue, New York, NY 10010, USA}
\author{Srivatsan Chakram}\email{schakram@physics.rutgers.edu}
  \affiliation{Department of Physics and Astronomy, 
Rutgers University, Piscataway, NJ 08854, USA}
\date{\today}
\begin{abstract}
    We propose a route to study monitored many-body dynamics in multimode bosonic systems using circuit quantum electrodynamics. In this experimental setting, we construct several bosonic models comprising brickwork circuits built from beam-splitter gates, local parity measurements, and optional on-site Hubbard interactions, and diagnose their monitored dynamics via ancilla purification and a learnability-based probe. Under parity measurements, generic gate sets exhibit behavior that is largely consistent with a conventional measurement-induced phase transition, while a special class of beam-splitter circuits shows an apparent critical-like high-measurement regime in which purification times scale linearly with system size. We show that for realistic noise, gate, and measurement rates, these signatures are observable with near-term circuit QED hardware. 
\end{abstract}
 \maketitle

Rapid experimental progress in quantum devices over the past two decades has produced a range of potential quantum computing platforms---including superconducting circuits, neutral Rydberg-atom arrays, and trapped ions---that can implement coherent unitary dynamics, mid-circuit measurements, and real-time feedback, enabling quantum error correction and heralding the development of future fault-tolerant machines~\cite{PhysRevLett.74.4091, Saffman_2010, arute2019quantum, noel2022measurement, google2023measurement}. Among these, bosonic circuit quantum electrodynamic (cQED) systems have shown particular promise for hardware-efficient quantum error correction~\cite{ofek2016extending, hu2019quantum, campagne2020quantum}, leveraging long coherence times~\cite{reagor_quantum_2016,milul_superconducting_2023}, large local bosonic Hilbert spaces, fast high-fidelity universal control~\cite{heeres2017implementing,rosenblum2018cnot,chakram_seamless_2021,chakram_multimode_2022, gao2019entanglement,chapman_high--off-ratio_2023}, high-fidelity quantum-non-demolition measurements~\cite{connolly2025full}, and fast feedback~\cite{ofek2016extending}.

With the incorporation of mid-circuit measurements and error correction in bosonic cQED platforms, measurement and feedback naturally become part of the dynamics. Monitored many-body dynamics has been extensively studied in qubit-based circuit models, leading to the discovery of the \textit{measurement-induced phase transition} (MIPT)~\cite{Li_2018, li2019measurement, Skinner_2019, Potter_2022,fisher2023random, Chan_2019, gullansDynamical2020, Bao_2020, Jian_2020, zabalo2020critical, gullans2020, Zabalo-2022}: a transition in the entanglement structure of the wavefunction driven by the competition between entangling unitary evolution and local measurements. Signatures of MIPT phenomena have also been probed experimentally, albeit only in qubit platforms to date, in superconducting circuits and trapped ion systems~\cite{noel2022measurement, google2023measurement, koh_measurement-induced_2023, Kamakari-2025}.

Monitored free bosons with Gaussian measurements have also begun to attract theoretical attention in recent years~\cite{Minoguchi_2022, Yokomizo_2025, Li_2025}.
However, in random brick-layer bosonic circuits with Gaussian unitaries interspersed with Gaussian measurements, no evidence is found for a MIPT at any finite measurement rate; instead, the system flows to an area-law–entangled phase~\cite{Tianciboson}.
This motivates studying monitored bosonic dynamics with \emph{non-Gaussian} ingredients naturally available in cQED, such as parity measurements and interacting gates.

Rapid advancements in the coherence and control of microwave cavity devices in circuit QED now enable access to monitored dynamics in bosonic systems. While most demonstrations of quantum operations have been limited to one or a few modes, recent experiments have also realized multimode bosonic memories that support universal control across many long-lived modes~\cite{chakram_seamless_2021, chakram_multimode_2022, huang2025fastsidebandcontrolweakly, li2025cascaded}. The large local bosonic Hilbert space presents an opportunity to discover physics that is distinct from the standard qubit case. The bunching of bosons leads to characteristic Hong-–Ou-–Mandel interference effects~\cite{gao2018programmable} and underlies boson sampling~\cite{aaronson2011computational}, where linear, number-conserving beam-splitter dynamics generate nontrivial many-body correlations even in the absence of interactions. The large local Hilbert space also expands the space of operators for measurements, ranging from parity to photon number.

In this work, we advance the understanding of monitored bosonic dynamics on two fronts: we extend numerical analyses to non-Gaussian monitored bosonic evolution, uncovering novel entanglement phases and bosonic MIPTs; and we propose an experimental roadmap for realizing such monitored dynamics on near-term multimode cQED platforms, including a detailed error budget analysis.  The circuits studied here are composed of a combination of Gaussian beam-splitter gates, non-Gaussian tunable on-site Hubbard interactions, and local parity measurements. We diagnose the resulting monitored dynamics via the purification of an initially entangled ancillary mode and an alternative learnability probe. For a certain class of beam-splitter unitaries, we uncover behavior unique to the bosonic setting and find numerical evidence for a critical-like monitored phase whose purification time grows linearly with system size.  In contrast, for all other types of gates studied here, the dynamics appear consistent with a conventional measurement-induced phase transition between volume- and area-law entanglement (equivalently, between slow and $\mathcal{O}(1)$ purification) as the measurement rate is increased, with a dynamical critical exponent $z \approx 1$. Counterintuitively, we find that increasing the scrambling power of the unitaries \emph{decreases} the purification time of the high-measurement phase, highlighting a rich interplay between unitary dynamics and parity measurements in bosonic systems, where scrambling appears to help measurements reveal information, instead of hiding it~\cite{Skinner_2019}.  Finally, we assess near-term experimental feasibility on a platform employing the recently-developed cascaded Random Access Quantum Memory (RAQM) architecture~\cite{li2025cascaded}.

\begin{figure}
    \includegraphics[width=1.05\linewidth]{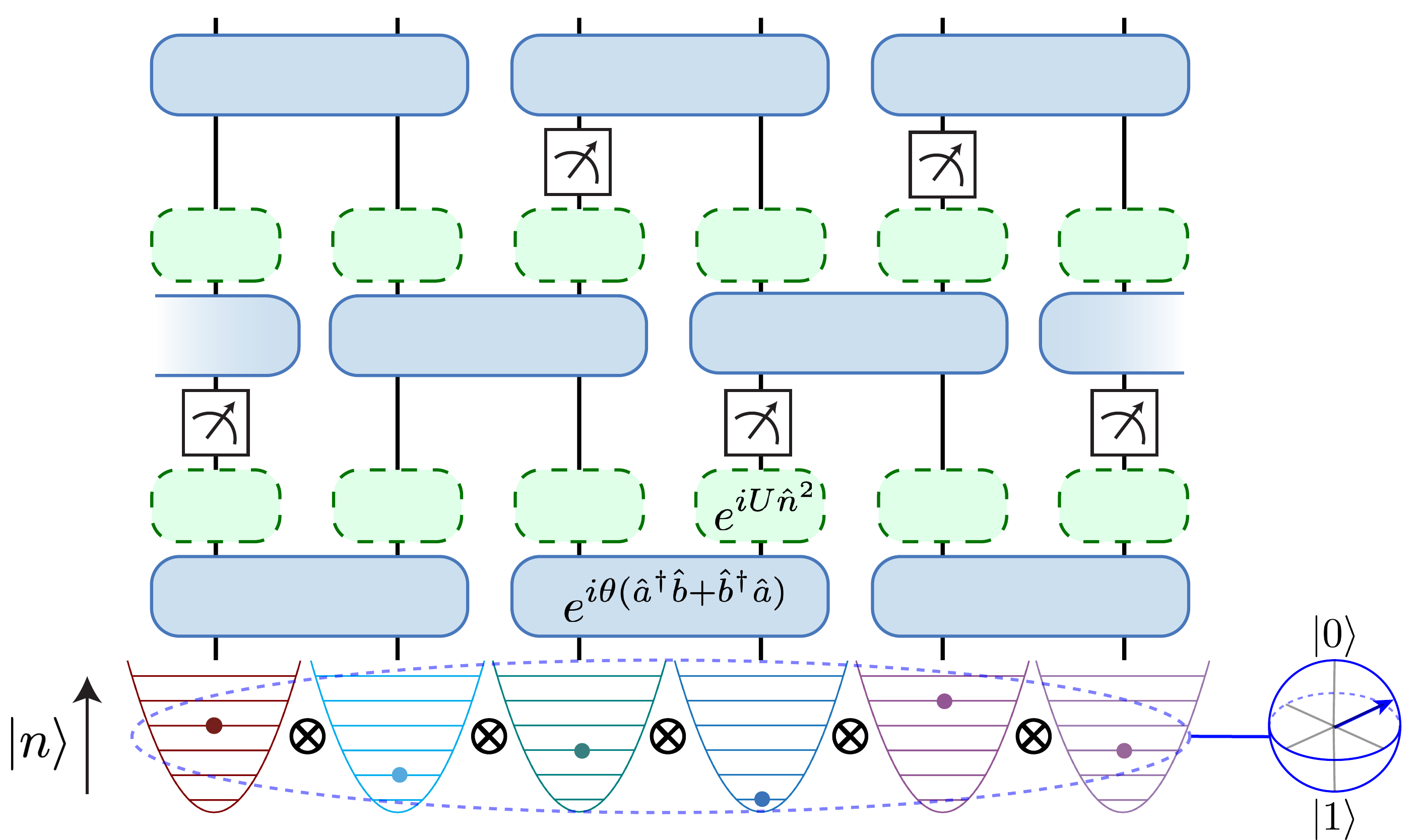}
    \caption{\textbf{Multimode bosonic analog of a brick-layer circuit for probing monitored dynamics}. 
    The resonant modes of low-loss superconducting cavities act as harmonic oscillators encoding bosonic degrees of freedom. 
    A reference qubit (dark blue Bloch sphere) is prepared in a globally entangled state with the modes and serves as a probe of the MIPT through its purification dynamics. 
    Each layer applies random beam-splitter gates (light blue), optional on-site Hubbard-like gates (green), and probabilistic mid-circuit measurements (squares with arrows) of parity or photon number.
    }
    \label{fig:Model}
\end{figure}

\noindent{\bf Monitored bosonic circuit-QED toolbox:}
The class of monitored circuits considered in this work is shown in Fig.~\ref{fig:Model}. 
They require a register of harmonic modes encoding bosonic degrees of freedom, entangling unitary gates (implemented here between nearest neighbors), and local quantum non-demolition (QND) mid-circuit measurements.
The unitary dynamics are chosen to be either (i) purely Gaussian beam-splitter operations or (ii) a combination of beam-splitter dynamics and a (non-Gaussian) local on-site Hubbard interaction.
These ingredients can be realized in circuit QED by encoding the bosonic modes in electromagnetic resonances of low-loss superconducting microwave cavities, and by coupling the cavity register to superconducting Josephson circuits that enable high-fidelity control, engineered mode--mode interactions, and QND readout.
The cavity resonances can exhibit coherence times ranging from milliseconds~\cite{reagor_quantum_2016, kudra2020high, chakram_seamless_2021} to tens of milliseconds in niobium cavities~\cite{romanenko2020three,milul_superconducting_2023,oriani2024niobium, kim2025}, and can be engineered to be multimode with tens of addressable resonances~\cite{chakram_seamless_2021, huang2025fastsidebandcontrolweakly, li2025cascaded}.  This combination of multimode capacity, long coherence, and high-fidelity control and measurement is crucial for realizing monitored dynamics at depths comparable to the system size. 

We implement beam-splitter operations in circuit QED by parametrically activating mode--mode interactions mediated by superconducting Josephson circuits. 
Such interactions can be engineered using a variety of tunable coupler circuits, which are driven either by a time-dependent voltage drive~\cite{frattini2017, chapman_high--off-ratio_2023} or a flux drive~\cite{lu2023high, li2025cascaded, maiti2025linear} at the difference frequency of the target cavity modes, realizing programmable beam-splitter gates with errors on the order of $\sim 10^{-3}$. 
In this way, one can implement nearest-neighbor hopping operations of the form
\begin{equation}
    V^{\rm hop}_{n,n+1}(\theta, \varphi)
    = \exp\!\left[i\theta\!\left(\hat a_n^\dagger \hat a_{n+1} e^{i\varphi}
    + \hat a_{n+1}^\dagger \hat a_n e^{-i\varphi}\right)\right],
\end{equation}
where $\hat a_n^{(\dagger)}$ are the annihilation (creation) operators on site $n$. 
The beam-splitter gates are applied for variable durations to implement a swap angle $\theta \in [0,2\pi]$, while the phase $\varphi$ is set by the phase of the drive tone. 
In what follows, it will be important to distinguish between evolutions generated from beam-splitter gates with fixed phases (BSFP) and those with randomized phases (BSRP), since randomizing $\varphi$ substantially increases the degree of mode mixing (and hence scrambling within a fixed-photon-number sector). 
This can be seen in the Heisenberg picture: $V^{\rm hop}_{n,n+1}(\theta,\varphi)$ implements an $\mathrm{SU(2)}$ rotation of the operator doublet $(\hat a_n^\dagger,\hat a_{n+1}^\dagger)$, with $\varphi$ selecting the rotation axis in this two-mode operator space. 
Fixed-phase gates therefore generate $\mathrm{SU(2)}$ rotations about a single axis, while randomizing $\varphi$ yields rotations about a range of axes, producing more generic mode mixing and state scrambling. 
We study both BSFP and BSRP circuits. 
In the presence of parity measurements, BSFP dynamics exhibits an apparent critical-like high-measurement regime, and we therefore focus on this case in the main text; results for BSRP circuits are discussed in the Supplementary Information.

To extend beyond Gaussian dynamics, we exploit the nonlinearity of a superconducting transmon circuit, which yields a dispersive interaction with the coupled cavity modes characterized by a dispersive shift $\chi$.
In the dispersive regime, the transmon frequency is shifted by $n\chi$ conditioned on the cavity photon number $n$.
This dispersive interaction enables a broad set of unitary control primitives for the joint transmon--bosonic-mode system~\cite{krastanov2015universal, heeres_cavity_2015, heeres2017implementing, eickbusch2021fast, huang2025fastsidebandcontrolweakly}, and provides a unified toolbox for accessing non-Gaussian physics in monitored circuits in two complementary ways:
(i) through unitary gates that implement effective nonlinear interactions, and (ii) through measurement backaction from QND measurements of nonlinear observables (e.g., parity or photon number), which makes individual quantum trajectories non-Gaussian.
 
We implement nonlinear on-site Hubbard-like interaction gates of the form
\begin{equation}\label{eq:interacting_unitary}
    V^{\mathrm{int}}_n(U) = e^{-iU(\hat a^\dagger_n \hat a_n)^2},
\end{equation}
where $U$ is a dimensionless parameter controlling the strength of the gate. This evolution can, in principle, be realized using a native self-Kerr interaction; however, in circuit-QED systems the self-Kerr is typically small ($\sim 10~\mathrm{kHz}$). We therefore implement these interacting gates via Selective Number-dependent Arbitrary Phase (SNAP) gates~\cite{krastanov2015universal, heeres_cavity_2015}, where photon-number--selective transmon rotations~\cite{schuster2007resolving} can be used to implement arbitrary photon-number-dependent geometric phases.
In particular, we realize a Hubbard interaction by imprinting a quadratic photon-number--dependent phase.
We use a recently developed optimal-control protocol~\cite{landgraf2024fastquantumcontrolcavities} to achieve this operation in a time $3.25/\chi$, where $\chi$ is the dispersive interaction strength. 

The same dispersive interaction further enables a range of quantum non-demolition (QND) measurements of the cavity modes.
In particular, it enables projections onto specific photon-number subspaces~\cite{schuster2007resolving} using number-resolved transmon $\pi$ rotations followed by transmon readout.
It also enables QND parity measurements~\cite{sun2013tracking} via a transmon Ramsey sequence with a dispersive interaction time $T = \pi/\chi$, which maps even and odd cavity parity onto distinct transmon states ($\ket{g}$ and $\ket{e}$).
This corresponds to the first bit of the binary representation of the photon number, and higher-order bits (the $m^{\mathrm{th}}$ bit) can be obtained by tuning the Ramsey idle time to $T = 1/(2^{m}\chi)$.
A series of adaptive Ramsey measurements with feedforward can then yield a QND photon-number ($\hat a^\dagger \hat a$) measurement, extracting one bit at a time~\cite{wang2020efficient} (see Supplementary Information).
Measurements of this type introduce non-Gaussianity into the trajectories; thus, all of the circuit models we study represent non-Gaussian monitored evolutions even when the unitary component is composed only of (Gaussian) beam-splitter gates, in contrast to previous works on monitored free bosons~\cite{Minoguchi_2022, Yokomizo_2025, Li_2025, Tianciboson}.
We focus on parity measurements in the main text because they incompletely collapse the local Hilbert space onto Fock basis states, which we will show leads to qualitative differences from the qubit case and novel measurement-induced phenomena.

\begin{figure*}[t]
    \centering
    \includegraphics[width=\linewidth]{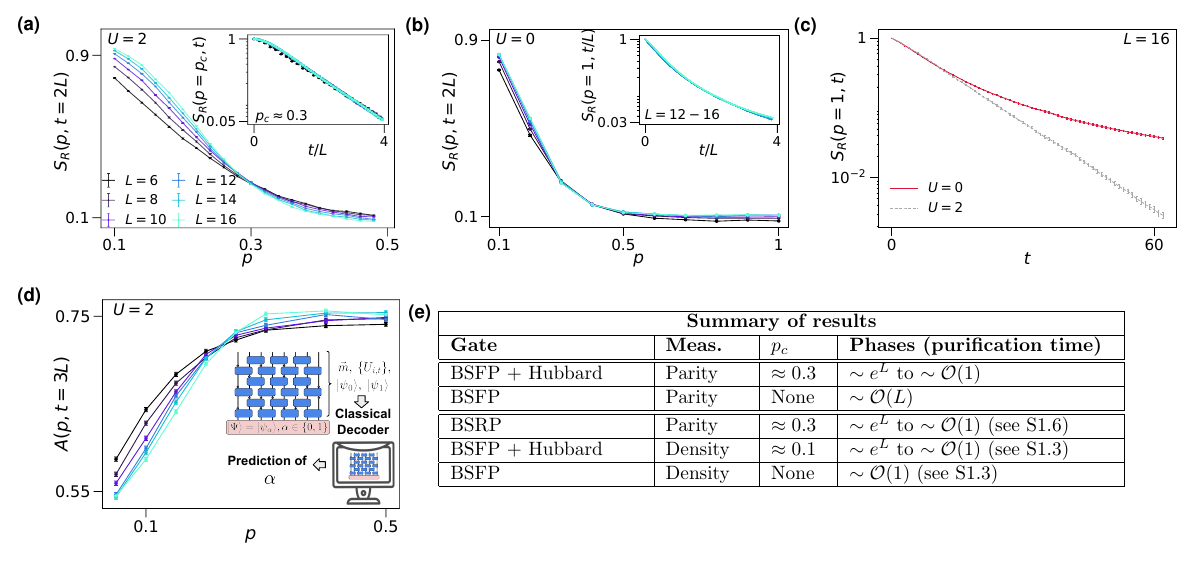}

    \caption{\textbf{Monitored dynamics with parity measurements.} Ancilla entanglement entropy $S_R$ for circuits composed of (a) both BSFP and on-site Hubbard gates of strength $U=2$ and (b) those composed of purely BSFP gates ($U=0$) for systems of size $L=4-16$, averaged over 10000 independent realizations.  For the $U=2$ case, both the clear crossing of the $S_R(p,t=2L)$ curves at $p_c \approx 0.3$ and the collapse of $S_R(p_c \approx 0.3, t/L)$ (inset) indicate the presence of a phase transition.  In contrast, the $U=0$ data shows no clear signature of a standard MIPT, with the ancilla not purifying at any value of $p$ at these time scales. In addition, the collapse of $S_R(p,t/L)$ curves for the presented times for $p \gtrsim 0.3$ for the largest numerically-accessible system sizes ($L=12-16$, inset) indicates that at higher measurement rates, the system is in a phase with a purification time that scales with system size.  (c) Comparison of $S_R(p=1,t)$ between the $U=2$ and $U=0$ cases. Turning on the on-site Hubbard gates takes the system from a phase in which the purification time scales with system size ($U=0$) to a phase with an $\mathcal{O} (1)$ purification time ($U=2)$.  \textbf{Learnability} (d) Simulated decoder accuracy $A(p)$ for monitored dynamics with BSFP and on-site Hubbard gates ($U=2$).  The accuracy $A(p)$ shows a similar transition to the ancilla entanglement entropy $S_R$ in (a) and may be used as an alternative post-selection free order parameter.  Inset: protocol for detecting the learnability transition.  A monitored circuit is run in the lab using an initial state of either $\ket{\psi_0}$ or $\ket{\psi_1}$, where $\braket{\psi_0 | \psi_1}=0$.  A classical decoder then outputs a prediction of the initial state given the measurement record $\vec m$, the gates, and knowledge of $\ket{\psi_0}, \ket{\psi_1}$. \textbf{Summary table} (e) Summary of all numerical results.  For non-BSFP gates with any measurement type, the data is consistent with a standard MIPT.  In contrast, the purely BSFP gates are consistent with either a critical phase at high $p$ (parity measurements) or a purified phase at any $p$ (density measurements); ``None'' refers to the lack of an MIPT  with $z=1$. }
    \label{fig:numerics}
\end{figure*}

{\bf Monitored Dynamics in Bosonic Circuits}:
Motivated by the capabilities of the circuit-QED toolbox described above, we now introduce several quantum-circuit models for monitored dynamics in multimode bosonic systems. We consider a brickwork circuit of $L$ bosonic modes (Fig.~\ref{fig:Model}) with a fixed total photon (particle) number $Q$, so that the Hilbert-space dimension in the fixed-$Q$ sector scales as $N=\binom{Q+L-1}{Q}$. Throughout this work, we fix the filling fraction $Q=L/2$ for chains of $L$ modes so that $Q$ scales extensively with system size and yields a large accessible state space. The dynamics proceeds in discrete layers: each timestep applies a brickwork pattern of two-mode unitary gates, followed by local mid-circuit measurements performed independently on each mode with probability $p$. The unitary layers are generated from beam-splitter gates with randomized swap angles. The beam-splitter phases are either kept fixed (BSFP) or also randomized (BSRP). In some models, the beam-splitter dynamics is supplemented with an on-site Hubbard interaction (implemented via a SNAP gate). The measurements are chosen to be either QND parity measurements or QND photon-number (density) measurements.

In the main text, we focus on two specific models that highlight both dynamics consistent with previously studied MIPTs in qubit circuits and monitored phenomena specific to bosonic circuits: (i) parity measurements with BSFP gates supplemented by on-site interactions, and
(ii) parity measurements with purely BSFP dynamics. In both cases, the unitary evolution conserves total photon number and is generated by repeated application of a two-mode gate of the form
\begin{equation}
    \hat{U}_{n,n+1}(\theta, U)
    = V^{\rm int}_{n+1}(U)\,V^{\rm int}_{n}(U)\,V^{\rm hop}_{n,n+1}(\theta,0),
    \label{eq:interacting_gate}
\end{equation}
where $U$ tunes the strength of the on-site Hubbard interaction, with $U=0$ corresponding to the non-interacting (pure beam-splitter) limit.

\emph{Probing MIPTs through entanglement:} The typical method of detecting a MIPT relies on observables that are nonlinear in the system's density matrix $\rho$ and conditioned on the mid-circuit measurement trajectories $\vec m$. One such quantity is the entropy of the entanglement of an ancillary reference qubit $R$ with the $L$ modes, which is obtained from preparing the initial state $\ket{\Psi} = \frac{1}{\sqrt{2}} ( \ket{\psi_0} \ket 0_R + \ket{\psi_1} \ket 1_R)$, where $\ket{\psi_0},\ket{\psi_1}$ are orthogonal Haar-random states in the space of states containing $Q=L/2$ bosons.  The MIPT is detectable as a transition in the entanglement entropy of $R$, given by $S_R = -\sum_{\vec m} p_{\vec m} \mathrm{Tr}[\rho_R \log_2(\rho_R)]$, where $\rho_R$ is the reduced density matrix of $R$ and $p_{\vec m}$ is the Born probability of trajectory $\vec m$~\cite{gullans2020}.  A second quantity that is frequently used to diagnose MIPTs is the bipartite entanglement entropy $S = -\mathrm{Tr}[\rho_A\log_2(\rho_A) ]$, where $A$ is the first $L/2$ sites of the system-- numerical results for the bipartite entanglement entropy are provided in S1.4 \cite{supp}. Since the experimental computation of both $S_R$ and $S$ requires post-selecting on individual measurement trajectories, the number of which grows exponentially with system size $L$, obtaining either quickly becomes experimentally unfeasible for even moderate system sizes.  

\emph{Learnability as a probe of MIPT:} To bypass the need for performing experimental post-selection, we also consider an alternative ``learnability''  perspective \cite{Barratt_2022, Dehghani_2023, akhtar2023, Li_2023, Ippoliti_2024, Agrawal_2024, McGinley_2024, kim2025} which views the MIPT as an information-theoretic transition rather than a transition in the entanglement properties of a quantum state.  To probe the learnability transition, we consider a classical decoder tasked with distinguishing between two initially orthogonal random states $\ket{\psi_0}, \ket{\psi_1}$ given a measurement record $\vec m$ of a monitored trajectory generated with the set of unitaries $\{U_{i,t}\}$, as illustrated in the inset of Fig.~\ref{fig:numerics}(d).  The transition appears in the accuracy of this decoder $A(p) = \frac{1}{N} \sum_n a(n,p)$, where $a(n,p) \in \{0,1\}$ is a binary variable that tracks the success of the decoder for the $n^{\rm th}$ record generated with measurement rate $p$-- see S1.2 for implementation details.  The advantage of the learnability perspective is that the exponential difficulty is shifted from an experimental task (post-selection) to a numerical task (the implementation of the decoder), which may permit moderate increases in experimentally-accessible system sizes.

\emph{Interacting circuits with BSFP and on-site Hubbard gates:} We first consider a brick layer monitored circuit evolving with both BSFP and on-site Hubbard gates ($U=2$) with parity measurements.  The (post-selected) ancilla entanglement entropy is shown in Fig.~\ref{fig:numerics}(a) for parity measurements for varying system sizes $L=8-16$. The $S_R$ curves are consistent with a transition with $z=1$ at $p_c \approx 0.3$, as demonstrated by the crossing in $S_R(p,t\sim L^z)$ and the collapse of the $S_R(p=p_c,t/L^z)$ curves for (Fig.~\ref{fig:numerics}a).  This behavior is similar to that of a standard MIPT, in which there is a transition from an exponentially-long to $\mathcal{O}(1)$ purification time in the ancilla $R$.  The results of the learnability probe are consistent with the $S_R$ data, as shown in S1.2.  

Qualitatively, the $S_R$ and learnability results appear to be representative of generic bosonic monitored dynamics.  For instance, choosing different values of $U>0$ or choosing to perform full number measurements may alter the location of $p_c$ but does not appear to affect the existence of a transition or the different phases.  The same appears to be true for circuits generated from purely beam-splitter with random $\varphi$ (BSRP) gates, indicating that non-Gaussian gates are not a requirement for observing a bosonic MIPT and the non-Gaussianity from measurements is sufficient-- see Table in Fig.~\ref{fig:numerics}(e).

\emph{Purely BSFP circuits:} In contrast to the $U=2$ case, the monitored dynamics generated from only BSFP gates ($U=0$) is strikingly different. Fig.~\ref{fig:numerics}(b) shows the data for the entanglement entropy of the ancilla qubit for evolution with BSFP gates together with parity measurement which is \textit{inconsistent} with a standard MIPT with $z=1$. The ancilla never purifies at any $p$ on timescales $t \sim L$, as most strikingly demonstrated at $p=1$.  Furthermore, we find that $S_R(p,t/L)$ collapses on time scales $t\sim L$ [Fig.~\ref{fig:numerics}(b) inset] for an entire range of large-$p$ values for the largest numerically-accessible system sizes ($L=12-16$).  This collapse is consistent with a novel phase in which the ancilla purifies on $\mathcal{O}(L)$ timescales-- given that this scaling is normally seen only at a single $p_c$, we refer to this phase as a \textit{critical phase}.  The contrast between the $U=0$ and $U=2$ curves for $S_R(p=1,t)$  (Fig. \ref{fig:numerics}(c)) demonstrates that the BSFP gate is a special case and that choosing a more generic gate restores the conventional high-measurement phase with ancilla purification on $\mathcal{O}(1)$ timescales. We note that this critical phase also requires the limited projective nature of the parity measurements-- if we instead consider density measurements, we find the ancilla always purifies at a time scale of $\mathcal{O}(1)$ (see S1.3 \cite{supp}).

While we find numerical evidence for a critical phase at high measurement rates from the $U=0$ data, we are unable to identify any clear signature of a transition from some other (possibly mixed) phase to this critical phase.  Two possibilities are that the system is that the critical phase endures across all values of $p$ or that there is a transition with $z>1$ (see S1.5 for preliminary discussion). As demonstrated by recent work on free-fermions, inferring a critical phase from numerical data from finite system sizes is challenging \cite{poboiko2023theory}-- it is possible that the apparent critical phase is an artifact of the small numerically-accessible system sizes.  Future analytical work is necessary to distinguish among these possibilities.

\emph{Scrambling with parity measurements:}  MIPTs are often understood as a competition between scrambling unitary dynamics, which dominates in the mixed phase, and the disentangling effect of measurements, which dominates in the purified phase.  With this intuition, choosing more scrambling gates would be expected to favor the mixed phase.  Our data displays the opposite trend: either by adding on-site Hubbard gates to a BSFP evolution or by using only BSRP gates, both of which are \textit{increase} scrambling compared to the purely BSFP case, the purification time of the high-measurement phase \textit{decreases} from $\mathcal{O}(L)$ to $\mathcal{O}(1)$.  This effect arises from the fact that parity measurements project onto a subspace of the local Hilbert space, rather than onto a single Fock basis state.  Therefore, scrambling in between repeated parity measurements creates more opportunities to collapse the state and thus reduce its entanglement entropy-- see S2 for further discussion \cite{supp}.

\emph{Bipartite entanglement entropy:} We provide results for the bipartite entanglement entropy $S$ for both the $U=0$ and $U=2$ cases in S1.4.  Intriguingly, for both cases, $S$ does not show any clear signature of a transition and is possibly consistent with critical $\sim \log L$ scaling at high measurement rates.  Although the coincidence of the apparent $\log L$ phase for $S$ and the critical phase for $S_R$ is consistent for the $U=0$ case, the $U=2$ case is more puzzling.  While inferring $\sim \log L$ scaling from numerics limited to small system sizes is challenging \cite{poboiko2023theory}, the data raises the intriguing possibility that there is a decoupling between the ancilla and bipartite entanglement transitions, a scenario which is left for future research.

\begin{figure*}\label{fig:setup}
    \centering
    \includegraphics[width=\linewidth]{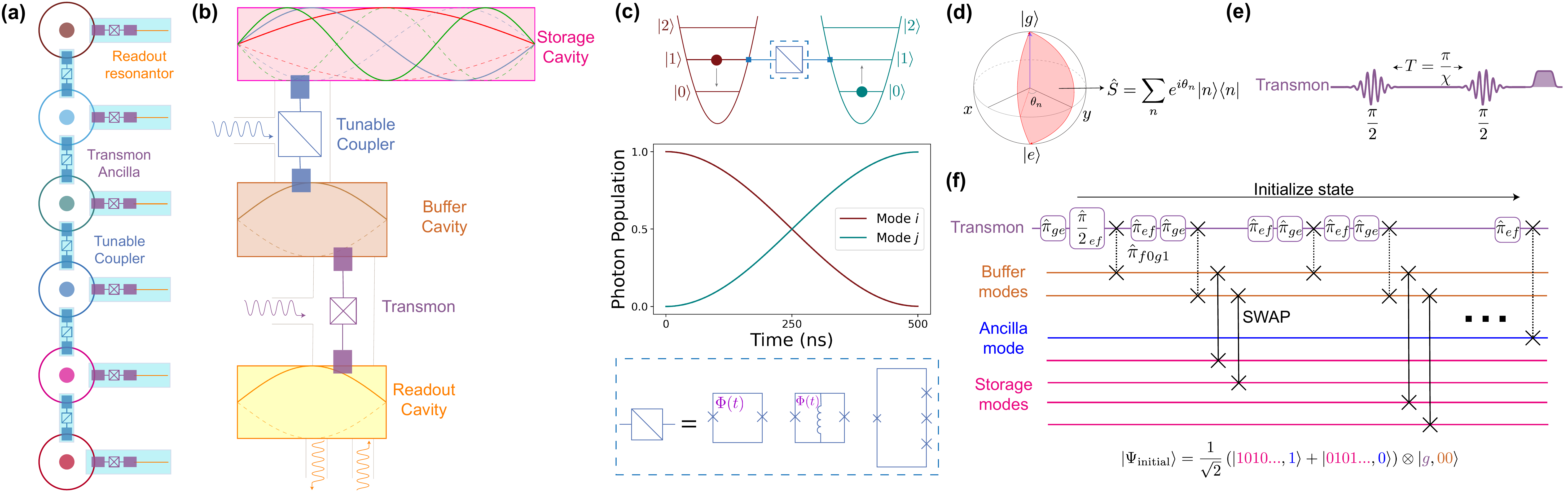}
    \caption{\textbf{Linear cavity array architecture} (a) Schematic of a 1D $\lambda/4$ cavity array coupled via tuneable couplers. Each cavity is also interfaced with a chip hosting a transmon and readout cavity. \textbf{Cascaded RAQM architecture} (b) Multimode cQED setup of a cascaded random access quantum memory (RAQM) architecture. A superconducting storage cavity coupled to a buffer cavity via a tunable coupler; the buffer cavity is interfaced with a readout resonator through a transmon. \textbf{Beam-splitter gate}(c) Cavity modes modeled as harmonic oscillators are coupled via a tunable coupler (ex. SQUID, LINC, or SNAIL) (dashed box). Using a SNAIL we realize a 250~ns beam-splitter gate creating an entangled $\frac{1}{\sqrt2}(\ket{10}+\ket{01})$ state between the modes, and 500~ns SWAP gate. \textbf{SNAP gate} (d)  SNAP operations, used to realize on-site Hubbard gates, apply geometric photon-number-dependent phases by driving the transmon qubit in a closed trajectory (shaded red) around the Bloch sphere in each photon number subspace. \textbf{Parity measurement} (e) Ramsey-based nondemolition photon-number parity readout using dispersive coupling, where two $\pi/2$-pulses are separated by some idle time $T=\pi/\chi$. \textbf{State preparation} (f) Pulse sequence for initializing a checkerboard-like state in the storage cavity entangled with a passive reference mode, using $\pi_{ge}/\pi_{ef}$ pulses, $f0g1$ sidebands, and coupler-mediated photon shuttling.}
\end{figure*}

{\bf Realizing Bosonic MIPT with Multimode Circuit QED}:
We consider two representative experimental platforms for realizing the theoretical bosonic circuit models. The first platform, shown in Fig.~\ref{fig:setup}(a), consists of a 1D array of coaxial $\lambda/4$ cavities~\cite{reagor_quantum_2016} coupled by tunable couplers~\cite{chapman_high--off-ratio_2023, lu2023high, li2025cascaded} that enable nearest-neighbor beam-splitter operations~\cite{chou_superconducting_2024, teoh2023dual}. Each cavity is also interfaced with its own transmon ancilla, coupled to a readout resonator, which can implement on-site interactions and mid-circuit measurements. This architecture supports parallel gate execution on distinct modes, but it is expensive in control hardware, requiring an ancilla qubit and a coupler element for each mode.

The second platform, shown in Fig.~\ref{fig:setup}(b), is a cascaded random access quantum memory (RAQM) architecture~\cite{li2025cascaded}, in which a high-coherence multimode storage cavity is coupled via a tunable coupler to a separate buffer cavity that serves as a cache memory and is, in turn, coupled to a transmon. The tunable coupler swaps states between storage modes and buffer modes. In the buffer cavity, the transmon enables universal cavity control (e.g., SNAP gates) and QND measurements of the mode, such as parity measurements, as shown in Fig.~\ref{fig:setup}(d,e). The tunable coupler also enables entangling beam-splitter operations between storage modes, as shown in Fig.~\ref{fig:setup}(c). By isolating the nonlinearity to the buffer/measurement cavity, this architecture suppresses spurious Kerr and cross-Kerr interactions among the storage modes, mitigating error propagation and reducing crosstalk during gate operations. Furthermore, we expect any residual interaction-induced phases to not qualitatively affect the monitored dynamics.

\emph{Preparation of the initial state:}
The numerical analysis of monitored bosonic circuits discussed earlier and shown in Fig.~\ref{fig:numerics} begins from a Haar-random state in the fixed-charge sector with $Q=L/2$ photons. We find that quantitatively similar results can also be obtained by starting from a ``checkerboard'' state that is globally entangled with a reference qubit and then scrambling with $\mathcal{O}(L)$ layers of beam-splitter and on-site Hubbard gates, as justified numerically in S1.1 and S.6~\cite{supp}. The checkerboard state $\ket{\Psi}=\frac{1}{\sqrt{2}}\left(\ket{101010\cdots}\ket{1}_R+\ket{010101\cdots}\ket{0}_R\right)$ entangles two orthogonal alternating-occupancy configurations of the storage-cavity modes with the $\ket{0}$ and $\ket{1}$ states of a reference qubit. In the RAQM architecture, this reference qubit is prepared in the transmon and then swapped into an additional ancillary storage mode (reference mode).

The checkerboard state can be prepared in the RAQM architecture using standard protocols used in circuit QED.  We harness the quartic nonlinearity of the transmon to implement a four-wave mixing parametric process that converts two excitations in the transmon (the $\ket{f}$ level) into a single photon in the dispersively coupled buffer cavity mode via an $\ket{f0}\!-\!\ket{g1}$ sideband transition~\cite{Zeytino_lu_2015, rosenblum2018cnot, huang2025fastsidebandcontrolweakly}. This is implemented by applying an external charge drive to the transmon at the difference frequency between the $\ket{f0}$ and $\ket{g1}$ states. By combining these sideband operations with transmon rotations on the $\ket{g}\!-\!\ket{e}$ and $\ket{e}\!-\!\ket{f}$ transitions, and temporary shelving of the population in the $\ket{e}$ manifold~\cite{huang2025fastsidebandcontrolweakly}, we generate an entangled state between the transmon and the $\{\ket{01},\ket{10}\}$ state of the pair of buffer modes. The photons in the buffer modes are then swapped into a pair of modes of the storage cavity via the tunable coupler, and this procedure is repeated to populate the storage cavity up to half-filling while maintaining entanglement with the transmon. Finally we swap the transmon state into the ancillary reference mode of storage cavity.  A schematic of the full state-preparation sequence is shown in Fig.~\ref{fig:setup}(f) and detailed further in S6. The execution time of the state-preparation protocol scales as $\sim L^2$ and is $4.5~\mu\mathrm{s}$ for $L=4$. After the checkerboard state is prepared, we apply $S=2L$ scrambling layers of nearest-neighbor beam-splitter gates followed by on-site Hubbard gates (implemented via SNAP gates) which provides an initial state that is sufficiently scrambled for the monitored dynamics.

\emph{Implementation of monitored bosonic dynamics:}
We now subject the scrambled input state to our monitored bosonic circuit model, as shown in Fig.~\ref{fig:Model}. We apply $M=2L$ monitored layers, where each layer consists of a set of BSFP gates and on-site Hubbard gates (for the $U=2$ case), followed by mid-circuit parity measurements performed independently on each mode with probability $p$. Measurement outcomes are obtained by reading out the transmon through its readout cavity and are recorded to define the measurement trajectory for each circuit run. All operations are performed sequentially in the RAQM architecture. For SNAP operations and parity measurements, which are transmon-mediated, we swap the target modes from the storage cavity to the buffer cavity, execute the operation, and then swap the modes back into storage. The transmon is reset to its ground state before the next operation. The total execution time of the monitored circuit scales as $\sim L^2$ and, for $L=4$ and $p=1$, is $\sim 100~\mu\mathrm{s}$ for the BSFP case and $\sim 150~\mu\mathrm{s}$ for the case with on-site Hubbard gates. Implementations of the gate operations and details of the protocol sequence on the RAQM architecture are described in S8.5.

\emph{Measuring the entanglement entropy of the reference mode:}
After completing the full sequence of gates and measurements, we extract the entanglement entropy of the reference mode, conditioned on the mid-circuit measurement record. Because the reference mode is restricted to the $\ket{0},\ket{1}$ subspace, its state can be swapped through the buffer cavity into the transmon and fully characterized using standard qubit tomography. Extensions to alternative encodings (e.g., using $\ket{0}$ and $\ket{2}$) enable error-detection schemes that can improve the fidelity of the density-matrix reconstruction, at the expense of additional shots~\cite{mai2026biased}. The reconstructed density matrix is then used to compute the entanglement entropy.

For a given measurement rate $p$, system size $L$, and random realization of beam-splitter angles, the total number of mid-circuit measurements scales as $m \sim L^2 p$, yielding $2^{m}$ possible measurement trajectories. Assuming state-of-the-art single-shot readout fidelity ($\sim 99\%$), we acquire $\sim 500$ shots per trajectory to reconstruct the ancilla reduced density matrix; this requires executing the monitored circuit $500 \times 2^{m}$ times for each of 200 independent realizations of the random unitaries, thereby sampling over both circuit instances and their associated measurement trajectories. Tomography results are binned by the observed measurement record, producing a reconstructed density matrix for each trajectory, from which the entanglement entropy is computed. After each run, the system is reset; tomography and reset are estimated to require approximately $2~\mu\mathrm{s}$ and $50~\mu\mathrm{s}$, respectively~\cite{huang2025fastsidebandcontrolweakly, li2025cascaded}. The total runtime of a single monitored-circuit execution, from state preparation to system reset, is approximately $160~\mu\mathrm{s}$ for the BSFP case and $210~\mu\mathrm{s}$ for the case with on-site Hubbard gates (for $p=1$ and $L=4$). For these parameters, this corresponds to a total acquisition time on the order of a few hundred hours.

\begin{figure}[t!]
    \centering
    \includegraphics[width=0.45\textwidth]{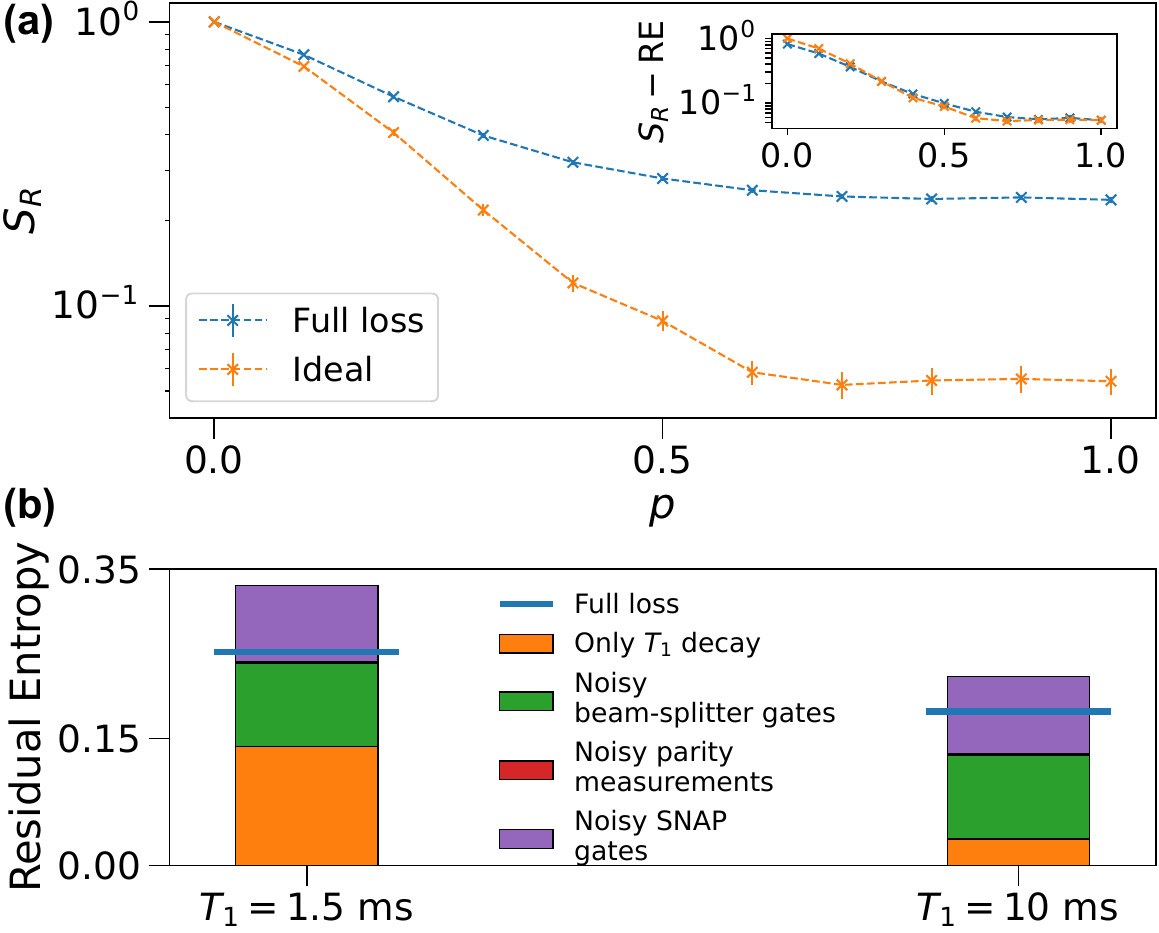}
    \caption{\textbf{Effects of noise in monitored bosonic circuits.}(a) Averaged entropy curves for a model containing both BSFP and on-site Hubbard gates with 8 scrambling and 8 monitored layers. Results are shown for ideal gates (orange) and for circuits with all noise sources active (blue). Inset: the same curves after subtracting the residual entropy (RE) from the full loss curve. (b) Breakdown of residual entropies generated at $p=1$ (for a worst case estimate), obtained from simulations in which a single noise source is active at a time.
    }
    \label{fig:error_budget}
\end{figure}

\emph{Feasibility of the experiment:}
We assess the feasibility of observing a MIPT in a realistic circuit-QED architecture in the presence of coherent and incoherent gate errors by performing end-to-end simulations of the proposed monitored bosonic circuits. We model noise at the level of elementary operations using Lindblad master-equation simulations that capture the dominant incoherent processes: cavity decay and thermal excitation, transmon relaxation and thermal excitation, and transmon dephasing. Each operation is simulated on the minimal local subsystem on which it acts: beam-splitter gates are simulated on a pair of cavity modes, while parity measurements and on-site Hubbard gates are simulated on a single cavity mode dispersively coupled to a two-level transmon. From these simulations, we extract quantum channels for each noisy operation, embed them into a larger composite Hilbert space, and apply them sequentially in a full circuit-level simulation. These circuit simulations do not include additional dephasing from many-body interactions, which we assume can be engineered to be sufficiently small~\cite{li2025cascaded}. The circuit and cavity parameters used in these simulations, which set the relevant gate time scales, are summarized in Tables~I and~II of S8.

We assume readily achievable thermal populations ($n_{\mathrm{th}}\sim 0.5\%$), coherence times for the transmon ($T_1, T_2 = 200, 50~\mu\mathrm{s}$) and tunable coupler ($T_1, T_2 = 50, 4.2~\mu\mathrm{s}$), and two cavity-coherence regimes, with mode lifetimes of $1.5~\mathrm{ms}$ achievable in aluminum cavities~\cite{reagor_quantum_2016, chakram_seamless_2021} and $10~\mathrm{ms}$ in niobium cavities~\cite{milul_superconducting_2023, kim2025}. At these coherence levels, beam-splitter gate fidelities exceed $0.999$ and can be achieved in $250$~ns, as has been demonstrated experimentally~\cite{chapman_high--off-ratio_2023, lu2023high, maiti2025linear}. We simulate circuits with $L=4$ cavity modes, $S=8$ scrambling layers, and $M=8$ monitored layers with random parity measurements. We model the modes as harmonic oscillators with a truncated local Hilbert-space dimension of $L/2+1=3$ to include all possible local configurations ($0$, $1$, or $2$ photons). The total dimension of the doubled Hilbert space for four modes, a reference qubit, and a transmon qubit is $\approx 10^5$, highlighting the substantial computational complexity of simulating the full open-system dynamics.

Using this noise model, we simulate beam-splitter circuits with and without on-site Hubbard gates during the monitored portion, and we identify the dominant error sources by selectively enabling individual noise channels. We verify that, before the noise channels are turned on, the numerical results for circuits with $L=4$ modes match the ideal simulation data generated in the same way as the $S_R$ curves in Fig.~\ref{fig:numerics}. In all cases, noise produces a finite residual entropy that lifts the saturation value of the ideal entropy curves at large measurement rates. For monitored circuits containing on-site Hubbard gates (implemented via SNAP), we find that cavity $T_1$ decay and imperfections in the BSFP and on-site Hubbard gates dominate the residual entropy. At lower cavity lifetimes ($T_1=1.5$~ms), idle decay of spectator modes and the reference mode during sequential gate applications is the leading error source; increasing the lifetime to $10$~ms suppresses this contribution, leaving BSFP and on-site Hubbard gate errors as the dominant source of residual entropy, as shown in Fig.~\ref{fig:error_budget}(b). Unlike the RAQM architecture, the linear cavity array discussed above supports parallel gate execution and can therefore mitigate this idling-$T_1$ contribution by reducing the time modes spend inactive between operations. Beam-splitter circuits without on-site Hubbard gates exhibit the same qualitative trends but with smaller residual entropies due to shorter experimental wall times, since they do not require SNAP gates. Additional details on the error budgets for all monitored circuits considered are provided in S8.5. We note that the qualitative behavior of the ideal dynamics can be approximately recovered by subtracting the noise-induced residual, as shown in the inset of Fig.~\ref{fig:error_budget}(a), similar to what has been done in qubit realizations of MIPTs~\cite{koh_measurement-induced_2023}.

{\bf Discussion}:
We have theoretically investigated monitored dynamics in bosonic systems and outlined feasible experimental protocols for probing MIPTs on current 3D circuit QED platforms.
For small system sizes, Lindbladian simulations incorporating decay, thermal excitation, and dephasing errors on both the cavity modes and the coupler and ancilla circuits indicate that the experiment is within reach of existing hardware~\cite{li2025cascaded}. Noise induces a excess residual entanglement entropy of the reference mode at high measurement rates, which can be subtracted off~\cite{koh_measurement-induced_2023}, allowing for the recovery of the underlying ideal behavior. Although our simulations do not include errors arising from state preparation or ancilla tomography, these errors are not expected to qualitatively alter the transition, but rather increase statistical overhead and experimental runtime. Ancilla tomography errors may be mitigated through error-detectable encoding schemes~\cite{rosenblum2018fault}.

By embedding existing schemes for high-fidelity gate operations driven by superconducting circuits ~\cite{landgraf2024fastquantumcontrolcavities, chapman_high--off-ratio_2023, lu2023high} into a multimode architecture with tens of high-coherence bosonic modes, this proposal enables access to a regime of monitored dynamics that is inaccessible on qubit-based platforms. This novelty arises from the bosonic statistics of the cavity modes, which permit bosons to bunch in individual modes and allow for the projective strength of measurements to be tuned from parity to full number resolution. As demonstrated in our results, this flexibility alters the nature of the bosonic MIPT. In the case of parity measurements, our numerical results indicate that there are standard purification transitions for most types of unitary gates, while dynamics generated from only BSFP gates exhibit a new phase at high measurement rates in which the ancilla purifies at timescales that scale as the system size. We identify the effect that choosing more scrambling gates enhances the purification process, an effect that has no analogue in qubit systems.  Furthermore, for generic, non-BSFP gates, we numerically observe an apparent co-existence between a phase in which the ancilla purifies on $\mathcal{O}(1)$ timescales and a phase in which the bipartite entanglement seems to scale logarithmically with system size, which raises a host of interesting questions.

Other interesting questions include how the filling fraction affects the nature (or existence) of the transition and whether ancilla purification and bipartite entanglement diagnose distinct phases for generic gates at high measurement rates.
As demonstrated in studies of both interacting and free fermions, numerically identifying transitions from finite-size data can be subtle~\cite{Poboiko_2025, PhysRevResearch.6.043246, poboiko2023theory, Guo_2025, cao2019entanglement, Fava_2023}, underscoring the need for complementary analytical understanding of monitored bosonic circuits.
Elucidating the mechanisms underlying bosonic MIPTs and their broader implications is an important direction for future research.
For example, the existence of a MIPT under beam-splitter dynamics with random phases (see S1.6)\textemdash closely related to boson-sampling dynamics and number-conserving gate operations in arrays of dual-rail qubits\textemdash motivates exploring possible connections between the observed transition and (i) the classical simulability of monitored beam-splitter dynamics, e.g.\ via the computational cost of sampling from the conditional output distribution, and (ii) fault-tolerance thresholds for dual-rail erasure codes.

\textbf{Acknowledgments:}  C.M. thanks Kabir Khanna for helpful conversations and acknowledges support from NSF GRFP-1938059. S.P acknowledges support from NSF GRFP-2233066. J.H., T.D., and S.C were supported by the Army Research Office under Grant Number W911NF-23-1-0096 and W911NF-23-1-0251. This work was partially supported by the U.S. Department of Energy, Office of Science, National Quantum Information Science Research Centers, Superconducting Quantum Materials and Systems Center (SQMS), under Contract No. 89243024CSC000002. This work was partially supported by the Army Research Office Grant No.~W911NF-23-1-0144 ( J.H.P.), the US-ONR grant No.~N00014-23-1-2357 (S.P. and J.H.P.).

\bibliography{references}

\begin{thebibliography}{16}%
\makeatletter
\providecommand \@ifxundefined [1]{%
 \@ifx{#1\undefined}
}%
\providecommand \@ifnum [1]{%
 \ifnum #1\expandafter \@firstoftwo
 \else \expandafter \@secondoftwo
 \fi
}%
\providecommand \@ifx [1]{%
 \ifx #1\expandafter \@firstoftwo
 \else \expandafter \@secondoftwo
 \fi
}%
\providecommand \natexlab [1]{#1}%
\providecommand \enquote  [1]{``#1''}%
\providecommand \bibnamefont  [1]{#1}%
\providecommand \bibfnamefont [1]{#1}%
\providecommand \citenamefont [1]{#1}%
\providecommand \href@noop [0]{\@secondoftwo}%
\providecommand \href [0]{\begingroup \@sanitize@url \@href}%
\providecommand \@href[1]{\@@startlink{#1}\@@href}%
\providecommand \@@href[1]{\endgroup#1\@@endlink}%
\providecommand \@sanitize@url [0]{\catcode `\\12\catcode `\$12\catcode `\&12\catcode `\#12\catcode `\^12\catcode `\_12\catcode `\%12\relax}%
\providecommand \@@startlink[1]{}%
\providecommand \@@endlink[0]{}%
\providecommand \url  [0]{\begingroup\@sanitize@url \@url }%
\providecommand \@url [1]{\endgroup\@href {#1}{\urlprefix }}%
\providecommand \urlprefix  [0]{URL }%
\providecommand \Eprint [0]{\href }%
\providecommand \doibase [0]{https://doi.org/}%
\providecommand \selectlanguage [0]{\@gobble}%
\providecommand \bibinfo  [0]{\@secondoftwo}%
\providecommand \bibfield  [0]{\@secondoftwo}%
\providecommand \translation [1]{[#1]}%
\providecommand \BibitemOpen [0]{}%
\providecommand \bibitemStop [0]{}%
\providecommand \bibitemNoStop [0]{.\EOS\space}%
\providecommand \EOS [0]{\spacefactor3000\relax}%
\providecommand \BibitemShut  [1]{\csname bibitem#1\endcsname}%
\let\auto@bib@innerbib\@empty
\bibitem [{\citenamefont {Huang}\ \emph {et~al.}(2025)\citenamefont {Huang}, \citenamefont {DiNapoli}, \citenamefont {Rockwood}, \citenamefont {Yuan}, \citenamefont {Narasimhan}, \citenamefont {Gupta}, \citenamefont {Bal}, \citenamefont {Crisa}, \citenamefont {Garattoni}, \citenamefont {Lu}, \citenamefont {Jiang},\ and\ \citenamefont {Chakram}}]{huang2025fastsidebandcontrolweakly}%
  \BibitemOpen
  \bibfield  {author} {\bibinfo {author} {\bibfnamefont {J.}~\bibnamefont {Huang}}, \bibinfo {author} {\bibfnamefont {T.~J.}\ \bibnamefont {DiNapoli}}, \bibinfo {author} {\bibfnamefont {G.}~\bibnamefont {Rockwood}}, \bibinfo {author} {\bibfnamefont {M.}~\bibnamefont {Yuan}}, \bibinfo {author} {\bibfnamefont {P.}~\bibnamefont {Narasimhan}}, \bibinfo {author} {\bibfnamefont {E.}~\bibnamefont {Gupta}}, \bibinfo {author} {\bibfnamefont {M.}~\bibnamefont {Bal}}, \bibinfo {author} {\bibfnamefont {F.}~\bibnamefont {Crisa}}, \bibinfo {author} {\bibfnamefont {S.}~\bibnamefont {Garattoni}}, \bibinfo {author} {\bibfnamefont {Y.}~\bibnamefont {Lu}}, \bibinfo {author} {\bibfnamefont {L.}~\bibnamefont {Jiang}},\ and\ \bibinfo {author} {\bibfnamefont {S.}~\bibnamefont {Chakram}},\ }\href {https://arxiv.org/abs/2503.10623} {\bibinfo {title} {Fast sideband control of a weakly coupled multimode bosonic memory}} (\bibinfo {year} {2025}),\ \Eprint {https://arxiv.org/abs/2503.10623} {arXiv:2503.10623 [quant-ph]} \BibitemShut
  {NoStop}%
\bibitem [{\citenamefont {Barratt}\ \emph {et~al.}(2022)\citenamefont {Barratt}, \citenamefont {Agrawal}, \citenamefont {Potter}, \citenamefont {Gopalakrishnan},\ and\ \citenamefont {Vasseur}}]{Barratt_2022}%
  \BibitemOpen
  \bibfield  {author} {\bibinfo {author} {\bibfnamefont {F.}~\bibnamefont {Barratt}}, \bibinfo {author} {\bibfnamefont {U.}~\bibnamefont {Agrawal}}, \bibinfo {author} {\bibfnamefont {A.~C.}\ \bibnamefont {Potter}}, \bibinfo {author} {\bibfnamefont {S.}~\bibnamefont {Gopalakrishnan}},\ and\ \bibinfo {author} {\bibfnamefont {R.}~\bibnamefont {Vasseur}},\ }\bibfield  {title} {\bibinfo {title} {Transitions in the learnability of global charges from local measurements},\ }\bibfield  {journal} {\bibinfo  {journal} {Physical Review Letters}\ }\textbf {\bibinfo {volume} {129}},\ \href {https://doi.org/10.1103/physrevlett.129.200602} {10.1103/physrevlett.129.200602} (\bibinfo {year} {2022})\BibitemShut {NoStop}%
\bibitem [{\citenamefont {Agrawal}\ \emph {et~al.}(2024)\citenamefont {Agrawal}, \citenamefont {Lopez-Piqueres}, \citenamefont {Vasseur}, \citenamefont {Gopalakrishnan},\ and\ \citenamefont {Potter}}]{Agrawal_2024}%
  \BibitemOpen
  \bibfield  {author} {\bibinfo {author} {\bibfnamefont {U.}~\bibnamefont {Agrawal}}, \bibinfo {author} {\bibfnamefont {J.}~\bibnamefont {Lopez-Piqueres}}, \bibinfo {author} {\bibfnamefont {R.}~\bibnamefont {Vasseur}}, \bibinfo {author} {\bibfnamefont {S.}~\bibnamefont {Gopalakrishnan}},\ and\ \bibinfo {author} {\bibfnamefont {A.~C.}\ \bibnamefont {Potter}},\ }\bibfield  {title} {\bibinfo {title} {Observing quantum measurement collapse as a learnability phase transition},\ }\bibfield  {journal} {\bibinfo  {journal} {Physical Review X}\ }\textbf {\bibinfo {volume} {14}},\ \href {https://doi.org/10.1103/physrevx.14.041012} {10.1103/physrevx.14.041012} (\bibinfo {year} {2024})\BibitemShut {NoStop}%
\bibitem [{\citenamefont {Singh}\ \emph {et~al.}(2025)\citenamefont {Singh}, \citenamefont {Vasseur}, \citenamefont {Potter},\ and\ \citenamefont {Gopalakrishnan}}]{singh2025}%
  \BibitemOpen
  \bibfield  {author} {\bibinfo {author} {\bibfnamefont {H.}~\bibnamefont {Singh}}, \bibinfo {author} {\bibfnamefont {R.}~\bibnamefont {Vasseur}}, \bibinfo {author} {\bibfnamefont {A.~C.}\ \bibnamefont {Potter}},\ and\ \bibinfo {author} {\bibfnamefont {S.}~\bibnamefont {Gopalakrishnan}},\ }\href {https://arxiv.org/abs/2503.10308} {\bibinfo {title} {Mixed-state learnability transitions in monitored noisy quantum dynamics}} (\bibinfo {year} {2025}),\ \Eprint {https://arxiv.org/abs/2503.10308} {arXiv:2503.10308 [quant-ph]} \BibitemShut {NoStop}%
\bibitem [{\citenamefont {Kim}\ \emph {et~al.}(2025)\citenamefont {Kim}, \citenamefont {Kumar}, \citenamefont {Zhou}, \citenamefont {Xu}, \citenamefont {Vasseur},\ and\ \citenamefont {Kim}}]{kim2025}%
  \BibitemOpen
  \bibfield  {author} {\bibinfo {author} {\bibfnamefont {H.}~\bibnamefont {Kim}}, \bibinfo {author} {\bibfnamefont {A.}~\bibnamefont {Kumar}}, \bibinfo {author} {\bibfnamefont {Y.}~\bibnamefont {Zhou}}, \bibinfo {author} {\bibfnamefont {Y.}~\bibnamefont {Xu}}, \bibinfo {author} {\bibfnamefont {R.}~\bibnamefont {Vasseur}},\ and\ \bibinfo {author} {\bibfnamefont {E.-A.}\ \bibnamefont {Kim}},\ }\href {https://arxiv.org/abs/2508.15895} {\bibinfo {title} {Learning measurement-induced phase transitions using attention}} (\bibinfo {year} {2025}),\ \Eprint {https://arxiv.org/abs/2508.15895} {arXiv:2508.15895 [quant-ph]} \BibitemShut {NoStop}%
\bibitem [{\citenamefont {Zabalo}\ \emph {et~al.}(2020)\citenamefont {Zabalo}, \citenamefont {Gullans}, \citenamefont {Wilson}, \citenamefont {Gopalakrishnan}, \citenamefont {Huse},\ and\ \citenamefont {Pixley}}]{zabalo2020critical}%
  \BibitemOpen
  \bibfield  {author} {\bibinfo {author} {\bibfnamefont {A.}~\bibnamefont {Zabalo}}, \bibinfo {author} {\bibfnamefont {M.~J.}\ \bibnamefont {Gullans}}, \bibinfo {author} {\bibfnamefont {J.~H.}\ \bibnamefont {Wilson}}, \bibinfo {author} {\bibfnamefont {S.}~\bibnamefont {Gopalakrishnan}}, \bibinfo {author} {\bibfnamefont {D.~A.}\ \bibnamefont {Huse}},\ and\ \bibinfo {author} {\bibfnamefont {J.}~\bibnamefont {Pixley}},\ }\bibfield  {title} {\bibinfo {title} {Critical properties of the measurement-induced transition in random quantum circuits},\ }\href@noop {} {\bibfield  {journal} {\bibinfo  {journal} {Physical Review B}\ }\textbf {\bibinfo {volume} {101}},\ \bibinfo {pages} {060301} (\bibinfo {year} {2020})}\BibitemShut {NoStop}%
\bibitem [{\citenamefont {Skinner}\ \emph {et~al.}(2019)\citenamefont {Skinner}, \citenamefont {Ruhman},\ and\ \citenamefont {Nahum}}]{Skinner_2019}%
  \BibitemOpen
  \bibfield  {author} {\bibinfo {author} {\bibfnamefont {B.}~\bibnamefont {Skinner}}, \bibinfo {author} {\bibfnamefont {J.}~\bibnamefont {Ruhman}},\ and\ \bibinfo {author} {\bibfnamefont {A.}~\bibnamefont {Nahum}},\ }\bibfield  {title} {\bibinfo {title} {Measurement-induced phase transitions in the dynamics of entanglement},\ }\bibfield  {journal} {\bibinfo  {journal} {Physical Review X}\ }\textbf {\bibinfo {volume} {9}},\ \href {https://doi.org/10.1103/physrevx.9.031009} {10.1103/physrevx.9.031009} (\bibinfo {year} {2019})\BibitemShut {NoStop}%
\bibitem [{\citenamefont {Jian}\ \emph {et~al.}(2020)\citenamefont {Jian}, \citenamefont {You}, \citenamefont {Vasseur},\ and\ \citenamefont {Ludwig}}]{Jian_2020}%
  \BibitemOpen
  \bibfield  {author} {\bibinfo {author} {\bibfnamefont {C.-M.}\ \bibnamefont {Jian}}, \bibinfo {author} {\bibfnamefont {Y.-Z.}\ \bibnamefont {You}}, \bibinfo {author} {\bibfnamefont {R.}~\bibnamefont {Vasseur}},\ and\ \bibinfo {author} {\bibfnamefont {A.~W.~W.}\ \bibnamefont {Ludwig}},\ }\bibfield  {title} {\bibinfo {title} {Measurement-induced criticality in random quantum circuits},\ }\bibfield  {journal} {\bibinfo  {journal} {Physical Review B}\ }\textbf {\bibinfo {volume} {101}},\ \href {https://doi.org/10.1103/physrevb.101.104302} {10.1103/physrevb.101.104302} (\bibinfo {year} {2020})\BibitemShut {NoStop}%
\bibitem [{\citenamefont {Chapman}\ \emph {et~al.}()\citenamefont {Chapman}, \citenamefont {De~Graaf}, \citenamefont {Xue}, \citenamefont {Zhang}, \citenamefont {Teoh}, \citenamefont {Curtis}, \citenamefont {Tsunoda}, \citenamefont {Eickbusch}, \citenamefont {Read}, \citenamefont {Koottandavida}, \citenamefont {Mundhada}, \citenamefont {Frunzio}, \citenamefont {Devoret}, \citenamefont {Girvin},\ and\ \citenamefont {Schoelkopf}}]{chapman_high--off-ratio_2023}%
  \BibitemOpen
  \bibfield  {author} {\bibinfo {author} {\bibfnamefont {B.~J.}\ \bibnamefont {Chapman}}, \bibinfo {author} {\bibfnamefont {S.~J.}\ \bibnamefont {De~Graaf}}, \bibinfo {author} {\bibfnamefont {S.~H.}\ \bibnamefont {Xue}}, \bibinfo {author} {\bibfnamefont {Y.}~\bibnamefont {Zhang}}, \bibinfo {author} {\bibfnamefont {J.}~\bibnamefont {Teoh}}, \bibinfo {author} {\bibfnamefont {J.~C.}\ \bibnamefont {Curtis}}, \bibinfo {author} {\bibfnamefont {T.}~\bibnamefont {Tsunoda}}, \bibinfo {author} {\bibfnamefont {A.}~\bibnamefont {Eickbusch}}, \bibinfo {author} {\bibfnamefont {A.~P.}\ \bibnamefont {Read}}, \bibinfo {author} {\bibfnamefont {A.}~\bibnamefont {Koottandavida}}, \bibinfo {author} {\bibfnamefont {S.~O.}\ \bibnamefont {Mundhada}}, \bibinfo {author} {\bibfnamefont {L.}~\bibnamefont {Frunzio}}, \bibinfo {author} {\bibfnamefont {M.}~\bibnamefont {Devoret}}, \bibinfo {author} {\bibfnamefont {S.}~\bibnamefont {Girvin}},\ and\ \bibinfo {author} {\bibfnamefont {R.}~\bibnamefont {Schoelkopf}},\ }\bibfield  {title}
  {\bibinfo {title} {High-on-off-ratio beam-splitter interaction for gates on bosonically encoded qubits},\ }\href {https://doi.org/10.1103/PRXQuantum.4.020355} {\ \textbf {\bibinfo {volume} {4}},\ \bibinfo {pages} {020355}}\BibitemShut {NoStop}%
\bibitem [{\citenamefont {Lu}\ \emph {et~al.}(2023)\citenamefont {Lu}, \citenamefont {Maiti}, \citenamefont {Garmon}, \citenamefont {Ganjam}, \citenamefont {Zhang}, \citenamefont {Claes}, \citenamefont {Frunzio}, \citenamefont {Girvin},\ and\ \citenamefont {Schoelkopf}}]{lu2023high}%
  \BibitemOpen
  \bibfield  {author} {\bibinfo {author} {\bibfnamefont {Y.}~\bibnamefont {Lu}}, \bibinfo {author} {\bibfnamefont {A.}~\bibnamefont {Maiti}}, \bibinfo {author} {\bibfnamefont {J.~W.}\ \bibnamefont {Garmon}}, \bibinfo {author} {\bibfnamefont {S.}~\bibnamefont {Ganjam}}, \bibinfo {author} {\bibfnamefont {Y.}~\bibnamefont {Zhang}}, \bibinfo {author} {\bibfnamefont {J.}~\bibnamefont {Claes}}, \bibinfo {author} {\bibfnamefont {L.}~\bibnamefont {Frunzio}}, \bibinfo {author} {\bibfnamefont {S.~M.}\ \bibnamefont {Girvin}},\ and\ \bibinfo {author} {\bibfnamefont {R.~J.}\ \bibnamefont {Schoelkopf}},\ }\bibfield  {title} {\bibinfo {title} {High-fidelity parametric beamsplitting with a parity-protected converter},\ }\href@noop {} {\bibfield  {journal} {\bibinfo  {journal} {nature communications}\ }\textbf {\bibinfo {volume} {14}},\ \bibinfo {pages} {5767} (\bibinfo {year} {2023})}\BibitemShut {NoStop}%
\bibitem [{\citenamefont {Frattini}\ \emph {et~al.}(2017)\citenamefont {Frattini}, \citenamefont {Vool}, \citenamefont {Shankar}, \citenamefont {Narla}, \citenamefont {Sliwa},\ and\ \citenamefont {Devoret}}]{frattini2017}%
  \BibitemOpen
  \bibfield  {author} {\bibinfo {author} {\bibfnamefont {N.}~\bibnamefont {Frattini}}, \bibinfo {author} {\bibfnamefont {U.}~\bibnamefont {Vool}}, \bibinfo {author} {\bibfnamefont {S.}~\bibnamefont {Shankar}}, \bibinfo {author} {\bibfnamefont {A.}~\bibnamefont {Narla}}, \bibinfo {author} {\bibfnamefont {K.}~\bibnamefont {Sliwa}},\ and\ \bibinfo {author} {\bibfnamefont {M.}~\bibnamefont {Devoret}},\ }\bibfield  {title} {\bibinfo {title} {3-wave mixing josephson dipole element},\ }\href@noop {} {\bibfield  {journal} {\bibinfo  {journal} {Applied Physics Letters}\ }\textbf {\bibinfo {volume} {110}},\ \bibinfo {pages} {222603} (\bibinfo {year} {2017})}\BibitemShut {NoStop}%
\bibitem [{\citenamefont {Wang}\ \emph {et~al.}(2020)\citenamefont {Wang}, \citenamefont {Curtis}, \citenamefont {Lester}, \citenamefont {Zhang}, \citenamefont {Gao}, \citenamefont {Freeze}, \citenamefont {Batista}, \citenamefont {Vaccaro}, \citenamefont {Chuang}, \citenamefont {Frunzio}, \citenamefont {Jiang}, \citenamefont {Girvin},\ and\ \citenamefont {Schoelkopf}}]{Wang_2020}%
  \BibitemOpen
  \bibfield  {author} {\bibinfo {author} {\bibfnamefont {C.~S.}\ \bibnamefont {Wang}}, \bibinfo {author} {\bibfnamefont {J.~C.}\ \bibnamefont {Curtis}}, \bibinfo {author} {\bibfnamefont {B.~J.}\ \bibnamefont {Lester}}, \bibinfo {author} {\bibfnamefont {Y.}~\bibnamefont {Zhang}}, \bibinfo {author} {\bibfnamefont {Y.~Y.}\ \bibnamefont {Gao}}, \bibinfo {author} {\bibfnamefont {J.}~\bibnamefont {Freeze}}, \bibinfo {author} {\bibfnamefont {V.~S.}\ \bibnamefont {Batista}}, \bibinfo {author} {\bibfnamefont {P.~H.}\ \bibnamefont {Vaccaro}}, \bibinfo {author} {\bibfnamefont {I.~L.}\ \bibnamefont {Chuang}}, \bibinfo {author} {\bibfnamefont {L.}~\bibnamefont {Frunzio}}, \bibinfo {author} {\bibfnamefont {L.}~\bibnamefont {Jiang}}, \bibinfo {author} {\bibfnamefont {S.}~\bibnamefont {Girvin}},\ and\ \bibinfo {author} {\bibfnamefont {R.~J.}\ \bibnamefont {Schoelkopf}},\ }\bibfield  {title} {\bibinfo {title} {Efficient multiphoton sampling of molecular vibronic spectra on a superconducting bosonic processor},\ }\bibfield
  {journal} {\bibinfo  {journal} {Physical Review X}\ }\textbf {\bibinfo {volume} {10}},\ \href {https://doi.org/10.1103/physrevx.10.021060} {10.1103/physrevx.10.021060} (\bibinfo {year} {2020})\BibitemShut {NoStop}%
\bibitem [{\citenamefont {Landgraf}\ \emph {et~al.}(2024)\citenamefont {Landgraf}, \citenamefont {Flühmann}, \citenamefont {Fösel}, \citenamefont {Marquardt},\ and\ \citenamefont {Schoelkopf}}]{landgraf2024fastquantumcontrolcavities}%
  \BibitemOpen
  \bibfield  {author} {\bibinfo {author} {\bibfnamefont {J.}~\bibnamefont {Landgraf}}, \bibinfo {author} {\bibfnamefont {C.}~\bibnamefont {Flühmann}}, \bibinfo {author} {\bibfnamefont {T.}~\bibnamefont {Fösel}}, \bibinfo {author} {\bibfnamefont {F.}~\bibnamefont {Marquardt}},\ and\ \bibinfo {author} {\bibfnamefont {R.~J.}\ \bibnamefont {Schoelkopf}},\ }\href {https://arxiv.org/abs/2310.10498} {\bibinfo {title} {Fast quantum control of cavities using an improved protocol without coherent errors}} (\bibinfo {year} {2024}),\ \Eprint {https://arxiv.org/abs/2310.10498} {arXiv:2310.10498 [quant-ph]} \BibitemShut {NoStop}%
\bibitem [{\citenamefont {Lambert}\ \emph {et~al.}(2026)\citenamefont {Lambert}, \citenamefont {Gigu{`e}re}, \citenamefont {Menczel}, \citenamefont {Li}, \citenamefont {Hopf}, \citenamefont {Su{'a}rez}, \citenamefont {Gali}, \citenamefont {Lishman}, \citenamefont {Gadhvi}, \citenamefont {Agarwal}, \citenamefont {Galicia}, \citenamefont {Shammah}, \citenamefont {Nation}, \citenamefont {Johansson}, \citenamefont {Ahmed}, \citenamefont {Cross}, \citenamefont {Pitchford},\ and\ \citenamefont {Nori}}]{qutip5}%
  \BibitemOpen
  \bibfield  {author} {\bibinfo {author} {\bibfnamefont {N.}~\bibnamefont {Lambert}}, \bibinfo {author} {\bibfnamefont {E.}~\bibnamefont {Gigu{`e}re}}, \bibinfo {author} {\bibfnamefont {P.}~\bibnamefont {Menczel}}, \bibinfo {author} {\bibfnamefont {B.}~\bibnamefont {Li}}, \bibinfo {author} {\bibfnamefont {P.}~\bibnamefont {Hopf}}, \bibinfo {author} {\bibfnamefont {G.}~\bibnamefont {Su{'a}rez}}, \bibinfo {author} {\bibfnamefont {M.}~\bibnamefont {Gali}}, \bibinfo {author} {\bibfnamefont {J.}~\bibnamefont {Lishman}}, \bibinfo {author} {\bibfnamefont {R.}~\bibnamefont {Gadhvi}}, \bibinfo {author} {\bibfnamefont {R.}~\bibnamefont {Agarwal}}, \bibinfo {author} {\bibfnamefont {A.}~\bibnamefont {Galicia}}, \bibinfo {author} {\bibfnamefont {N.}~\bibnamefont {Shammah}}, \bibinfo {author} {\bibfnamefont {P.}~\bibnamefont {Nation}}, \bibinfo {author} {\bibfnamefont {J.~R.}\ \bibnamefont {Johansson}}, \bibinfo {author} {\bibfnamefont {S.}~\bibnamefont {Ahmed}}, \bibinfo {author} {\bibfnamefont {S.}~\bibnamefont {Cross}},
  \bibinfo {author} {\bibfnamefont {A.}~\bibnamefont {Pitchford}},\ and\ \bibinfo {author} {\bibfnamefont {F.}~\bibnamefont {Nori}},\ }\bibfield  {title} {\bibinfo {title} {Qutip 5: The quantum toolbox in {Python}},\ }\href {https://doi.org/10.1016/j.physrep.2025.10.001} {\bibfield  {journal} {\bibinfo  {journal} {Physics Reports}\ }\textbf {\bibinfo {volume} {1153}},\ \bibinfo {pages} {1} (\bibinfo {year} {2026})}\BibitemShut {NoStop}%
\bibitem [{\citenamefont {Johansson}\ \emph {et~al.}(2013)\citenamefont {Johansson}, \citenamefont {Nation},\ and\ \citenamefont {Nori}}]{johansson2013qutip}%
  \BibitemOpen
  \bibfield  {author} {\bibinfo {author} {\bibfnamefont {J.}~\bibnamefont {Johansson}}, \bibinfo {author} {\bibfnamefont {P.}~\bibnamefont {Nation}},\ and\ \bibinfo {author} {\bibfnamefont {F.}~\bibnamefont {Nori}},\ }\bibfield  {title} {\bibinfo {title} {Qutip 2: A python framework for the dynamics of open quantum systems},\ }\href {https://doi.org/https://doi.org/10.1016/j.cpc.2012.11.019} {\bibfield  {journal} {\bibinfo  {journal} {Computer Physics Communications}\ }\textbf {\bibinfo {volume} {184}},\ \bibinfo {pages} {1234} (\bibinfo {year} {2013})}\BibitemShut {NoStop}%
\bibitem [{\citenamefont {Johansson}\ \emph {et~al.}(2012)\citenamefont {Johansson}, \citenamefont {Nation},\ and\ \citenamefont {Nori}}]{johansson2012qutip}%
  \BibitemOpen
  \bibfield  {author} {\bibinfo {author} {\bibfnamefont {J.}~\bibnamefont {Johansson}}, \bibinfo {author} {\bibfnamefont {P.}~\bibnamefont {Nation}},\ and\ \bibinfo {author} {\bibfnamefont {F.}~\bibnamefont {Nori}},\ }\bibfield  {title} {\bibinfo {title} {Qutip: An open-source python framework for the dynamics of open quantum systems},\ }\href {https://doi.org/https://doi.org/10.1016/j.cpc.2012.02.021} {\bibfield  {journal} {\bibinfo  {journal} {Computer Physics Communications}\ }\textbf {\bibinfo {volume} {183}},\ \bibinfo {pages} {1760} (\bibinfo {year} {2012})}\BibitemShut {NoStop}%
\end{thebibliography}%


\begin{thebibliography}{72}%
\makeatletter
\providecommand \@ifxundefined [1]{%
 \@ifx{#1\undefined}
}%
\providecommand \@ifnum [1]{%
 \ifnum #1\expandafter \@firstoftwo
 \else \expandafter \@secondoftwo
 \fi
}%
\providecommand \@ifx [1]{%
 \ifx #1\expandafter \@firstoftwo
 \else \expandafter \@secondoftwo
 \fi
}%
\providecommand \natexlab [1]{#1}%
\providecommand \enquote  [1]{``#1''}%
\providecommand \bibnamefont  [1]{#1}%
\providecommand \bibfnamefont [1]{#1}%
\providecommand \citenamefont [1]{#1}%
\providecommand \href@noop [0]{\@secondoftwo}%
\providecommand \href [0]{\begingroup \@sanitize@url \@href}%
\providecommand \@href[1]{\@@startlink{#1}\@@href}%
\providecommand \@@href[1]{\endgroup#1\@@endlink}%
\providecommand \@sanitize@url [0]{\catcode `\\12\catcode `\$12\catcode `\&12\catcode `\#12\catcode `\^12\catcode `\_12\catcode `\%12\relax}%
\providecommand \@@startlink[1]{}%
\providecommand \@@endlink[0]{}%
\providecommand \url  [0]{\begingroup\@sanitize@url \@url }%
\providecommand \@url [1]{\endgroup\@href {#1}{\urlprefix }}%
\providecommand \urlprefix  [0]{URL }%
\providecommand \Eprint [0]{\href }%
\providecommand \doibase [0]{http://dx.doi.org/}%
\providecommand \selectlanguage [0]{\@gobble}%
\providecommand \bibinfo  [0]{\@secondoftwo}%
\providecommand \bibfield  [0]{\@secondoftwo}%
\providecommand \translation [1]{[#1]}%
\providecommand \BibitemOpen [0]{}%
\providecommand \bibitemStop [0]{}%
\providecommand \bibitemNoStop [0]{.\EOS\space}%
\providecommand \EOS [0]{\spacefactor3000\relax}%
\providecommand \BibitemShut  [1]{\csname bibitem#1\endcsname}%
\let\auto@bib@innerbib\@empty
\bibitem [{\citenamefont {Cirac}\ and\ \citenamefont {Zoller}(1995)}]{PhysRevLett.74.4091}%
  \BibitemOpen
  \bibfield  {author} {\bibinfo {author} {\bibfnamefont {J.~I.}\ \bibnamefont {Cirac}}\ and\ \bibinfo {author} {\bibfnamefont {P.}~\bibnamefont {Zoller}},\ }\bibfield  {title} {\enquote {\bibinfo {title} {Quantum computations with cold trapped ions},}\ }\href {\doibase 10.1103/PhysRevLett.74.4091} {\bibfield  {journal} {\bibinfo  {journal} {Phys. Rev. Lett.}\ }\textbf {\bibinfo {volume} {74}},\ \bibinfo {pages} {4091--4094} (\bibinfo {year} {1995})}\BibitemShut {NoStop}%
\bibitem [{\citenamefont {Saffman}\ \emph {et~al.}(2010)\citenamefont {Saffman}, \citenamefont {Walker},\ and\ \citenamefont {Mølmer}}]{Saffman_2010}%
  \BibitemOpen
  \bibfield  {author} {\bibinfo {author} {\bibfnamefont {M.}~\bibnamefont {Saffman}}, \bibinfo {author} {\bibfnamefont {T.~G.}\ \bibnamefont {Walker}}, \ and\ \bibinfo {author} {\bibfnamefont {K.}~\bibnamefont {Mølmer}},\ }\bibfield  {title} {\enquote {\bibinfo {title} {Quantum information with rydberg atoms},}\ }\href {\doibase 10.1103/revmodphys.82.2313} {\bibfield  {journal} {\bibinfo  {journal} {Reviews of Modern Physics}\ }\textbf {\bibinfo {volume} {82}},\ \bibinfo {pages} {2313–2363} (\bibinfo {year} {2010})}\BibitemShut {NoStop}%
\bibitem [{\citenamefont {Arute}\ \emph {et~al.}(2019)\citenamefont {Arute}, \citenamefont {Arya}, \citenamefont {Babbush}, \citenamefont {Bacon}, \citenamefont {Bardin}, \citenamefont {Barends}, \citenamefont {Biswas}, \citenamefont {Boixo}, \citenamefont {Brandao}, \citenamefont {Buell} \emph {et~al.}}]{arute2019quantum}%
  \BibitemOpen
  \bibfield  {author} {\bibinfo {author} {\bibfnamefont {Frank}\ \bibnamefont {Arute}}, \bibinfo {author} {\bibfnamefont {Kunal}\ \bibnamefont {Arya}}, \bibinfo {author} {\bibfnamefont {Ryan}\ \bibnamefont {Babbush}}, \bibinfo {author} {\bibfnamefont {Dave}\ \bibnamefont {Bacon}}, \bibinfo {author} {\bibfnamefont {Joseph~C}\ \bibnamefont {Bardin}}, \bibinfo {author} {\bibfnamefont {Rami}\ \bibnamefont {Barends}}, \bibinfo {author} {\bibfnamefont {Rupak}\ \bibnamefont {Biswas}}, \bibinfo {author} {\bibfnamefont {Sergio}\ \bibnamefont {Boixo}}, \bibinfo {author} {\bibfnamefont {Fernando~GSL}\ \bibnamefont {Brandao}}, \bibinfo {author} {\bibfnamefont {David~A}\ \bibnamefont {Buell}},  \emph {et~al.},\ }\bibfield  {title} {\enquote {\bibinfo {title} {Quantum supremacy using a programmable superconducting processor},}\ }\href@noop {} {\bibfield  {journal} {\bibinfo  {journal} {Nature}\ }\textbf {\bibinfo {volume} {574}},\ \bibinfo {pages} {505--510} (\bibinfo {year} {2019})}\BibitemShut {NoStop}%
\bibitem [{\citenamefont {Noel}\ \emph {et~al.}(2022)\citenamefont {Noel}, \citenamefont {Niroula}, \citenamefont {Zhu}, \citenamefont {Risinger}, \citenamefont {Egan}, \citenamefont {Biswas}, \citenamefont {Cetina}, \citenamefont {Gorshkov}, \citenamefont {Gullans}, \citenamefont {Huse} \emph {et~al.}}]{noel2022measurement}%
  \BibitemOpen
  \bibfield  {author} {\bibinfo {author} {\bibfnamefont {Crystal}\ \bibnamefont {Noel}}, \bibinfo {author} {\bibfnamefont {Pradeep}\ \bibnamefont {Niroula}}, \bibinfo {author} {\bibfnamefont {Daiwei}\ \bibnamefont {Zhu}}, \bibinfo {author} {\bibfnamefont {Andrew}\ \bibnamefont {Risinger}}, \bibinfo {author} {\bibfnamefont {Laird}\ \bibnamefont {Egan}}, \bibinfo {author} {\bibfnamefont {Debopriyo}\ \bibnamefont {Biswas}}, \bibinfo {author} {\bibfnamefont {Marko}\ \bibnamefont {Cetina}}, \bibinfo {author} {\bibfnamefont {Alexey~V}\ \bibnamefont {Gorshkov}}, \bibinfo {author} {\bibfnamefont {Michael~J}\ \bibnamefont {Gullans}}, \bibinfo {author} {\bibfnamefont {David~A}\ \bibnamefont {Huse}},  \emph {et~al.},\ }\bibfield  {title} {\enquote {\bibinfo {title} {Measurement-induced quantum phases realized in a trapped-ion quantum computer},}\ }\href@noop {} {\bibfield  {journal} {\bibinfo  {journal} {Nature Physics}\ }\textbf {\bibinfo {volume} {18}},\ \bibinfo {pages} {760--764} (\bibinfo {year}
  {2022})}\BibitemShut {NoStop}%
\bibitem [{\citenamefont {{Google Quantum AI and Collaborators}}(2023)}]{google2023measurement}%
  \BibitemOpen
  \bibfield  {author} {\bibinfo {author} {\bibnamefont {{Google Quantum AI and Collaborators}}},\ }\bibfield  {title} {\enquote {\bibinfo {title} {Measurement-induced entanglement and teleportation on a noisy quantum processor},}\ }\href@noop {} {\bibfield  {journal} {\bibinfo  {journal} {Nature}\ }\textbf {\bibinfo {volume} {622}},\ \bibinfo {pages} {481--486} (\bibinfo {year} {2023})}\BibitemShut {NoStop}%
\bibitem [{\citenamefont {Ofek}\ \emph {et~al.}(2016)\citenamefont {Ofek}, \citenamefont {Petrenko}, \citenamefont {Heeres}, \citenamefont {Reinhold}, \citenamefont {Leghtas}, \citenamefont {Vlastakis}, \citenamefont {Liu}, \citenamefont {Frunzio}, \citenamefont {Girvin}, \citenamefont {Jiang} \emph {et~al.}}]{ofek2016extending}%
  \BibitemOpen
  \bibfield  {author} {\bibinfo {author} {\bibfnamefont {Nissim}\ \bibnamefont {Ofek}}, \bibinfo {author} {\bibfnamefont {Andrei}\ \bibnamefont {Petrenko}}, \bibinfo {author} {\bibfnamefont {Reinier}\ \bibnamefont {Heeres}}, \bibinfo {author} {\bibfnamefont {Philip}\ \bibnamefont {Reinhold}}, \bibinfo {author} {\bibfnamefont {Zaki}\ \bibnamefont {Leghtas}}, \bibinfo {author} {\bibfnamefont {Brian}\ \bibnamefont {Vlastakis}}, \bibinfo {author} {\bibfnamefont {Yehan}\ \bibnamefont {Liu}}, \bibinfo {author} {\bibfnamefont {Luigi}\ \bibnamefont {Frunzio}}, \bibinfo {author} {\bibfnamefont {SM}~\bibnamefont {Girvin}}, \bibinfo {author} {\bibfnamefont {L}~\bibnamefont {Jiang}},  \emph {et~al.},\ }\bibfield  {title} {\enquote {\bibinfo {title} {Extending the lifetime of a quantum bit with error correction in superconducting circuits},}\ }\href@noop {} {\bibfield  {journal} {\bibinfo  {journal} {Nature}\ }\textbf {\bibinfo {volume} {536}},\ \bibinfo {pages} {441--445} (\bibinfo {year} {2016})}\BibitemShut {NoStop}%
\bibitem [{\citenamefont {Hu}\ \emph {et~al.}(2019)\citenamefont {Hu}, \citenamefont {Ma}, \citenamefont {Cai}, \citenamefont {Mu}, \citenamefont {Xu}, \citenamefont {Wang}, \citenamefont {Wu}, \citenamefont {Wang}, \citenamefont {Song}, \citenamefont {Zou} \emph {et~al.}}]{hu2019quantum}%
  \BibitemOpen
  \bibfield  {author} {\bibinfo {author} {\bibfnamefont {Ling}\ \bibnamefont {Hu}}, \bibinfo {author} {\bibfnamefont {Yuwei}\ \bibnamefont {Ma}}, \bibinfo {author} {\bibfnamefont {Weizhou}\ \bibnamefont {Cai}}, \bibinfo {author} {\bibfnamefont {Xianghao}\ \bibnamefont {Mu}}, \bibinfo {author} {\bibfnamefont {Yuan}\ \bibnamefont {Xu}}, \bibinfo {author} {\bibfnamefont {Weiting}\ \bibnamefont {Wang}}, \bibinfo {author} {\bibfnamefont {Yukai}\ \bibnamefont {Wu}}, \bibinfo {author} {\bibfnamefont {Haiyan}\ \bibnamefont {Wang}}, \bibinfo {author} {\bibfnamefont {YP}~\bibnamefont {Song}}, \bibinfo {author} {\bibfnamefont {C-L}\ \bibnamefont {Zou}},  \emph {et~al.},\ }\bibfield  {title} {\enquote {\bibinfo {title} {Quantum error correction and universal gate set operation on a binomial bosonic logical qubit},}\ }\href@noop {} {\bibfield  {journal} {\bibinfo  {journal} {Nature Physics}\ }\textbf {\bibinfo {volume} {15}},\ \bibinfo {pages} {503--508} (\bibinfo {year} {2019})}\BibitemShut {NoStop}%
\bibitem [{\citenamefont {Campagne-Ibarcq}\ \emph {et~al.}(2020)\citenamefont {Campagne-Ibarcq}, \citenamefont {Eickbusch}, \citenamefont {Touzard}, \citenamefont {Zalys-Geller}, \citenamefont {Frattini}, \citenamefont {Sivak}, \citenamefont {Reinhold}, \citenamefont {Puri}, \citenamefont {Shankar}, \citenamefont {Schoelkopf} \emph {et~al.}}]{campagne2020quantum}%
  \BibitemOpen
  \bibfield  {author} {\bibinfo {author} {\bibfnamefont {Philippe}\ \bibnamefont {Campagne-Ibarcq}}, \bibinfo {author} {\bibfnamefont {Alec}\ \bibnamefont {Eickbusch}}, \bibinfo {author} {\bibfnamefont {Steven}\ \bibnamefont {Touzard}}, \bibinfo {author} {\bibfnamefont {Evan}\ \bibnamefont {Zalys-Geller}}, \bibinfo {author} {\bibfnamefont {Nicholas~E}\ \bibnamefont {Frattini}}, \bibinfo {author} {\bibfnamefont {Volodymyr~V}\ \bibnamefont {Sivak}}, \bibinfo {author} {\bibfnamefont {Philip}\ \bibnamefont {Reinhold}}, \bibinfo {author} {\bibfnamefont {Shruti}\ \bibnamefont {Puri}}, \bibinfo {author} {\bibfnamefont {Shyam}\ \bibnamefont {Shankar}}, \bibinfo {author} {\bibfnamefont {Robert~J}\ \bibnamefont {Schoelkopf}},  \emph {et~al.},\ }\bibfield  {title} {\enquote {\bibinfo {title} {Quantum error correction of a qubit encoded in grid states of an oscillator},}\ }\href@noop {} {\bibfield  {journal} {\bibinfo  {journal} {Nature}\ }\textbf {\bibinfo {volume} {584}},\ \bibinfo {pages} {368--372} (\bibinfo {year}
  {2020})}\BibitemShut {NoStop}%
\bibitem [{\citenamefont {Reagor}\ \emph {et~al.}()\citenamefont {Reagor}, \citenamefont {Pfaff}, \citenamefont {Axline}, \citenamefont {Heeres}, \citenamefont {Ofek}, \citenamefont {Sliwa}, \citenamefont {Holland}, \citenamefont {Wang}, \citenamefont {Blumoff}, \citenamefont {Chou}, \citenamefont {Hatridge}, \citenamefont {Frunzio}, \citenamefont {Devoret}, \citenamefont {Jiang},\ and\ \citenamefont {Schoelkopf}}]{reagor_quantum_2016}%
  \BibitemOpen
  \bibfield  {author} {\bibinfo {author} {\bibfnamefont {Matthew}\ \bibnamefont {Reagor}}, \bibinfo {author} {\bibfnamefont {Wolfgang}\ \bibnamefont {Pfaff}}, \bibinfo {author} {\bibfnamefont {Christopher}\ \bibnamefont {Axline}}, \bibinfo {author} {\bibfnamefont {Reinier~W.}\ \bibnamefont {Heeres}}, \bibinfo {author} {\bibfnamefont {Nissim}\ \bibnamefont {Ofek}}, \bibinfo {author} {\bibfnamefont {Katrina}\ \bibnamefont {Sliwa}}, \bibinfo {author} {\bibfnamefont {Eric}\ \bibnamefont {Holland}}, \bibinfo {author} {\bibfnamefont {Chen}\ \bibnamefont {Wang}}, \bibinfo {author} {\bibfnamefont {Jacob}\ \bibnamefont {Blumoff}}, \bibinfo {author} {\bibfnamefont {Kevin}\ \bibnamefont {Chou}}, \bibinfo {author} {\bibfnamefont {Michael~J.}\ \bibnamefont {Hatridge}}, \bibinfo {author} {\bibfnamefont {Luigi}\ \bibnamefont {Frunzio}}, \bibinfo {author} {\bibfnamefont {Michel~H.}\ \bibnamefont {Devoret}}, \bibinfo {author} {\bibfnamefont {Liang}\ \bibnamefont {Jiang}}, \ and\ \bibinfo {author} {\bibfnamefont {Robert~J.}\
  \bibnamefont {Schoelkopf}},\ }\bibfield  {title} {\enquote {\bibinfo {title} {Quantum memory with millisecond coherence in circuit {QED}},}\ }\href {\doibase 10.1103/PhysRevB.94.014506} {\ \textbf {\bibinfo {volume} {94}},\ \bibinfo {pages} {014506}},\ \bibinfo {note} {publisher: American Physical Society}\BibitemShut {NoStop}%
\bibitem [{\citenamefont {Milul}\ \emph {et~al.}()\citenamefont {Milul}, \citenamefont {Guttel}, \citenamefont {Goldblatt}, \citenamefont {Hazanov}, \citenamefont {Joshi}, \citenamefont {Chausovsky}, \citenamefont {Kahn}, \citenamefont {Çiftyürek}, \citenamefont {Lafont},\ and\ \citenamefont {Rosenblum}}]{milul_superconducting_2023}%
  \BibitemOpen
  \bibfield  {author} {\bibinfo {author} {\bibfnamefont {Ofir}\ \bibnamefont {Milul}}, \bibinfo {author} {\bibfnamefont {Barkay}\ \bibnamefont {Guttel}}, \bibinfo {author} {\bibfnamefont {Uri}\ \bibnamefont {Goldblatt}}, \bibinfo {author} {\bibfnamefont {Sergey}\ \bibnamefont {Hazanov}}, \bibinfo {author} {\bibfnamefont {Lalit~M.}\ \bibnamefont {Joshi}}, \bibinfo {author} {\bibfnamefont {Daniel}\ \bibnamefont {Chausovsky}}, \bibinfo {author} {\bibfnamefont {Nitzan}\ \bibnamefont {Kahn}}, \bibinfo {author} {\bibfnamefont {Engin}\ \bibnamefont {Çiftyürek}}, \bibinfo {author} {\bibfnamefont {Fabien}\ \bibnamefont {Lafont}}, \ and\ \bibinfo {author} {\bibfnamefont {Serge}\ \bibnamefont {Rosenblum}},\ }\bibfield  {title} {\enquote {\bibinfo {title} {Superconducting cavity qubit with tens of milliseconds single-photon coherence time},}\ }\href {\doibase 10.1103/PRXQuantum.4.030336} {\ \textbf {\bibinfo {volume} {4}},\ \bibinfo {pages} {030336}},\ \bibinfo {note} {publisher: American Physical Society}\BibitemShut
  {NoStop}%
\bibitem [{\citenamefont {Heeres}\ \emph {et~al.}(2017)\citenamefont {Heeres}, \citenamefont {Reinhold}, \citenamefont {Ofek}, \citenamefont {Frunzio}, \citenamefont {Jiang}, \citenamefont {Devoret},\ and\ \citenamefont {Schoelkopf}}]{heeres2017implementing}%
  \BibitemOpen
  \bibfield  {author} {\bibinfo {author} {\bibfnamefont {Reinier~W}\ \bibnamefont {Heeres}}, \bibinfo {author} {\bibfnamefont {Philip}\ \bibnamefont {Reinhold}}, \bibinfo {author} {\bibfnamefont {Nissim}\ \bibnamefont {Ofek}}, \bibinfo {author} {\bibfnamefont {Luigi}\ \bibnamefont {Frunzio}}, \bibinfo {author} {\bibfnamefont {Liang}\ \bibnamefont {Jiang}}, \bibinfo {author} {\bibfnamefont {Michel~H}\ \bibnamefont {Devoret}}, \ and\ \bibinfo {author} {\bibfnamefont {Robert~J}\ \bibnamefont {Schoelkopf}},\ }\bibfield  {title} {\enquote {\bibinfo {title} {Implementing a universal gate set on a logical qubit encoded in an oscillator},}\ }\href@noop {} {\bibfield  {journal} {\bibinfo  {journal} {Nature Communications}\ }\textbf {\bibinfo {volume} {8}},\ \bibinfo {pages} {94} (\bibinfo {year} {2017})}\BibitemShut {NoStop}%
\bibitem [{\citenamefont {Rosenblum}\ \emph {et~al.}(2018{\natexlab{a}})\citenamefont {Rosenblum}, \citenamefont {Gao}, \citenamefont {Reinhold}, \citenamefont {Wang}, \citenamefont {Axline}, \citenamefont {Frunzio}, \citenamefont {Girvin}, \citenamefont {Jiang}, \citenamefont {Mirrahimi}, \citenamefont {Devoret} \emph {et~al.}}]{rosenblum2018cnot}%
  \BibitemOpen
  \bibfield  {author} {\bibinfo {author} {\bibfnamefont {Serge}\ \bibnamefont {Rosenblum}}, \bibinfo {author} {\bibfnamefont {Yvonne~Y}\ \bibnamefont {Gao}}, \bibinfo {author} {\bibfnamefont {Philip}\ \bibnamefont {Reinhold}}, \bibinfo {author} {\bibfnamefont {Chen}\ \bibnamefont {Wang}}, \bibinfo {author} {\bibfnamefont {Christopher~J}\ \bibnamefont {Axline}}, \bibinfo {author} {\bibfnamefont {Luigi}\ \bibnamefont {Frunzio}}, \bibinfo {author} {\bibfnamefont {Steven~M}\ \bibnamefont {Girvin}}, \bibinfo {author} {\bibfnamefont {Liang}\ \bibnamefont {Jiang}}, \bibinfo {author} {\bibfnamefont {Mazyar}\ \bibnamefont {Mirrahimi}}, \bibinfo {author} {\bibfnamefont {Michel~H}\ \bibnamefont {Devoret}},  \emph {et~al.},\ }\bibfield  {title} {\enquote {\bibinfo {title} {A cnot gate between multiphoton qubits encoded in two cavities},}\ }\href@noop {} {\bibfield  {journal} {\bibinfo  {journal} {Nature communications}\ }\textbf {\bibinfo {volume} {9}},\ \bibinfo {pages} {652} (\bibinfo {year}
  {2018}{\natexlab{a}})}\BibitemShut {NoStop}%
\bibitem [{\citenamefont {Chakram}\ \emph {et~al.}({\natexlab{a}})\citenamefont {Chakram}, \citenamefont {Oriani}, \citenamefont {Naik}, \citenamefont {Dixit}, \citenamefont {He}, \citenamefont {Agrawal}, \citenamefont {Kwon},\ and\ \citenamefont {Schuster}}]{chakram_seamless_2021}%
  \BibitemOpen
  \bibfield  {author} {\bibinfo {author} {\bibfnamefont {Srivatsan}\ \bibnamefont {Chakram}}, \bibinfo {author} {\bibfnamefont {Andrew~E.}\ \bibnamefont {Oriani}}, \bibinfo {author} {\bibfnamefont {Ravi~K.}\ \bibnamefont {Naik}}, \bibinfo {author} {\bibfnamefont {Akash~V.}\ \bibnamefont {Dixit}}, \bibinfo {author} {\bibfnamefont {Kevin}\ \bibnamefont {He}}, \bibinfo {author} {\bibfnamefont {Ankur}\ \bibnamefont {Agrawal}}, \bibinfo {author} {\bibfnamefont {Hyeokshin}\ \bibnamefont {Kwon}}, \ and\ \bibinfo {author} {\bibfnamefont {David~I.}\ \bibnamefont {Schuster}},\ }\bibfield  {title} {\enquote {\bibinfo {title} {Seamless high- q microwave cavities for multimode circuit quantum electrodynamics},}\ }\href {\doibase 10.1103/PhysRevLett.127.107701} {\ \textbf {\bibinfo {volume} {127}},\ \bibinfo {pages} {107701} ({\natexlab{a}})}\BibitemShut {NoStop}%
\bibitem [{\citenamefont {Chakram}\ \emph {et~al.}({\natexlab{b}})\citenamefont {Chakram}, \citenamefont {He}, \citenamefont {Dixit}, \citenamefont {Oriani}, \citenamefont {Naik}, \citenamefont {Leung}, \citenamefont {Kwon}, \citenamefont {Ma}, \citenamefont {Jiang},\ and\ \citenamefont {Schuster}}]{chakram_multimode_2022}%
  \BibitemOpen
  \bibfield  {author} {\bibinfo {author} {\bibfnamefont {Srivatsan}\ \bibnamefont {Chakram}}, \bibinfo {author} {\bibfnamefont {Kevin}\ \bibnamefont {He}}, \bibinfo {author} {\bibfnamefont {Akash~V.}\ \bibnamefont {Dixit}}, \bibinfo {author} {\bibfnamefont {Andrew~E.}\ \bibnamefont {Oriani}}, \bibinfo {author} {\bibfnamefont {Ravi~K.}\ \bibnamefont {Naik}}, \bibinfo {author} {\bibfnamefont {Nelson}\ \bibnamefont {Leung}}, \bibinfo {author} {\bibfnamefont {Hyeokshin}\ \bibnamefont {Kwon}}, \bibinfo {author} {\bibfnamefont {Wen-Long}\ \bibnamefont {Ma}}, \bibinfo {author} {\bibfnamefont {Liang}\ \bibnamefont {Jiang}}, \ and\ \bibinfo {author} {\bibfnamefont {David~I.}\ \bibnamefont {Schuster}},\ }\bibfield  {title} {\enquote {\bibinfo {title} {Multimode photon blockade},}\ }\href {\doibase 10.1038/s41567-022-01630-y} {\ \textbf {\bibinfo {volume} {18}},\ \bibinfo {pages} {879--884} ({\natexlab{b}})},\ \bibinfo {note} {publisher: Nature Publishing Group}\BibitemShut {NoStop}%
\bibitem [{\citenamefont {Gao}\ \emph {et~al.}(2019)\citenamefont {Gao}, \citenamefont {Lester}, \citenamefont {Chou}, \citenamefont {Frunzio}, \citenamefont {Devoret}, \citenamefont {Jiang}, \citenamefont {Girvin},\ and\ \citenamefont {Schoelkopf}}]{gao2019entanglement}%
  \BibitemOpen
  \bibfield  {author} {\bibinfo {author} {\bibfnamefont {Yvonne~Y}\ \bibnamefont {Gao}}, \bibinfo {author} {\bibfnamefont {Brian~J}\ \bibnamefont {Lester}}, \bibinfo {author} {\bibfnamefont {Kevin~S}\ \bibnamefont {Chou}}, \bibinfo {author} {\bibfnamefont {Luigi}\ \bibnamefont {Frunzio}}, \bibinfo {author} {\bibfnamefont {Michel~H}\ \bibnamefont {Devoret}}, \bibinfo {author} {\bibfnamefont {Liang}\ \bibnamefont {Jiang}}, \bibinfo {author} {\bibfnamefont {SM}~\bibnamefont {Girvin}}, \ and\ \bibinfo {author} {\bibfnamefont {Robert~J}\ \bibnamefont {Schoelkopf}},\ }\bibfield  {title} {\enquote {\bibinfo {title} {Entanglement of bosonic modes through an engineered exchange interaction},}\ }\href@noop {} {\bibfield  {journal} {\bibinfo  {journal} {Nature}\ }\textbf {\bibinfo {volume} {566}},\ \bibinfo {pages} {509} (\bibinfo {year} {2019})}\BibitemShut {NoStop}%
\bibitem [{\citenamefont {Chapman}\ \emph {et~al.}()\citenamefont {Chapman}, \citenamefont {De~Graaf}, \citenamefont {Xue}, \citenamefont {Zhang}, \citenamefont {Teoh}, \citenamefont {Curtis}, \citenamefont {Tsunoda}, \citenamefont {Eickbusch}, \citenamefont {Read}, \citenamefont {Koottandavida}, \citenamefont {Mundhada}, \citenamefont {Frunzio}, \citenamefont {Devoret}, \citenamefont {Girvin},\ and\ \citenamefont {Schoelkopf}}]{chapman_high--off-ratio_2023}%
  \BibitemOpen
  \bibfield  {author} {\bibinfo {author} {\bibfnamefont {Benjamin~J.}\ \bibnamefont {Chapman}}, \bibinfo {author} {\bibfnamefont {Stijn~J.}\ \bibnamefont {De~Graaf}}, \bibinfo {author} {\bibfnamefont {Sophia~H.}\ \bibnamefont {Xue}}, \bibinfo {author} {\bibfnamefont {Yaxing}\ \bibnamefont {Zhang}}, \bibinfo {author} {\bibfnamefont {James}\ \bibnamefont {Teoh}}, \bibinfo {author} {\bibfnamefont {Jacob~C.}\ \bibnamefont {Curtis}}, \bibinfo {author} {\bibfnamefont {Takahiro}\ \bibnamefont {Tsunoda}}, \bibinfo {author} {\bibfnamefont {Alec}\ \bibnamefont {Eickbusch}}, \bibinfo {author} {\bibfnamefont {Alexander~P.}\ \bibnamefont {Read}}, \bibinfo {author} {\bibfnamefont {Akshay}\ \bibnamefont {Koottandavida}}, \bibinfo {author} {\bibfnamefont {Shantanu~O.}\ \bibnamefont {Mundhada}}, \bibinfo {author} {\bibfnamefont {Luigi}\ \bibnamefont {Frunzio}}, \bibinfo {author} {\bibfnamefont {M.H.}\ \bibnamefont {Devoret}}, \bibinfo {author} {\bibfnamefont {S.M.}\ \bibnamefont {Girvin}}, \ and\ \bibinfo {author}
  {\bibfnamefont {R.J.}\ \bibnamefont {Schoelkopf}},\ }\bibfield  {title} {\enquote {\bibinfo {title} {High-on-off-ratio beam-splitter interaction for gates on bosonically encoded qubits},}\ }\href {\doibase 10.1103/PRXQuantum.4.020355} {\ \textbf {\bibinfo {volume} {4}},\ \bibinfo {pages} {020355}}\BibitemShut {NoStop}%
\bibitem [{\citenamefont {Connolly}\ \emph {et~al.}(2025)\citenamefont {Connolly}, \citenamefont {Kurilovich}, \citenamefont {Kurilovich}, \citenamefont {B{\o}ttcher}, \citenamefont {Hazra}, \citenamefont {Dai}, \citenamefont {Ding}, \citenamefont {Joshi}, \citenamefont {Nho}, \citenamefont {Diamond} \emph {et~al.}}]{connolly2025full}%
  \BibitemOpen
  \bibfield  {author} {\bibinfo {author} {\bibfnamefont {Thomas}\ \bibnamefont {Connolly}}, \bibinfo {author} {\bibfnamefont {Pavel~D}\ \bibnamefont {Kurilovich}}, \bibinfo {author} {\bibfnamefont {Vladislav~D}\ \bibnamefont {Kurilovich}}, \bibinfo {author} {\bibfnamefont {Charlotte~GL}\ \bibnamefont {B{\o}ttcher}}, \bibinfo {author} {\bibfnamefont {Sumeru}\ \bibnamefont {Hazra}}, \bibinfo {author} {\bibfnamefont {Wei}\ \bibnamefont {Dai}}, \bibinfo {author} {\bibfnamefont {Andy~Z}\ \bibnamefont {Ding}}, \bibinfo {author} {\bibfnamefont {Vidul~R}\ \bibnamefont {Joshi}}, \bibinfo {author} {\bibfnamefont {Heekun}\ \bibnamefont {Nho}}, \bibinfo {author} {\bibfnamefont {Spencer}\ \bibnamefont {Diamond}},  \emph {et~al.},\ }\bibfield  {title} {\enquote {\bibinfo {title} {Full characterization of measurement-induced transitions of a superconducting qubit},}\ }\href@noop {} {\bibfield  {journal} {\bibinfo  {journal} {arXiv preprint arXiv:2506.05306}\ } (\bibinfo {year} {2025})}\BibitemShut {NoStop}%
\bibitem [{\citenamefont {Li}\ \emph {et~al.}(2018)\citenamefont {Li}, \citenamefont {Chen},\ and\ \citenamefont {Fisher}}]{Li_2018}%
  \BibitemOpen
  \bibfield  {author} {\bibinfo {author} {\bibfnamefont {Yaodong}\ \bibnamefont {Li}}, \bibinfo {author} {\bibfnamefont {Xiao}\ \bibnamefont {Chen}}, \ and\ \bibinfo {author} {\bibfnamefont {Matthew P.~A.}\ \bibnamefont {Fisher}},\ }\bibfield  {title} {\enquote {\bibinfo {title} {Quantum zeno effect and the many-body entanglement transition},}\ }\href {\doibase 10.1103/physrevb.98.205136} {\bibfield  {journal} {\bibinfo  {journal} {Physical Review B}\ }\textbf {\bibinfo {volume} {98}} (\bibinfo {year} {2018}),\ 10.1103/physrevb.98.205136}\BibitemShut {NoStop}%
\bibitem [{\citenamefont {Li}\ \emph {et~al.}(2019)\citenamefont {Li}, \citenamefont {Chen},\ and\ \citenamefont {Fisher}}]{li2019measurement}%
  \BibitemOpen
  \bibfield  {author} {\bibinfo {author} {\bibfnamefont {Yaodong}\ \bibnamefont {Li}}, \bibinfo {author} {\bibfnamefont {Xiao}\ \bibnamefont {Chen}}, \ and\ \bibinfo {author} {\bibfnamefont {Matthew~PA}\ \bibnamefont {Fisher}},\ }\bibfield  {title} {\enquote {\bibinfo {title} {Measurement-driven entanglement transition in hybrid quantum circuits},}\ }\href@noop {} {\bibfield  {journal} {\bibinfo  {journal} {Physical Review B}\ }\textbf {\bibinfo {volume} {100}},\ \bibinfo {pages} {134306} (\bibinfo {year} {2019})}\BibitemShut {NoStop}%
\bibitem [{\citenamefont {Skinner}\ \emph {et~al.}(2019)\citenamefont {Skinner}, \citenamefont {Ruhman},\ and\ \citenamefont {Nahum}}]{Skinner_2019}%
  \BibitemOpen
  \bibfield  {author} {\bibinfo {author} {\bibfnamefont {Brian}\ \bibnamefont {Skinner}}, \bibinfo {author} {\bibfnamefont {Jonathan}\ \bibnamefont {Ruhman}}, \ and\ \bibinfo {author} {\bibfnamefont {Adam}\ \bibnamefont {Nahum}},\ }\bibfield  {title} {\enquote {\bibinfo {title} {Measurement-induced phase transitions in the dynamics of entanglement},}\ }\href {\doibase 10.1103/physrevx.9.031009} {\bibfield  {journal} {\bibinfo  {journal} {Physical Review X}\ }\textbf {\bibinfo {volume} {9}} (\bibinfo {year} {2019}),\ 10.1103/physrevx.9.031009}\BibitemShut {NoStop}%
\bibitem [{\citenamefont {Potter}\ and\ \citenamefont {Vasseur}(2022)}]{Potter_2022}%
  \BibitemOpen
  \bibfield  {author} {\bibinfo {author} {\bibfnamefont {Andrew~C.}\ \bibnamefont {Potter}}\ and\ \bibinfo {author} {\bibfnamefont {Romain}\ \bibnamefont {Vasseur}},\ }\enquote {\bibinfo {title} {Entanglement dynamics in hybrid quantum circuits},}\ in\ \href {\doibase 10.1007/978-3-031-03998-0_9} {\emph {\bibinfo {booktitle} {Entanglement in Spin Chains}}}\ (\bibinfo  {publisher} {Springer International Publishing},\ \bibinfo {year} {2022})\ p.\ \bibinfo {pages} {211–249}\BibitemShut {NoStop}%
\bibitem [{\citenamefont {Fisher}\ \emph {et~al.}(2023)\citenamefont {Fisher}, \citenamefont {Khemani}, \citenamefont {Nahum},\ and\ \citenamefont {Vijay}}]{fisher2023random}%
  \BibitemOpen
  \bibfield  {author} {\bibinfo {author} {\bibfnamefont {Matthew~PA}\ \bibnamefont {Fisher}}, \bibinfo {author} {\bibfnamefont {Vedika}\ \bibnamefont {Khemani}}, \bibinfo {author} {\bibfnamefont {Adam}\ \bibnamefont {Nahum}}, \ and\ \bibinfo {author} {\bibfnamefont {Sagar}\ \bibnamefont {Vijay}},\ }\bibfield  {title} {\enquote {\bibinfo {title} {Random quantum circuits},}\ }\href@noop {} {\bibfield  {journal} {\bibinfo  {journal} {Annual Review of Condensed Matter Physics}\ }\textbf {\bibinfo {volume} {14}},\ \bibinfo {pages} {335--379} (\bibinfo {year} {2023})}\BibitemShut {NoStop}%
\bibitem [{\citenamefont {Chan}\ \emph {et~al.}(2019)\citenamefont {Chan}, \citenamefont {Nandkishore}, \citenamefont {Pretko},\ and\ \citenamefont {Smith}}]{Chan_2019}%
  \BibitemOpen
  \bibfield  {author} {\bibinfo {author} {\bibfnamefont {Amos}\ \bibnamefont {Chan}}, \bibinfo {author} {\bibfnamefont {Rahul~M.}\ \bibnamefont {Nandkishore}}, \bibinfo {author} {\bibfnamefont {Michael}\ \bibnamefont {Pretko}}, \ and\ \bibinfo {author} {\bibfnamefont {Graeme}\ \bibnamefont {Smith}},\ }\bibfield  {title} {\enquote {\bibinfo {title} {Unitary-projective entanglement dynamics},}\ }\href {\doibase 10.1103/physrevb.99.224307} {\bibfield  {journal} {\bibinfo  {journal} {Physical Review B}\ }\textbf {\bibinfo {volume} {99}} (\bibinfo {year} {2019}),\ 10.1103/physrevb.99.224307}\BibitemShut {NoStop}%
\bibitem [{\citenamefont {Gullans}\ and\ \citenamefont {Huse}(2020{\natexlab{a}})}]{gullansDynamical2020}%
  \BibitemOpen
  \bibfield  {author} {\bibinfo {author} {\bibfnamefont {Michael~J.}\ \bibnamefont {Gullans}}\ and\ \bibinfo {author} {\bibfnamefont {David~A.}\ \bibnamefont {Huse}},\ }\bibfield  {title} {\enquote {\bibinfo {title} {Dynamical purification phase transition induced by quantum measurements},}\ }\href {\doibase 10.1103/physrevx.10.041020} {\bibfield  {journal} {\bibinfo  {journal} {Physical Review X}\ }\textbf {\bibinfo {volume} {10}} (\bibinfo {year} {2020}{\natexlab{a}}),\ 10.1103/physrevx.10.041020}\BibitemShut {NoStop}%
\bibitem [{\citenamefont {Bao}\ \emph {et~al.}(2020)\citenamefont {Bao}, \citenamefont {Choi},\ and\ \citenamefont {Altman}}]{Bao_2020}%
  \BibitemOpen
  \bibfield  {author} {\bibinfo {author} {\bibfnamefont {Yimu}\ \bibnamefont {Bao}}, \bibinfo {author} {\bibfnamefont {Soonwon}\ \bibnamefont {Choi}}, \ and\ \bibinfo {author} {\bibfnamefont {Ehud}\ \bibnamefont {Altman}},\ }\bibfield  {title} {\enquote {\bibinfo {title} {Theory of the phase transition in random unitary circuits with measurements},}\ }\href {\doibase 10.1103/physrevb.101.104301} {\bibfield  {journal} {\bibinfo  {journal} {Physical Review B}\ }\textbf {\bibinfo {volume} {101}} (\bibinfo {year} {2020}),\ 10.1103/physrevb.101.104301}\BibitemShut {NoStop}%
\bibitem [{\citenamefont {Jian}\ \emph {et~al.}(2020)\citenamefont {Jian}, \citenamefont {You}, \citenamefont {Vasseur},\ and\ \citenamefont {Ludwig}}]{Jian_2020}%
  \BibitemOpen
  \bibfield  {author} {\bibinfo {author} {\bibfnamefont {Chao-Ming}\ \bibnamefont {Jian}}, \bibinfo {author} {\bibfnamefont {Yi-Zhuang}\ \bibnamefont {You}}, \bibinfo {author} {\bibfnamefont {Romain}\ \bibnamefont {Vasseur}}, \ and\ \bibinfo {author} {\bibfnamefont {Andreas W.~W.}\ \bibnamefont {Ludwig}},\ }\bibfield  {title} {\enquote {\bibinfo {title} {Measurement-induced criticality in random quantum circuits},}\ }\href {\doibase 10.1103/physrevb.101.104302} {\bibfield  {journal} {\bibinfo  {journal} {Physical Review B}\ }\textbf {\bibinfo {volume} {101}} (\bibinfo {year} {2020}),\ 10.1103/physrevb.101.104302}\BibitemShut {NoStop}%
\bibitem [{\citenamefont {Zabalo}\ \emph {et~al.}(2020)\citenamefont {Zabalo}, \citenamefont {Gullans}, \citenamefont {Wilson}, \citenamefont {Gopalakrishnan}, \citenamefont {Huse},\ and\ \citenamefont {Pixley}}]{zabalo2020critical}%
  \BibitemOpen
  \bibfield  {author} {\bibinfo {author} {\bibfnamefont {Aidan}\ \bibnamefont {Zabalo}}, \bibinfo {author} {\bibfnamefont {Michael~J}\ \bibnamefont {Gullans}}, \bibinfo {author} {\bibfnamefont {Justin~H}\ \bibnamefont {Wilson}}, \bibinfo {author} {\bibfnamefont {Sarang}\ \bibnamefont {Gopalakrishnan}}, \bibinfo {author} {\bibfnamefont {David~A}\ \bibnamefont {Huse}}, \ and\ \bibinfo {author} {\bibfnamefont {JH}~\bibnamefont {Pixley}},\ }\bibfield  {title} {\enquote {\bibinfo {title} {Critical properties of the measurement-induced transition in random quantum circuits},}\ }\href@noop {} {\bibfield  {journal} {\bibinfo  {journal} {Physical Review B}\ }\textbf {\bibinfo {volume} {101}},\ \bibinfo {pages} {060301} (\bibinfo {year} {2020})}\BibitemShut {NoStop}%
\bibitem [{\citenamefont {Gullans}\ and\ \citenamefont {Huse}(2020{\natexlab{b}})}]{gullans2020}%
  \BibitemOpen
  \bibfield  {author} {\bibinfo {author} {\bibfnamefont {Michael~J.}\ \bibnamefont {Gullans}}\ and\ \bibinfo {author} {\bibfnamefont {David~A.}\ \bibnamefont {Huse}},\ }\bibfield  {title} {\enquote {\bibinfo {title} {Scalable probes of measurement-induced criticality},}\ }\href {\doibase 10.1103/PhysRevLett.125.070606} {\bibfield  {journal} {\bibinfo  {journal} {Phys. Rev. Lett.}\ }\textbf {\bibinfo {volume} {125}},\ \bibinfo {pages} {070606} (\bibinfo {year} {2020}{\natexlab{b}})}\BibitemShut {NoStop}%
\bibitem [{\citenamefont {Zabalo}\ \emph {et~al.}(2022)\citenamefont {Zabalo}, \citenamefont {Gullans}, \citenamefont {Wilson}, \citenamefont {Vasseur}, \citenamefont {Ludwig}, \citenamefont {Gopalakrishnan}, \citenamefont {Huse},\ and\ \citenamefont {Pixley}}]{Zabalo-2022}%
  \BibitemOpen
  \bibfield  {author} {\bibinfo {author} {\bibfnamefont {A.}~\bibnamefont {Zabalo}}, \bibinfo {author} {\bibfnamefont {M.~J.}\ \bibnamefont {Gullans}}, \bibinfo {author} {\bibfnamefont {J.~H.}\ \bibnamefont {Wilson}}, \bibinfo {author} {\bibfnamefont {R.}~\bibnamefont {Vasseur}}, \bibinfo {author} {\bibfnamefont {A.~W.~W.}\ \bibnamefont {Ludwig}}, \bibinfo {author} {\bibfnamefont {S.}~\bibnamefont {Gopalakrishnan}}, \bibinfo {author} {\bibfnamefont {David~A.}\ \bibnamefont {Huse}}, \ and\ \bibinfo {author} {\bibfnamefont {J.~H.}\ \bibnamefont {Pixley}},\ }\bibfield  {title} {\enquote {\bibinfo {title} {Operator scaling dimensions and multifractality at measurement-induced transitions},}\ }\href {\doibase 10.1103/PhysRevLett.128.050602} {\bibfield  {journal} {\bibinfo  {journal} {Phys. Rev. Lett.}\ }\textbf {\bibinfo {volume} {128}},\ \bibinfo {pages} {050602} (\bibinfo {year} {2022})}\BibitemShut {NoStop}%
\bibitem [{\citenamefont {Koh}\ \emph {et~al.}()\citenamefont {Koh}, \citenamefont {Sun}, \citenamefont {Motta},\ and\ \citenamefont {Minnich}}]{koh_measurement-induced_2023}%
  \BibitemOpen
  \bibfield  {author} {\bibinfo {author} {\bibfnamefont {Jin~Ming}\ \bibnamefont {Koh}}, \bibinfo {author} {\bibfnamefont {Shi-Ning}\ \bibnamefont {Sun}}, \bibinfo {author} {\bibfnamefont {Mario}\ \bibnamefont {Motta}}, \ and\ \bibinfo {author} {\bibfnamefont {Austin~J.}\ \bibnamefont {Minnich}},\ }\bibfield  {title} {\enquote {\bibinfo {title} {Measurement-induced entanglement phase transition on a superconducting quantum processor with mid-circuit readout},}\ }\href {\doibase 10.1038/s41567-023-02076-6} {\ \textbf {\bibinfo {volume} {19}},\ \bibinfo {pages} {1314--1319}},\ \bibinfo {note} {number: 9 Publisher: Nature Publishing Group}\BibitemShut {NoStop}%
\bibitem [{\citenamefont {Kamakari}\ \emph {et~al.}(2025)\citenamefont {Kamakari}, \citenamefont {Sun}, \citenamefont {Li}, \citenamefont {Thio}, \citenamefont {Gujarati}, \citenamefont {Fisher}, \citenamefont {Motta},\ and\ \citenamefont {Minnich}}]{Kamakari-2025}%
  \BibitemOpen
  \bibfield  {author} {\bibinfo {author} {\bibfnamefont {Hirsh}\ \bibnamefont {Kamakari}}, \bibinfo {author} {\bibfnamefont {Jiace}\ \bibnamefont {Sun}}, \bibinfo {author} {\bibfnamefont {Yaodong}\ \bibnamefont {Li}}, \bibinfo {author} {\bibfnamefont {Jonathan~J.}\ \bibnamefont {Thio}}, \bibinfo {author} {\bibfnamefont {Tanvi~P.}\ \bibnamefont {Gujarati}}, \bibinfo {author} {\bibfnamefont {Matthew P.~A.}\ \bibnamefont {Fisher}}, \bibinfo {author} {\bibfnamefont {Mario}\ \bibnamefont {Motta}}, \ and\ \bibinfo {author} {\bibfnamefont {Austin~J.}\ \bibnamefont {Minnich}},\ }\bibfield  {title} {\enquote {\bibinfo {title} {Experimental demonstration of scalable cross-entropy benchmarking to detect measurement-induced phase transitions on a superconducting quantum processor},}\ }\href {\doibase 10.1103/PhysRevLett.134.120401} {\bibfield  {journal} {\bibinfo  {journal} {Phys. Rev. Lett.}\ }\textbf {\bibinfo {volume} {134}},\ \bibinfo {pages} {120401} (\bibinfo {year} {2025})}\BibitemShut {NoStop}%
\bibitem [{\citenamefont {Minoguchi}\ \emph {et~al.}(2022)\citenamefont {Minoguchi}, \citenamefont {Rabl},\ and\ \citenamefont {Buchhold}}]{Minoguchi_2022}%
  \BibitemOpen
  \bibfield  {author} {\bibinfo {author} {\bibfnamefont {Yuri}\ \bibnamefont {Minoguchi}}, \bibinfo {author} {\bibfnamefont {Peter}\ \bibnamefont {Rabl}}, \ and\ \bibinfo {author} {\bibfnamefont {Michael}\ \bibnamefont {Buchhold}},\ }\bibfield  {title} {\enquote {\bibinfo {title} {Continuous gaussian measurements of the free boson cft: A model for exactly solvable and detectable measurement-induced dynamics},}\ }\href {\doibase 10.21468/scipostphys.12.1.009} {\bibfield  {journal} {\bibinfo  {journal} {SciPost Physics}\ }\textbf {\bibinfo {volume} {12}} (\bibinfo {year} {2022}),\ 10.21468/scipostphys.12.1.009}\BibitemShut {NoStop}%
\bibitem [{\citenamefont {Yokomizo}\ and\ \citenamefont {Ashida}(2025)}]{Yokomizo_2025}%
  \BibitemOpen
  \bibfield  {author} {\bibinfo {author} {\bibfnamefont {Kazuki}\ \bibnamefont {Yokomizo}}\ and\ \bibinfo {author} {\bibfnamefont {Yuto}\ \bibnamefont {Ashida}},\ }\bibfield  {title} {\enquote {\bibinfo {title} {Measurement-induced phase transition in free bosons},}\ }\href {\doibase 10.1103/y5r3-tv78} {\bibfield  {journal} {\bibinfo  {journal} {Physical Review B}\ }\textbf {\bibinfo {volume} {111}} (\bibinfo {year} {2025}),\ 10.1103/y5r3-tv78}\BibitemShut {NoStop}%
\bibitem [{\citenamefont {Li}\ \emph {et~al.}(2025{\natexlab{a}})\citenamefont {Li}, \citenamefont {Delmonte},\ and\ \citenamefont {Fazio}}]{Li_2025}%
  \BibitemOpen
  \bibfield  {author} {\bibinfo {author} {\bibfnamefont {Zejian}\ \bibnamefont {Li}}, \bibinfo {author} {\bibfnamefont {Anna}\ \bibnamefont {Delmonte}}, \ and\ \bibinfo {author} {\bibfnamefont {Rosario}\ \bibnamefont {Fazio}},\ }\bibfield  {title} {\enquote {\bibinfo {title} {Emergent deterministic entanglement dynamics in monitored infinite-range bosonic systems},}\ }\href {\doibase 10.1103/bhst-127b} {\bibfield  {journal} {\bibinfo  {journal} {Physical Review B}\ }\textbf {\bibinfo {volume} {112}} (\bibinfo {year} {2025}{\natexlab{a}}),\ 10.1103/bhst-127b}\BibitemShut {NoStop}%
\bibitem [{\citenamefont {Zhou}\ and\ \citenamefont {Chen}(2021)}]{Tianciboson}%
  \BibitemOpen
  \bibfield  {author} {\bibinfo {author} {\bibfnamefont {Tianci}\ \bibnamefont {Zhou}}\ and\ \bibinfo {author} {\bibfnamefont {Xiao}\ \bibnamefont {Chen}},\ }\bibfield  {title} {\enquote {\bibinfo {title} {Nonunitary entanglement dynamics in continuous-variable systems},}\ }\href {\doibase 10.1103/PhysRevB.104.L180301} {\bibfield  {journal} {\bibinfo  {journal} {Phys. Rev. B}\ }\textbf {\bibinfo {volume} {104}},\ \bibinfo {pages} {L180301} (\bibinfo {year} {2021})}\BibitemShut {NoStop}%
\bibitem [{\citenamefont {Huang}\ \emph {et~al.}(2025)\citenamefont {Huang}, \citenamefont {DiNapoli}, \citenamefont {Rockwood}, \citenamefont {Yuan}, \citenamefont {Narasimhan}, \citenamefont {Gupta}, \citenamefont {Bal}, \citenamefont {Crisa}, \citenamefont {Garattoni}, \citenamefont {Lu}, \citenamefont {Jiang},\ and\ \citenamefont {Chakram}}]{huang2025fastsidebandcontrolweakly}%
  \BibitemOpen
  \bibfield  {author} {\bibinfo {author} {\bibfnamefont {Jordan}\ \bibnamefont {Huang}}, \bibinfo {author} {\bibfnamefont {Thomas~J.}\ \bibnamefont {DiNapoli}}, \bibinfo {author} {\bibfnamefont {Gavin}\ \bibnamefont {Rockwood}}, \bibinfo {author} {\bibfnamefont {Ming}\ \bibnamefont {Yuan}}, \bibinfo {author} {\bibfnamefont {Prathyankara}\ \bibnamefont {Narasimhan}}, \bibinfo {author} {\bibfnamefont {Eesh}\ \bibnamefont {Gupta}}, \bibinfo {author} {\bibfnamefont {Mustafa}\ \bibnamefont {Bal}}, \bibinfo {author} {\bibfnamefont {Francesco}\ \bibnamefont {Crisa}}, \bibinfo {author} {\bibfnamefont {Sabrina}\ \bibnamefont {Garattoni}}, \bibinfo {author} {\bibfnamefont {Yao}\ \bibnamefont {Lu}}, \bibinfo {author} {\bibfnamefont {Liang}\ \bibnamefont {Jiang}}, \ and\ \bibinfo {author} {\bibfnamefont {Srivatsan}\ \bibnamefont {Chakram}},\ }\href {https://arxiv.org/abs/2503.10623} {\enquote {\bibinfo {title} {Fast sideband control of a weakly coupled multimode bosonic memory},}\ } (\bibinfo {year} {2025}),\ \Eprint
  {http://arxiv.org/abs/2503.10623} {arXiv:2503.10623 [quant-ph]} \BibitemShut {NoStop}%
\bibitem [{\citenamefont {Li}\ \emph {et~al.}(2025{\natexlab{b}})\citenamefont {Li}, \citenamefont {Gupta}, \citenamefont {Zhao}, \citenamefont {Banerjee}, \citenamefont {Lu}, \citenamefont {Roy}, \citenamefont {Oriani}, \citenamefont {Vrajitoarea}, \citenamefont {Chakram},\ and\ \citenamefont {Schuster}}]{li2025cascaded}%
  \BibitemOpen
  \bibfield  {author} {\bibinfo {author} {\bibfnamefont {Ziqian}\ \bibnamefont {Li}}, \bibinfo {author} {\bibfnamefont {Eesh}\ \bibnamefont {Gupta}}, \bibinfo {author} {\bibfnamefont {Fang}\ \bibnamefont {Zhao}}, \bibinfo {author} {\bibfnamefont {Riju}\ \bibnamefont {Banerjee}}, \bibinfo {author} {\bibfnamefont {Yao}\ \bibnamefont {Lu}}, \bibinfo {author} {\bibfnamefont {Tanay}\ \bibnamefont {Roy}}, \bibinfo {author} {\bibfnamefont {Andrew}\ \bibnamefont {Oriani}}, \bibinfo {author} {\bibfnamefont {Andrei}\ \bibnamefont {Vrajitoarea}}, \bibinfo {author} {\bibfnamefont {Srivatsan}\ \bibnamefont {Chakram}}, \ and\ \bibinfo {author} {\bibfnamefont {David~I.}\ \bibnamefont {Schuster}},\ }\href {https://arxiv.org/abs/2503.13953} {\enquote {\bibinfo {title} {A cascaded random access quantum memory},}\ } (\bibinfo {year} {2025}{\natexlab{b}}),\ \Eprint {http://arxiv.org/abs/2503.13953} {arXiv:2503.13953 [quant-ph]} \BibitemShut {NoStop}%
\bibitem [{\citenamefont {Gao}\ \emph {et~al.}(2018)\citenamefont {Gao}, \citenamefont {Lester}, \citenamefont {Zhang}, \citenamefont {Wang}, \citenamefont {Rosenblum}, \citenamefont {Frunzio}, \citenamefont {Jiang}, \citenamefont {Girvin},\ and\ \citenamefont {Schoelkopf}}]{gao2018programmable}%
  \BibitemOpen
  \bibfield  {author} {\bibinfo {author} {\bibfnamefont {Yvonne~Y}\ \bibnamefont {Gao}}, \bibinfo {author} {\bibfnamefont {Brian~J}\ \bibnamefont {Lester}}, \bibinfo {author} {\bibfnamefont {Yaxing}\ \bibnamefont {Zhang}}, \bibinfo {author} {\bibfnamefont {Chen}\ \bibnamefont {Wang}}, \bibinfo {author} {\bibfnamefont {Serge}\ \bibnamefont {Rosenblum}}, \bibinfo {author} {\bibfnamefont {Luigi}\ \bibnamefont {Frunzio}}, \bibinfo {author} {\bibfnamefont {Liang}\ \bibnamefont {Jiang}}, \bibinfo {author} {\bibfnamefont {SM}~\bibnamefont {Girvin}}, \ and\ \bibinfo {author} {\bibfnamefont {Robert~J}\ \bibnamefont {Schoelkopf}},\ }\bibfield  {title} {\enquote {\bibinfo {title} {Programmable interference between two microwave quantum memories},}\ }\href@noop {} {\bibfield  {journal} {\bibinfo  {journal} {Physical Review X}\ }\textbf {\bibinfo {volume} {8}},\ \bibinfo {pages} {021073} (\bibinfo {year} {2018})}\BibitemShut {NoStop}%
\bibitem [{\citenamefont {Aaronson}\ and\ \citenamefont {Arkhipov}(2011)}]{aaronson2011computational}%
  \BibitemOpen
  \bibfield  {author} {\bibinfo {author} {\bibfnamefont {Scott}\ \bibnamefont {Aaronson}}\ and\ \bibinfo {author} {\bibfnamefont {Alex}\ \bibnamefont {Arkhipov}},\ }\bibfield  {title} {\enquote {\bibinfo {title} {The computational complexity of linear optics},}\ }in\ \href@noop {} {\emph {\bibinfo {booktitle} {Proceedings of the forty-third annual ACM symposium on Theory of computing}}}\ (\bibinfo {year} {2011})\ pp.\ \bibinfo {pages} {333--342}\BibitemShut {NoStop}%
\bibitem [{\citenamefont {Kudra}\ \emph {et~al.}(2020)\citenamefont {Kudra}, \citenamefont {Bizn{\'a}rov{\'a}}, \citenamefont {Fadavi~Roudsari}, \citenamefont {Burnett}, \citenamefont {Niepce}, \citenamefont {Gasparinetti}, \citenamefont {Wickman},\ and\ \citenamefont {Delsing}}]{kudra2020high}%
  \BibitemOpen
  \bibfield  {author} {\bibinfo {author} {\bibfnamefont {M}~\bibnamefont {Kudra}}, \bibinfo {author} {\bibfnamefont {J}~\bibnamefont {Bizn{\'a}rov{\'a}}}, \bibinfo {author} {\bibfnamefont {A}~\bibnamefont {Fadavi~Roudsari}}, \bibinfo {author} {\bibfnamefont {JJ}~\bibnamefont {Burnett}}, \bibinfo {author} {\bibfnamefont {D}~\bibnamefont {Niepce}}, \bibinfo {author} {\bibfnamefont {S}~\bibnamefont {Gasparinetti}}, \bibinfo {author} {\bibfnamefont {B}~\bibnamefont {Wickman}}, \ and\ \bibinfo {author} {\bibfnamefont {P}~\bibnamefont {Delsing}},\ }\bibfield  {title} {\enquote {\bibinfo {title} {High quality three-dimensional aluminum microwave cavities},}\ }\href@noop {} {\bibfield  {journal} {\bibinfo  {journal} {Applied Physics Letters}\ }\textbf {\bibinfo {volume} {117}} (\bibinfo {year} {2020})}\BibitemShut {NoStop}%
\bibitem [{\citenamefont {Romanenko}\ \emph {et~al.}(2020)\citenamefont {Romanenko}, \citenamefont {Pilipenko}, \citenamefont {Zorzetti}, \citenamefont {Frolov}, \citenamefont {Awida}, \citenamefont {Belomestnykh}, \citenamefont {Posen},\ and\ \citenamefont {Grassellino}}]{romanenko2020three}%
  \BibitemOpen
  \bibfield  {author} {\bibinfo {author} {\bibfnamefont {A}~\bibnamefont {Romanenko}}, \bibinfo {author} {\bibfnamefont {R}~\bibnamefont {Pilipenko}}, \bibinfo {author} {\bibfnamefont {S}~\bibnamefont {Zorzetti}}, \bibinfo {author} {\bibfnamefont {D}~\bibnamefont {Frolov}}, \bibinfo {author} {\bibfnamefont {M}~\bibnamefont {Awida}}, \bibinfo {author} {\bibfnamefont {S}~\bibnamefont {Belomestnykh}}, \bibinfo {author} {\bibfnamefont {S}~\bibnamefont {Posen}}, \ and\ \bibinfo {author} {\bibfnamefont {A}~\bibnamefont {Grassellino}},\ }\bibfield  {title} {\enquote {\bibinfo {title} {Three-dimensional superconducting resonators at t< 20 mk with photon lifetimes up to $\tau$= 2 s},}\ }\href@noop {} {\bibfield  {journal} {\bibinfo  {journal} {Physical Review Applied}\ }\textbf {\bibinfo {volume} {13}},\ \bibinfo {pages} {034032} (\bibinfo {year} {2020})}\BibitemShut {NoStop}%
\bibitem [{\citenamefont {Oriani}\ \emph {et~al.}(2024)\citenamefont {Oriani}, \citenamefont {Zhao}, \citenamefont {Roy}, \citenamefont {Anferov}, \citenamefont {He}, \citenamefont {Agrawal}, \citenamefont {Banerjee}, \citenamefont {Chakram},\ and\ \citenamefont {Schuster}}]{oriani2024niobium}%
  \BibitemOpen
  \bibfield  {author} {\bibinfo {author} {\bibfnamefont {Andrew~E}\ \bibnamefont {Oriani}}, \bibinfo {author} {\bibfnamefont {Fang}\ \bibnamefont {Zhao}}, \bibinfo {author} {\bibfnamefont {Tanay}\ \bibnamefont {Roy}}, \bibinfo {author} {\bibfnamefont {Alexander}\ \bibnamefont {Anferov}}, \bibinfo {author} {\bibfnamefont {Kevin}\ \bibnamefont {He}}, \bibinfo {author} {\bibfnamefont {Ankur}\ \bibnamefont {Agrawal}}, \bibinfo {author} {\bibfnamefont {Riju}\ \bibnamefont {Banerjee}}, \bibinfo {author} {\bibfnamefont {Srivatsan}\ \bibnamefont {Chakram}}, \ and\ \bibinfo {author} {\bibfnamefont {David~I}\ \bibnamefont {Schuster}},\ }\bibfield  {title} {\enquote {\bibinfo {title} {Niobium coaxial cavities with internal quality factors exceeding 1.5 billion for circuit quantum electrodynamics},}\ }\href@noop {} {\bibfield  {journal} {\bibinfo  {journal} {arXiv preprint arXiv:2403.00286}\ } (\bibinfo {year} {2024})}\BibitemShut {NoStop}%
\bibitem [{\citenamefont {Kim}\ \emph {et~al.}(2025)\citenamefont {Kim}, \citenamefont {Kumar}, \citenamefont {Zhou}, \citenamefont {Xu}, \citenamefont {Vasseur},\ and\ \citenamefont {Kim}}]{kim2025}%
  \BibitemOpen
  \bibfield  {author} {\bibinfo {author} {\bibfnamefont {Hyejin}\ \bibnamefont {Kim}}, \bibinfo {author} {\bibfnamefont {Abhishek}\ \bibnamefont {Kumar}}, \bibinfo {author} {\bibfnamefont {Yiqing}\ \bibnamefont {Zhou}}, \bibinfo {author} {\bibfnamefont {Yichen}\ \bibnamefont {Xu}}, \bibinfo {author} {\bibfnamefont {Romain}\ \bibnamefont {Vasseur}}, \ and\ \bibinfo {author} {\bibfnamefont {Eun-Ah}\ \bibnamefont {Kim}},\ }\href {https://arxiv.org/abs/2508.15895} {\enquote {\bibinfo {title} {Learning measurement-induced phase transitions using attention},}\ } (\bibinfo {year} {2025}),\ \Eprint {http://arxiv.org/abs/2508.15895} {arXiv:2508.15895 [quant-ph]} \BibitemShut {NoStop}%
\bibitem [{\citenamefont {Frattini}\ \emph {et~al.}(2017)\citenamefont {Frattini}, \citenamefont {Vool}, \citenamefont {Shankar}, \citenamefont {Narla}, \citenamefont {Sliwa},\ and\ \citenamefont {Devoret}}]{frattini2017}%
  \BibitemOpen
  \bibfield  {author} {\bibinfo {author} {\bibfnamefont {NE}~\bibnamefont {Frattini}}, \bibinfo {author} {\bibfnamefont {U}~\bibnamefont {Vool}}, \bibinfo {author} {\bibfnamefont {S}~\bibnamefont {Shankar}}, \bibinfo {author} {\bibfnamefont {A}~\bibnamefont {Narla}}, \bibinfo {author} {\bibfnamefont {KM}~\bibnamefont {Sliwa}}, \ and\ \bibinfo {author} {\bibfnamefont {MH}~\bibnamefont {Devoret}},\ }\bibfield  {title} {\enquote {\bibinfo {title} {3-wave mixing josephson dipole element},}\ }\href@noop {} {\bibfield  {journal} {\bibinfo  {journal} {Applied Physics Letters}\ }\textbf {\bibinfo {volume} {110}},\ \bibinfo {pages} {222603} (\bibinfo {year} {2017})}\BibitemShut {NoStop}%
\bibitem [{\citenamefont {Lu}\ \emph {et~al.}(2023)\citenamefont {Lu}, \citenamefont {Maiti}, \citenamefont {Garmon}, \citenamefont {Ganjam}, \citenamefont {Zhang}, \citenamefont {Claes}, \citenamefont {Frunzio}, \citenamefont {Girvin},\ and\ \citenamefont {Schoelkopf}}]{lu2023high}%
  \BibitemOpen
  \bibfield  {author} {\bibinfo {author} {\bibfnamefont {Yao}\ \bibnamefont {Lu}}, \bibinfo {author} {\bibfnamefont {Aniket}\ \bibnamefont {Maiti}}, \bibinfo {author} {\bibfnamefont {John~WO}\ \bibnamefont {Garmon}}, \bibinfo {author} {\bibfnamefont {Suhas}\ \bibnamefont {Ganjam}}, \bibinfo {author} {\bibfnamefont {Yaxing}\ \bibnamefont {Zhang}}, \bibinfo {author} {\bibfnamefont {Jahan}\ \bibnamefont {Claes}}, \bibinfo {author} {\bibfnamefont {Luigi}\ \bibnamefont {Frunzio}}, \bibinfo {author} {\bibfnamefont {Steven~M}\ \bibnamefont {Girvin}}, \ and\ \bibinfo {author} {\bibfnamefont {Robert~J}\ \bibnamefont {Schoelkopf}},\ }\bibfield  {title} {\enquote {\bibinfo {title} {High-fidelity parametric beamsplitting with a parity-protected converter},}\ }\href@noop {} {\bibfield  {journal} {\bibinfo  {journal} {nature communications}\ }\textbf {\bibinfo {volume} {14}},\ \bibinfo {pages} {5767} (\bibinfo {year} {2023})}\BibitemShut {NoStop}%
\bibitem [{\citenamefont {Maiti}\ \emph {et~al.}(2025)\citenamefont {Maiti}, \citenamefont {Garmon}, \citenamefont {Lu}, \citenamefont {Miano}, \citenamefont {Frunzio},\ and\ \citenamefont {Schoelkopf}}]{maiti2025linear}%
  \BibitemOpen
  \bibfield  {author} {\bibinfo {author} {\bibfnamefont {Aniket}\ \bibnamefont {Maiti}}, \bibinfo {author} {\bibfnamefont {John~WO}\ \bibnamefont {Garmon}}, \bibinfo {author} {\bibfnamefont {Yao}\ \bibnamefont {Lu}}, \bibinfo {author} {\bibfnamefont {Alessandro}\ \bibnamefont {Miano}}, \bibinfo {author} {\bibfnamefont {Luigi}\ \bibnamefont {Frunzio}}, \ and\ \bibinfo {author} {\bibfnamefont {Robert~J}\ \bibnamefont {Schoelkopf}},\ }\bibfield  {title} {\enquote {\bibinfo {title} {Linear quantum coupler for clean bosonic control},}\ }\href@noop {} {\bibfield  {journal} {\bibinfo  {journal} {PRX Quantum}\ }\textbf {\bibinfo {volume} {6}},\ \bibinfo {pages} {040326} (\bibinfo {year} {2025})}\BibitemShut {NoStop}%
\bibitem [{\citenamefont {Krastanov}\ \emph {et~al.}(2015)\citenamefont {Krastanov}, \citenamefont {Albert}, \citenamefont {Shen}, \citenamefont {Zou}, \citenamefont {Heeres}, \citenamefont {Vlastakis}, \citenamefont {Schoelkopf},\ and\ \citenamefont {Jiang}}]{krastanov2015universal}%
  \BibitemOpen
  \bibfield  {author} {\bibinfo {author} {\bibfnamefont {Stefan}\ \bibnamefont {Krastanov}}, \bibinfo {author} {\bibfnamefont {Victor~V}\ \bibnamefont {Albert}}, \bibinfo {author} {\bibfnamefont {Chao}\ \bibnamefont {Shen}}, \bibinfo {author} {\bibfnamefont {Chang-Ling}\ \bibnamefont {Zou}}, \bibinfo {author} {\bibfnamefont {Reinier~W}\ \bibnamefont {Heeres}}, \bibinfo {author} {\bibfnamefont {Brian}\ \bibnamefont {Vlastakis}}, \bibinfo {author} {\bibfnamefont {Robert~J}\ \bibnamefont {Schoelkopf}}, \ and\ \bibinfo {author} {\bibfnamefont {Liang}\ \bibnamefont {Jiang}},\ }\bibfield  {title} {\enquote {\bibinfo {title} {Universal control of an oscillator with dispersive coupling to a qubit},}\ }\href@noop {} {\bibfield  {journal} {\bibinfo  {journal} {Physical Review A}\ }\textbf {\bibinfo {volume} {92}},\ \bibinfo {pages} {040303} (\bibinfo {year} {2015})}\BibitemShut {NoStop}%
\bibitem [{\citenamefont {Heeres}\ \emph {et~al.}()\citenamefont {Heeres}, \citenamefont {Vlastakis}, \citenamefont {Holland}, \citenamefont {Krastanov}, \citenamefont {Albert}, \citenamefont {Frunzio}, \citenamefont {Jiang},\ and\ \citenamefont {Schoelkopf}}]{heeres_cavity_2015}%
  \BibitemOpen
  \bibfield  {author} {\bibinfo {author} {\bibfnamefont {Reinier~W.}\ \bibnamefont {Heeres}}, \bibinfo {author} {\bibfnamefont {Brian}\ \bibnamefont {Vlastakis}}, \bibinfo {author} {\bibfnamefont {Eric}\ \bibnamefont {Holland}}, \bibinfo {author} {\bibfnamefont {Stefan}\ \bibnamefont {Krastanov}}, \bibinfo {author} {\bibfnamefont {Victor~V.}\ \bibnamefont {Albert}}, \bibinfo {author} {\bibfnamefont {Luigi}\ \bibnamefont {Frunzio}}, \bibinfo {author} {\bibfnamefont {Liang}\ \bibnamefont {Jiang}}, \ and\ \bibinfo {author} {\bibfnamefont {Robert~J.}\ \bibnamefont {Schoelkopf}},\ }\bibfield  {title} {\enquote {\bibinfo {title} {Cavity state manipulation using photon-number selective phase gates},}\ }\href {\doibase 10.1103/PhysRevLett.115.137002} {\ \textbf {\bibinfo {volume} {115}},\ \bibinfo {pages} {137002}},\ \bibinfo {note} {publisher: American Physical Society}\BibitemShut {NoStop}%
\bibitem [{\citenamefont {Eickbusch}\ \emph {et~al.}(2022)\citenamefont {Eickbusch}, \citenamefont {Sivak}, \citenamefont {Ding}, \citenamefont {Elder}, \citenamefont {Jha}, \citenamefont {Venkatraman}, \citenamefont {Royer}, \citenamefont {Girvin}, \citenamefont {Schoelkopf},\ and\ \citenamefont {Devoret}}]{eickbusch2021fast}%
  \BibitemOpen
  \bibfield  {author} {\bibinfo {author} {\bibfnamefont {Alec}\ \bibnamefont {Eickbusch}}, \bibinfo {author} {\bibfnamefont {Volodymyr}\ \bibnamefont {Sivak}}, \bibinfo {author} {\bibfnamefont {Andy~Z}\ \bibnamefont {Ding}}, \bibinfo {author} {\bibfnamefont {Salvatore~S}\ \bibnamefont {Elder}}, \bibinfo {author} {\bibfnamefont {Shantanu~R}\ \bibnamefont {Jha}}, \bibinfo {author} {\bibfnamefont {Jayameenakshi}\ \bibnamefont {Venkatraman}}, \bibinfo {author} {\bibfnamefont {Baptiste}\ \bibnamefont {Royer}}, \bibinfo {author} {\bibfnamefont {Steven~M}\ \bibnamefont {Girvin}}, \bibinfo {author} {\bibfnamefont {Robert~J}\ \bibnamefont {Schoelkopf}}, \ and\ \bibinfo {author} {\bibfnamefont {Michel~H}\ \bibnamefont {Devoret}},\ }\bibfield  {title} {\enquote {\bibinfo {title} {Fast universal control of an oscillator with weak dispersive coupling to a qubit},}\ }\href@noop {} {\bibfield  {journal} {\bibinfo  {journal} {Nature Physics}\ }\textbf {\bibinfo {volume} {18}},\ \bibinfo {pages} {1464--1469} (\bibinfo {year}
  {2022})}\BibitemShut {NoStop}%
\bibitem [{\citenamefont {Schuster}\ \emph {et~al.}(2007)\citenamefont {Schuster}, \citenamefont {Houck}, \citenamefont {Schreier}, \citenamefont {Wallraff}, \citenamefont {Gambetta}, \citenamefont {Blais}, \citenamefont {Frunzio}, \citenamefont {Majer}, \citenamefont {Johnson}, \citenamefont {Devoret} \emph {et~al.}}]{schuster2007resolving}%
  \BibitemOpen
  \bibfield  {author} {\bibinfo {author} {\bibfnamefont {DI}~\bibnamefont {Schuster}}, \bibinfo {author} {\bibfnamefont {Andrew~Addison}\ \bibnamefont {Houck}}, \bibinfo {author} {\bibfnamefont {JA}~\bibnamefont {Schreier}}, \bibinfo {author} {\bibfnamefont {A}~\bibnamefont {Wallraff}}, \bibinfo {author} {\bibfnamefont {JM}~\bibnamefont {Gambetta}}, \bibinfo {author} {\bibfnamefont {A}~\bibnamefont {Blais}}, \bibinfo {author} {\bibfnamefont {L}~\bibnamefont {Frunzio}}, \bibinfo {author} {\bibfnamefont {J}~\bibnamefont {Majer}}, \bibinfo {author} {\bibfnamefont {B}~\bibnamefont {Johnson}}, \bibinfo {author} {\bibfnamefont {MH}~\bibnamefont {Devoret}},  \emph {et~al.},\ }\bibfield  {title} {\enquote {\bibinfo {title} {Resolving photon number states in a superconducting circuit},}\ }\href@noop {} {\bibfield  {journal} {\bibinfo  {journal} {Nature}\ }\textbf {\bibinfo {volume} {445}},\ \bibinfo {pages} {515--518} (\bibinfo {year} {2007})}\BibitemShut {NoStop}%
\bibitem [{\citenamefont {Landgraf}\ \emph {et~al.}(2024)\citenamefont {Landgraf}, \citenamefont {Flühmann}, \citenamefont {Fösel}, \citenamefont {Marquardt},\ and\ \citenamefont {Schoelkopf}}]{landgraf2024fastquantumcontrolcavities}%
  \BibitemOpen
  \bibfield  {author} {\bibinfo {author} {\bibfnamefont {Jonas}\ \bibnamefont {Landgraf}}, \bibinfo {author} {\bibfnamefont {Christa}\ \bibnamefont {Flühmann}}, \bibinfo {author} {\bibfnamefont {Thomas}\ \bibnamefont {Fösel}}, \bibinfo {author} {\bibfnamefont {Florian}\ \bibnamefont {Marquardt}}, \ and\ \bibinfo {author} {\bibfnamefont {Robert~J.}\ \bibnamefont {Schoelkopf}},\ }\href {https://arxiv.org/abs/2310.10498} {\enquote {\bibinfo {title} {Fast quantum control of cavities using an improved protocol without coherent errors},}\ } (\bibinfo {year} {2024}),\ \Eprint {http://arxiv.org/abs/2310.10498} {arXiv:2310.10498 [quant-ph]} \BibitemShut {NoStop}%
\bibitem [{\citenamefont {Sun}\ \emph {et~al.}(2013)\citenamefont {Sun}, \citenamefont {Petrenko}, \citenamefont {Leghtas}, \citenamefont {Vlastakis}, \citenamefont {Kirchmair}, \citenamefont {Sliwa}, \citenamefont {Narla}, \citenamefont {Hatridge}, \citenamefont {Shankar}, \citenamefont {Blumoff} \emph {et~al.}}]{sun2013tracking}%
  \BibitemOpen
  \bibfield  {author} {\bibinfo {author} {\bibfnamefont {Luyan}\ \bibnamefont {Sun}}, \bibinfo {author} {\bibfnamefont {Andrei}\ \bibnamefont {Petrenko}}, \bibinfo {author} {\bibfnamefont {Zaki}\ \bibnamefont {Leghtas}}, \bibinfo {author} {\bibfnamefont {Brian}\ \bibnamefont {Vlastakis}}, \bibinfo {author} {\bibfnamefont {Gerhard}\ \bibnamefont {Kirchmair}}, \bibinfo {author} {\bibfnamefont {KM}~\bibnamefont {Sliwa}}, \bibinfo {author} {\bibfnamefont {Aniruth}\ \bibnamefont {Narla}}, \bibinfo {author} {\bibfnamefont {Michael}\ \bibnamefont {Hatridge}}, \bibinfo {author} {\bibfnamefont {Shyam}\ \bibnamefont {Shankar}}, \bibinfo {author} {\bibfnamefont {Jacob}\ \bibnamefont {Blumoff}},  \emph {et~al.},\ }\bibfield  {title} {\enquote {\bibinfo {title} {Tracking photon jumps with repeated quantum non-demolition parity measurements},}\ }\href@noop {} {\bibfield  {journal} {\bibinfo  {journal} {arXiv preprint arXiv:1311.2534}\ } (\bibinfo {year} {2013})}\BibitemShut {NoStop}%
\bibitem [{\citenamefont {Wang}\ \emph {et~al.}(2020)\citenamefont {Wang}, \citenamefont {Curtis}, \citenamefont {Lester}, \citenamefont {Zhang}, \citenamefont {Gao}, \citenamefont {Freeze}, \citenamefont {Batista}, \citenamefont {Vaccaro}, \citenamefont {Chuang}, \citenamefont {Frunzio} \emph {et~al.}}]{wang2020efficient}%
  \BibitemOpen
  \bibfield  {author} {\bibinfo {author} {\bibfnamefont {Christopher~S}\ \bibnamefont {Wang}}, \bibinfo {author} {\bibfnamefont {Jacob~C}\ \bibnamefont {Curtis}}, \bibinfo {author} {\bibfnamefont {Brian~J}\ \bibnamefont {Lester}}, \bibinfo {author} {\bibfnamefont {Yaxing}\ \bibnamefont {Zhang}}, \bibinfo {author} {\bibfnamefont {Yvonne~Y}\ \bibnamefont {Gao}}, \bibinfo {author} {\bibfnamefont {Jessica}\ \bibnamefont {Freeze}}, \bibinfo {author} {\bibfnamefont {Victor~S}\ \bibnamefont {Batista}}, \bibinfo {author} {\bibfnamefont {Patrick~H}\ \bibnamefont {Vaccaro}}, \bibinfo {author} {\bibfnamefont {Isaac~L}\ \bibnamefont {Chuang}}, \bibinfo {author} {\bibfnamefont {Luigi}\ \bibnamefont {Frunzio}},  \emph {et~al.},\ }\bibfield  {title} {\enquote {\bibinfo {title} {Efficient multiphoton sampling of molecular vibronic spectra on a superconducting bosonic processor},}\ }\href@noop {} {\bibfield  {journal} {\bibinfo  {journal} {Physical Review X}\ }\textbf {\bibinfo {volume} {10}},\ \bibinfo {pages} {021060}
  (\bibinfo {year} {2020})}\BibitemShut {NoStop}%
\bibitem [{sup()}]{supp}%
  \BibitemOpen
  \href@noop {} {\bibinfo  {journal} {See Supplemental Material for additional details}\ }\BibitemShut {NoStop}%
\bibitem [{\citenamefont {Barratt}\ \emph {et~al.}(2022)\citenamefont {Barratt}, \citenamefont {Agrawal}, \citenamefont {Potter}, \citenamefont {Gopalakrishnan},\ and\ \citenamefont {Vasseur}}]{Barratt_2022}%
  \BibitemOpen
\bibfield  {journal} {  }\bibfield  {author} {\bibinfo {author} {\bibfnamefont {Fergus}\ \bibnamefont {Barratt}}, \bibinfo {author} {\bibfnamefont {Utkarsh}\ \bibnamefont {Agrawal}}, \bibinfo {author} {\bibfnamefont {Andrew~C.}\ \bibnamefont {Potter}}, \bibinfo {author} {\bibfnamefont {Sarang}\ \bibnamefont {Gopalakrishnan}}, \ and\ \bibinfo {author} {\bibfnamefont {Romain}\ \bibnamefont {Vasseur}},\ }\bibfield  {title} {\enquote {\bibinfo {title} {Transitions in the learnability of global charges from local measurements},}\ }\href {\doibase 10.1103/physrevlett.129.200602} {\bibfield  {journal} {\bibinfo  {journal} {Physical Review Letters}\ }\textbf {\bibinfo {volume} {129}} (\bibinfo {year} {2022}),\ 10.1103/physrevlett.129.200602}\BibitemShut {NoStop}%
\bibitem [{\citenamefont {Dehghani}\ \emph {et~al.}(2023)\citenamefont {Dehghani}, \citenamefont {Lavasani}, \citenamefont {Hafezi},\ and\ \citenamefont {Gullans}}]{Dehghani_2023}%
  \BibitemOpen
  \bibfield  {author} {\bibinfo {author} {\bibfnamefont {Hossein}\ \bibnamefont {Dehghani}}, \bibinfo {author} {\bibfnamefont {Ali}\ \bibnamefont {Lavasani}}, \bibinfo {author} {\bibfnamefont {Mohammad}\ \bibnamefont {Hafezi}}, \ and\ \bibinfo {author} {\bibfnamefont {Michael~J.}\ \bibnamefont {Gullans}},\ }\bibfield  {title} {\enquote {\bibinfo {title} {Neural-network decoders for measurement induced phase transitions},}\ }\href {\doibase 10.1038/s41467-023-37902-1} {\bibfield  {journal} {\bibinfo  {journal} {Nature Communications}\ }\textbf {\bibinfo {volume} {14}} (\bibinfo {year} {2023}),\ 10.1038/s41467-023-37902-1}\BibitemShut {NoStop}%
\bibitem [{\citenamefont {Akhtar}\ \emph {et~al.}(2023)\citenamefont {Akhtar}, \citenamefont {Hu},\ and\ \citenamefont {You}}]{akhtar2023}%
  \BibitemOpen
  \bibfield  {author} {\bibinfo {author} {\bibfnamefont {Ahmed~A.}\ \bibnamefont {Akhtar}}, \bibinfo {author} {\bibfnamefont {Hong-Ye}\ \bibnamefont {Hu}}, \ and\ \bibinfo {author} {\bibfnamefont {Yi-Zhuang}\ \bibnamefont {You}},\ }\href {https://arxiv.org/abs/2308.01653} {\enquote {\bibinfo {title} {Measurement-induced criticality is tomographically optimal},}\ } (\bibinfo {year} {2023}),\ \Eprint {http://arxiv.org/abs/2308.01653} {arXiv:2308.01653 [quant-ph]} \BibitemShut {NoStop}%
\bibitem [{\citenamefont {Li}\ \emph {et~al.}(2023)\citenamefont {Li}, \citenamefont {Zou}, \citenamefont {Glorioso}, \citenamefont {Altman},\ and\ \citenamefont {Fisher}}]{Li_2023}%
  \BibitemOpen
  \bibfield  {author} {\bibinfo {author} {\bibfnamefont {Yaodong}\ \bibnamefont {Li}}, \bibinfo {author} {\bibfnamefont {Yijian}\ \bibnamefont {Zou}}, \bibinfo {author} {\bibfnamefont {Paolo}\ \bibnamefont {Glorioso}}, \bibinfo {author} {\bibfnamefont {Ehud}\ \bibnamefont {Altman}}, \ and\ \bibinfo {author} {\bibfnamefont {Matthew~P. A.}\ \bibnamefont {Fisher}},\ }\bibfield  {title} {\enquote {\bibinfo {title} {Cross entropy benchmark for measurement-induced phase transitions},}\ }\href {\doibase 10.1103/physrevlett.130.220404} {\bibfield  {journal} {\bibinfo  {journal} {Physical Review Letters}\ }\textbf {\bibinfo {volume} {130}} (\bibinfo {year} {2023}),\ 10.1103/physrevlett.130.220404}\BibitemShut {NoStop}%
\bibitem [{\citenamefont {Ippoliti}\ and\ \citenamefont {Khemani}(2024)}]{Ippoliti_2024}%
  \BibitemOpen
  \bibfield  {author} {\bibinfo {author} {\bibfnamefont {Matteo}\ \bibnamefont {Ippoliti}}\ and\ \bibinfo {author} {\bibfnamefont {Vedika}\ \bibnamefont {Khemani}},\ }\bibfield  {title} {\enquote {\bibinfo {title} {Learnability transitions in monitored quantum dynamics via eavesdropper’s classical shadows},}\ }\href {\doibase 10.1103/prxquantum.5.020304} {\bibfield  {journal} {\bibinfo  {journal} {PRX Quantum}\ }\textbf {\bibinfo {volume} {5}} (\bibinfo {year} {2024}),\ 10.1103/prxquantum.5.020304}\BibitemShut {NoStop}%
\bibitem [{\citenamefont {Agrawal}\ \emph {et~al.}(2024)\citenamefont {Agrawal}, \citenamefont {Lopez-Piqueres}, \citenamefont {Vasseur}, \citenamefont {Gopalakrishnan},\ and\ \citenamefont {Potter}}]{Agrawal_2024}%
  \BibitemOpen
  \bibfield  {author} {\bibinfo {author} {\bibfnamefont {Utkarsh}\ \bibnamefont {Agrawal}}, \bibinfo {author} {\bibfnamefont {Javier}\ \bibnamefont {Lopez-Piqueres}}, \bibinfo {author} {\bibfnamefont {Romain}\ \bibnamefont {Vasseur}}, \bibinfo {author} {\bibfnamefont {Sarang}\ \bibnamefont {Gopalakrishnan}}, \ and\ \bibinfo {author} {\bibfnamefont {Andrew~C.}\ \bibnamefont {Potter}},\ }\bibfield  {title} {\enquote {\bibinfo {title} {Observing quantum measurement collapse as a learnability phase transition},}\ }\href {\doibase 10.1103/physrevx.14.041012} {\bibfield  {journal} {\bibinfo  {journal} {Physical Review X}\ }\textbf {\bibinfo {volume} {14}} (\bibinfo {year} {2024}),\ 10.1103/physrevx.14.041012}\BibitemShut {NoStop}%
\bibitem [{\citenamefont {McGinley}(2024)}]{McGinley_2024}%
  \BibitemOpen
  \bibfield  {author} {\bibinfo {author} {\bibfnamefont {Max}\ \bibnamefont {McGinley}},\ }\bibfield  {title} {\enquote {\bibinfo {title} {Postselection-free learning of measurement-induced quantum dynamics},}\ }\href {\doibase 10.1103/prxquantum.5.020347} {\bibfield  {journal} {\bibinfo  {journal} {PRX Quantum}\ }\textbf {\bibinfo {volume} {5}} (\bibinfo {year} {2024}),\ 10.1103/prxquantum.5.020347}\BibitemShut {NoStop}%
\bibitem [{\citenamefont {Poboiko}\ \emph {et~al.}(2023)\citenamefont {Poboiko}, \citenamefont {P{\"o}pperl}, \citenamefont {Gornyi},\ and\ \citenamefont {Mirlin}}]{poboiko2023theory}%
  \BibitemOpen
  \bibfield  {author} {\bibinfo {author} {\bibfnamefont {Igor}\ \bibnamefont {Poboiko}}, \bibinfo {author} {\bibfnamefont {Paul}\ \bibnamefont {P{\"o}pperl}}, \bibinfo {author} {\bibfnamefont {Igor~V}\ \bibnamefont {Gornyi}}, \ and\ \bibinfo {author} {\bibfnamefont {Alexander~D}\ \bibnamefont {Mirlin}},\ }\bibfield  {title} {\enquote {\bibinfo {title} {Theory of free fermions under random projective measurements},}\ }\href@noop {} {\bibfield  {journal} {\bibinfo  {journal} {Physical Review X}\ }\textbf {\bibinfo {volume} {13}},\ \bibinfo {pages} {041046} (\bibinfo {year} {2023})}\BibitemShut {NoStop}%
\bibitem [{\citenamefont {Chou}\ \emph {et~al.}(2024)\citenamefont {Chou}, \citenamefont {Shemma}, \citenamefont {McCarrick}, \citenamefont {Chien}, \citenamefont {Teoh}, \citenamefont {Winkel}, \citenamefont {Anderson}, \citenamefont {Chen}, \citenamefont {Curtis}, \citenamefont {de~Graaf}, \citenamefont {Garmon}, \citenamefont {Gudlewski}, \citenamefont {Kalfus}, \citenamefont {Keen}, \citenamefont {Khedkar}, \citenamefont {Lei}, \citenamefont {Liu}, \citenamefont {Lu}, \citenamefont {Lu}, \citenamefont {Maiti}, \citenamefont {Mastalli-Kelly}, \citenamefont {Mehta}, \citenamefont {Mundhada}, \citenamefont {Narla}, \citenamefont {Noh}, \citenamefont {Tsunoda}, \citenamefont {Xue}, \citenamefont {Yuan}, \citenamefont {Frunzio}, \citenamefont {Aumentado}, \citenamefont {Puri}, \citenamefont {Girvin}, \citenamefont {Moseley},\ and\ \citenamefont {Schoelkopf}}]{chou_superconducting_2024}%
  \BibitemOpen
  \bibfield  {author} {\bibinfo {author} {\bibfnamefont {Kevin~S.}\ \bibnamefont {Chou}}, \bibinfo {author} {\bibfnamefont {Tali}\ \bibnamefont {Shemma}}, \bibinfo {author} {\bibfnamefont {Heather}\ \bibnamefont {McCarrick}}, \bibinfo {author} {\bibfnamefont {Tzu-Chiao}\ \bibnamefont {Chien}}, \bibinfo {author} {\bibfnamefont {James~D.}\ \bibnamefont {Teoh}}, \bibinfo {author} {\bibfnamefont {Patrick}\ \bibnamefont {Winkel}}, \bibinfo {author} {\bibfnamefont {Amos}\ \bibnamefont {Anderson}}, \bibinfo {author} {\bibfnamefont {Jonathan}\ \bibnamefont {Chen}}, \bibinfo {author} {\bibfnamefont {Jacob~C.}\ \bibnamefont {Curtis}}, \bibinfo {author} {\bibfnamefont {Stijn~J.}\ \bibnamefont {de~Graaf}}, \bibinfo {author} {\bibfnamefont {John W.~O.}\ \bibnamefont {Garmon}}, \bibinfo {author} {\bibfnamefont {Benjamin}\ \bibnamefont {Gudlewski}}, \bibinfo {author} {\bibfnamefont {William~D.}\ \bibnamefont {Kalfus}}, \bibinfo {author} {\bibfnamefont {Trevor}\ \bibnamefont {Keen}}, \bibinfo {author} {\bibfnamefont
  {Nishaad}\ \bibnamefont {Khedkar}}, \bibinfo {author} {\bibfnamefont {Chan~U.}\ \bibnamefont {Lei}}, \bibinfo {author} {\bibfnamefont {Gangqiang}\ \bibnamefont {Liu}}, \bibinfo {author} {\bibfnamefont {Pinlei}\ \bibnamefont {Lu}}, \bibinfo {author} {\bibfnamefont {Yao}\ \bibnamefont {Lu}}, \bibinfo {author} {\bibfnamefont {Aniket}\ \bibnamefont {Maiti}}, \bibinfo {author} {\bibfnamefont {Luke}\ \bibnamefont {Mastalli-Kelly}}, \bibinfo {author} {\bibfnamefont {Nitish}\ \bibnamefont {Mehta}}, \bibinfo {author} {\bibfnamefont {Shantanu~O.}\ \bibnamefont {Mundhada}}, \bibinfo {author} {\bibfnamefont {Anirudh}\ \bibnamefont {Narla}}, \bibinfo {author} {\bibfnamefont {Taewan}\ \bibnamefont {Noh}}, \bibinfo {author} {\bibfnamefont {Takahiro}\ \bibnamefont {Tsunoda}}, \bibinfo {author} {\bibfnamefont {Sophia~H.}\ \bibnamefont {Xue}}, \bibinfo {author} {\bibfnamefont {Joseph~O.}\ \bibnamefont {Yuan}}, \bibinfo {author} {\bibfnamefont {Luigi}\ \bibnamefont {Frunzio}}, \bibinfo {author} {\bibfnamefont {José}\
  \bibnamefont {Aumentado}}, \bibinfo {author} {\bibfnamefont {Shruti}\ \bibnamefont {Puri}}, \bibinfo {author} {\bibfnamefont {Steven~M.}\ \bibnamefont {Girvin}}, \bibinfo {author} {\bibfnamefont {S.~Harvey}\ \bibnamefont {Moseley}}, \ and\ \bibinfo {author} {\bibfnamefont {Robert~J.}\ \bibnamefont {Schoelkopf}},\ }\bibfield  {title} {{\selectlanguage {english}\enquote {\bibinfo {title} {A superconducting dual-rail cavity qubit with erasure-detected logical measurements},}\ }}\href {\doibase 10.1038/s41567-024-02539-4} {\bibfield  {journal} {\bibinfo  {journal} {Nature Physics}\ }\textbf {\bibinfo {volume} {20}},\ \bibinfo {pages} {1454--1460} (\bibinfo {year} {2024})},\ \bibinfo {note} {publisher: Nature Publishing Group}\BibitemShut {NoStop}%
\bibitem [{\citenamefont {Teoh}\ \emph {et~al.}(2023)\citenamefont {Teoh}, \citenamefont {Winkel}, \citenamefont {Babla}, \citenamefont {Chapman}, \citenamefont {Claes}, \citenamefont {de~Graaf}, \citenamefont {Garmon}, \citenamefont {Kalfus}, \citenamefont {Lu}, \citenamefont {Maiti} \emph {et~al.}}]{teoh2023dual}%
  \BibitemOpen
  \bibfield  {author} {\bibinfo {author} {\bibfnamefont {James~D}\ \bibnamefont {Teoh}}, \bibinfo {author} {\bibfnamefont {Patrick}\ \bibnamefont {Winkel}}, \bibinfo {author} {\bibfnamefont {Harshvardhan~K}\ \bibnamefont {Babla}}, \bibinfo {author} {\bibfnamefont {Benjamin~J}\ \bibnamefont {Chapman}}, \bibinfo {author} {\bibfnamefont {Jahan}\ \bibnamefont {Claes}}, \bibinfo {author} {\bibfnamefont {Stijn~J}\ \bibnamefont {de~Graaf}}, \bibinfo {author} {\bibfnamefont {John~WO}\ \bibnamefont {Garmon}}, \bibinfo {author} {\bibfnamefont {William~D}\ \bibnamefont {Kalfus}}, \bibinfo {author} {\bibfnamefont {Yao}\ \bibnamefont {Lu}}, \bibinfo {author} {\bibfnamefont {Aniket}\ \bibnamefont {Maiti}},  \emph {et~al.},\ }\bibfield  {title} {\enquote {\bibinfo {title} {Dual-rail encoding with superconducting cavities},}\ }\href@noop {} {\bibfield  {journal} {\bibinfo  {journal} {Proceedings of the National Academy of Sciences}\ }\textbf {\bibinfo {volume} {120}},\ \bibinfo {pages} {e2221736120} (\bibinfo {year}
  {2023})}\BibitemShut {NoStop}%
\bibitem [{\citenamefont {Zeytinoglu}\ \emph {et~al.}(2015)\citenamefont {Zeytinoglu}, \citenamefont {Pechal}, \citenamefont {Berger}, \citenamefont {Abdumalikov}, \citenamefont {Wallraff},\ and\ \citenamefont {Filipp}}]{Zeytino_lu_2015}%
  \BibitemOpen
  \bibfield  {author} {\bibinfo {author} {\bibfnamefont {S.}~\bibnamefont {Zeytinoglu}}, \bibinfo {author} {\bibfnamefont {M.}~\bibnamefont {Pechal}}, \bibinfo {author} {\bibfnamefont {S.}~\bibnamefont {Berger}}, \bibinfo {author} {\bibfnamefont {A.~A.}\ \bibnamefont {Abdumalikov}}, \bibinfo {author} {\bibfnamefont {A.}~\bibnamefont {Wallraff}}, \ and\ \bibinfo {author} {\bibfnamefont {S.}~\bibnamefont {Filipp}},\ }\bibfield  {title} {\enquote {\bibinfo {title} {Microwave-induced amplitude- and phase-tunable qubit-resonator coupling in circuit quantum electrodynamics},}\ }\href {\doibase 10.1103/physreva.91.043846} {\bibfield  {journal} {\bibinfo  {journal} {Physical Review A}\ }\textbf {\bibinfo {volume} {91}} (\bibinfo {year} {2015}),\ 10.1103/physreva.91.043846}\BibitemShut {NoStop}%
\bibitem [{\citenamefont {Mai}\ \emph {et~al.}(2026)\citenamefont {Mai}, \citenamefont {Liu}, \citenamefont {Deng}, \citenamefont {Cai}, \citenamefont {Ni}, \citenamefont {Zhang}, \citenamefont {Hu}, \citenamefont {Zheng}, \citenamefont {Liu}, \citenamefont {Xu} \emph {et~al.}}]{mai2026biased}%
  \BibitemOpen
  \bibfield  {author} {\bibinfo {author} {\bibfnamefont {Jiasheng}\ \bibnamefont {Mai}}, \bibinfo {author} {\bibfnamefont {Qiyu}\ \bibnamefont {Liu}}, \bibinfo {author} {\bibfnamefont {Xiaowei}\ \bibnamefont {Deng}}, \bibinfo {author} {\bibfnamefont {Yanyan}\ \bibnamefont {Cai}}, \bibinfo {author} {\bibfnamefont {Zhongchu}\ \bibnamefont {Ni}}, \bibinfo {author} {\bibfnamefont {Libo}\ \bibnamefont {Zhang}}, \bibinfo {author} {\bibfnamefont {Ling}\ \bibnamefont {Hu}}, \bibinfo {author} {\bibfnamefont {Pan}\ \bibnamefont {Zheng}}, \bibinfo {author} {\bibfnamefont {Song}\ \bibnamefont {Liu}}, \bibinfo {author} {\bibfnamefont {Yuan}\ \bibnamefont {Xu}},  \emph {et~al.},\ }\bibfield  {title} {\enquote {\bibinfo {title} {A biased-erasure cavity qubit with hardware-efficient quantum error detection},}\ }\href@noop {} {\bibfield  {journal} {\bibinfo  {journal} {arXiv preprint arXiv:2601.21616}\ } (\bibinfo {year} {2026})}\BibitemShut {NoStop}%
\bibitem [{\citenamefont {Rosenblum}\ \emph {et~al.}(2018{\natexlab{b}})\citenamefont {Rosenblum}, \citenamefont {Reinhold}, \citenamefont {Mirrahimi}, \citenamefont {Jiang}, \citenamefont {Frunzio},\ and\ \citenamefont {Schoelkopf}}]{rosenblum2018fault}%
  \BibitemOpen
  \bibfield  {author} {\bibinfo {author} {\bibfnamefont {Serge}\ \bibnamefont {Rosenblum}}, \bibinfo {author} {\bibfnamefont {P}~\bibnamefont {Reinhold}}, \bibinfo {author} {\bibfnamefont {Mazyar}\ \bibnamefont {Mirrahimi}}, \bibinfo {author} {\bibfnamefont {Liang}\ \bibnamefont {Jiang}}, \bibinfo {author} {\bibfnamefont {L}~\bibnamefont {Frunzio}}, \ and\ \bibinfo {author} {\bibfnamefont {RJ}~\bibnamefont {Schoelkopf}},\ }\bibfield  {title} {\enquote {\bibinfo {title} {Fault-tolerant detection of a quantum error},}\ }\href@noop {} {\bibfield  {journal} {\bibinfo  {journal} {Science}\ }\textbf {\bibinfo {volume} {361}},\ \bibinfo {pages} {266--270} (\bibinfo {year} {2018}{\natexlab{b}})}\BibitemShut {NoStop}%
\bibitem [{\citenamefont {Poboiko}\ \emph {et~al.}(2025)\citenamefont {Poboiko}, \citenamefont {Pöpperl}, \citenamefont {Gornyi},\ and\ \citenamefont {Mirlin}}]{Poboiko_2025}%
  \BibitemOpen
  \bibfield  {author} {\bibinfo {author} {\bibfnamefont {Igor}\ \bibnamefont {Poboiko}}, \bibinfo {author} {\bibfnamefont {Paul}\ \bibnamefont {Pöpperl}}, \bibinfo {author} {\bibfnamefont {Igor~V.}\ \bibnamefont {Gornyi}}, \ and\ \bibinfo {author} {\bibfnamefont {Alexander~D.}\ \bibnamefont {Mirlin}},\ }\bibfield  {title} {\enquote {\bibinfo {title} {Measurement-induced transitions for interacting fermions},}\ }\href {\doibase 10.1103/physrevb.111.024204} {\bibfield  {journal} {\bibinfo  {journal} {Physical Review B}\ }\textbf {\bibinfo {volume} {111}} (\bibinfo {year} {2025}),\ 10.1103/physrevb.111.024204}\BibitemShut {NoStop}%
\bibitem [{\citenamefont {Fava}\ \emph {et~al.}(2024)\citenamefont {Fava}, \citenamefont {Piroli}, \citenamefont {Bernard},\ and\ \citenamefont {Nahum}}]{PhysRevResearch.6.043246}%
  \BibitemOpen
  \bibfield  {author} {\bibinfo {author} {\bibfnamefont {Michele}\ \bibnamefont {Fava}}, \bibinfo {author} {\bibfnamefont {Lorenzo}\ \bibnamefont {Piroli}}, \bibinfo {author} {\bibfnamefont {Denis}\ \bibnamefont {Bernard}}, \ and\ \bibinfo {author} {\bibfnamefont {Adam}\ \bibnamefont {Nahum}},\ }\bibfield  {title} {\enquote {\bibinfo {title} {Monitored fermions with conserved $u(1)$ charge},}\ }\href {\doibase 10.1103/PhysRevResearch.6.043246} {\bibfield  {journal} {\bibinfo  {journal} {Phys. Rev. Res.}\ }\textbf {\bibinfo {volume} {6}},\ \bibinfo {pages} {043246} (\bibinfo {year} {2024})}\BibitemShut {NoStop}%
\bibitem [{\citenamefont {Guo}\ \emph {et~al.}(2025)\citenamefont {Guo}, \citenamefont {Foster}, \citenamefont {Jian},\ and\ \citenamefont {Ludwig}}]{Guo_2025}%
  \BibitemOpen
  \bibfield  {author} {\bibinfo {author} {\bibfnamefont {Haoyu}\ \bibnamefont {Guo}}, \bibinfo {author} {\bibfnamefont {Matthew~S.}\ \bibnamefont {Foster}}, \bibinfo {author} {\bibfnamefont {Chao-Ming}\ \bibnamefont {Jian}}, \ and\ \bibinfo {author} {\bibfnamefont {Andreas W.~W.}\ \bibnamefont {Ludwig}},\ }\bibfield  {title} {\enquote {\bibinfo {title} {Field theory of monitored interacting fermion dynamics with charge conservation},}\ }\href {\doibase 10.1103/gdxd-pw8v} {\bibfield  {journal} {\bibinfo  {journal} {Physical Review B}\ }\textbf {\bibinfo {volume} {112}} (\bibinfo {year} {2025}),\ 10.1103/gdxd-pw8v}\BibitemShut {NoStop}%
\bibitem [{\citenamefont {Cao}\ \emph {et~al.}(2019)\citenamefont {Cao}, \citenamefont {Tilloy},\ and\ \citenamefont {De~Luca}}]{cao2019entanglement}%
  \BibitemOpen
  \bibfield  {author} {\bibinfo {author} {\bibfnamefont {Xiangyu}\ \bibnamefont {Cao}}, \bibinfo {author} {\bibfnamefont {Antoine}\ \bibnamefont {Tilloy}}, \ and\ \bibinfo {author} {\bibfnamefont {Andrea}\ \bibnamefont {De~Luca}},\ }\bibfield  {title} {\enquote {\bibinfo {title} {Entanglement in a fermion chain under continuous monitoring},}\ }\href@noop {} {\bibfield  {journal} {\bibinfo  {journal} {SciPost Physics}\ }\textbf {\bibinfo {volume} {7}},\ \bibinfo {pages} {024} (\bibinfo {year} {2019})}\BibitemShut {NoStop}%
\bibitem [{\citenamefont {Fava}\ \emph {et~al.}(2023)\citenamefont {Fava}, \citenamefont {Piroli}, \citenamefont {Swann}, \citenamefont {Bernard},\ and\ \citenamefont {Nahum}}]{Fava_2023}%
  \BibitemOpen
  \bibfield  {author} {\bibinfo {author} {\bibfnamefont {Michele}\ \bibnamefont {Fava}}, \bibinfo {author} {\bibfnamefont {Lorenzo}\ \bibnamefont {Piroli}}, \bibinfo {author} {\bibfnamefont {Tobias}\ \bibnamefont {Swann}}, \bibinfo {author} {\bibfnamefont {Denis}\ \bibnamefont {Bernard}}, \ and\ \bibinfo {author} {\bibfnamefont {Adam}\ \bibnamefont {Nahum}},\ }\bibfield  {title} {\enquote {\bibinfo {title} {Nonlinear sigma models for monitored dynamics of free fermions},}\ }\href {\doibase 10.1103/physrevx.13.041045} {\bibfield  {journal} {\bibinfo  {journal} {Physical Review X}\ }\textbf {\bibinfo {volume} {13}} (\bibinfo {year} {2023}),\ 10.1103/physrevx.13.041045}\BibitemShut {NoStop}%
\end{thebibliography}%

 \end{document}


\title{Supplementary Material: Universal monitored dynamics in
multimode bosonic systems}

 \author{Shivam Patel}
  \affiliation{Department of Physics and Astronomy, 
Rutgers University, Piscataway, NJ 08854, USA}
\affiliation{Center for Materials Theory,  
Rutgers University, Piscataway, NJ 08854, USA}
\author{Catherine McCarthy}
\thanks{co-first author}
\affiliation{Department of Theoretical Physics, University of Geneva, 24 quai Ernest-Ansermet, 1211 Gen\`eve, Switzerland}
 \author{Ahana Chakraborty}
 \affiliation{Department of Physics and Astronomy, Louisiana State University, Baton Rouge, Louisiana 70803, USA}
  \affiliation{Department of Physics and Astronomy, 
Rutgers University, Piscataway, NJ 08854, USA}
\affiliation{Center for Materials Theory,  
Rutgers University, Piscataway, NJ 08854, USA}
\author{Jordan Huang}
  \affiliation{Department of Physics and Astronomy, 
Rutgers University, Piscataway, NJ 08854, USA}
\author{Thomas Dinapoli}
  \affiliation{Department of Physics and Astronomy, 
Rutgers University, Piscataway, NJ 08854, USA}
\author{Romain Vasseur}
\affiliation{Department of Theoretical Physics, University of Geneva, 24 quai Ernest-Ansermet, 1211 Gen\`eve, Switzerland}
\author{J. H. Pixley}\email{jed.pixley@physics.rutgers.edu}
  \affiliation{Department of Physics and Astronomy, 
Rutgers University, Piscataway, NJ 08854, USA}
\affiliation{Center for Materials Theory,  
Rutgers University, Piscataway, NJ 08854, USA}
\affiliation{Center for Computational Quantum Physics, Flatiron Institute, 162 5th Avenue, New York, NY 10010, USA}
\author{Srivatsan Chakram}\email{schakram@physics.rutgers.edu}
  \affiliation{Department of Physics and Astronomy, 
Rutgers University, Piscataway, NJ 08854, USA}
\date{\today}

\definecolor{byzantine}{rgb}{0.74, 0.2, 0.64}
\newcommand{\catherine}[1]{{\color{byzantine} CM: #1}}

\maketitle

\section{Additional numerical results for monitored bosonic circuits} \label{sec:additional}

\subsection{Initial state preparation protocol} \label{sec:stateprep}

\begin{figure*}
    \centering
    \includegraphics[width=0.4\linewidth]{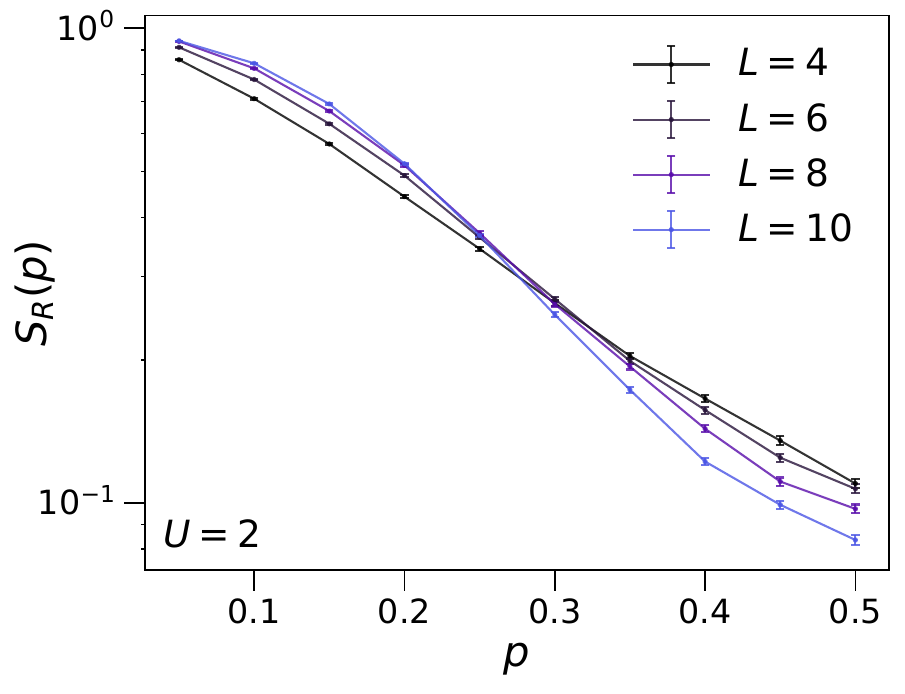}
    \hspace{0.025\linewidth}
    \includegraphics[width=0.4\linewidth]{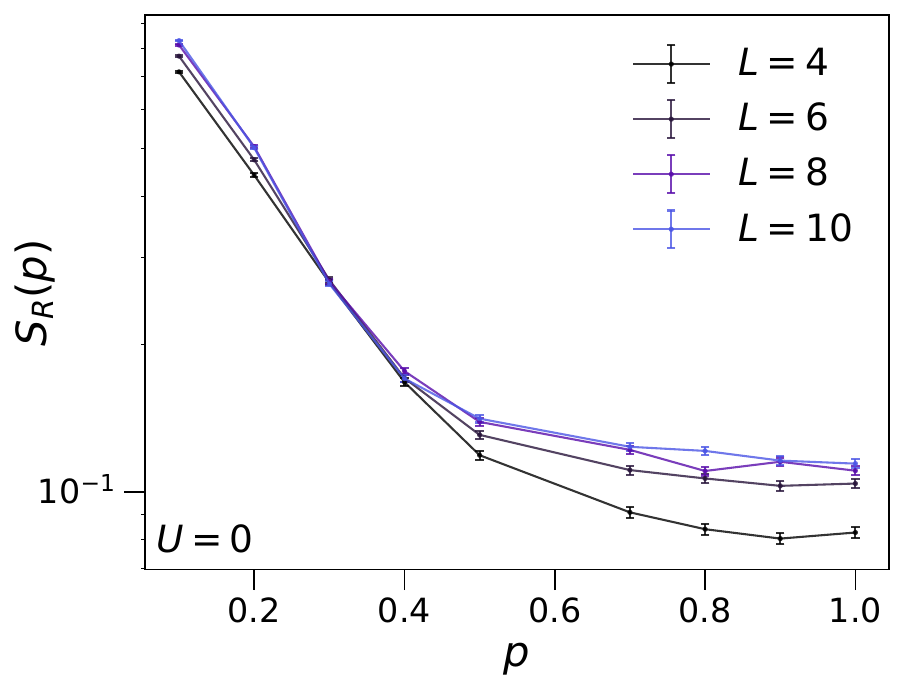}
    \caption{Ancilla entanglement entropy $S_R$ for monitored dynamics with BSFP gates both with (left) and without (right) on-site Hubbard gates.  The initial state is prepared by beginning from the ``checkerboard'' state $\ket{\Psi} = \frac{1}{\sqrt 2} \big[ \ket{0101...} \ket{0} + \ket{1010...} \ket{1} \big]$ and scrambling with $2L$ layers of BSFP and on-site Hubbard gates, before running $2L$ monitored layers.  Both cases reproduce the results beginning from a Haar-random state shown in the main text for experimentally accessible system sizes.}
    \label{fig:checkerboard}
\end{figure*}

fig:numericsHere we demonstrate the validity of the state preparation protocol discussed in the main text.  The numerics in the main text mostly start from a Haar-random initial state, which is not efficiently experimentally preparable.  In the experiment we instead begin with a ``checkerboard'' initial state $\ket{\Psi} = \frac{1}{\sqrt 2} \big[ \ket{0101...} \ket{0} + \ket{1010...} \ket{1} \big]$ and then scramble with $2L$ layers of interacting gates without measurements before beginning the monitored dynamics for $2L$ layers of gates.  For monitored dynamics both with and without on-site Hubbard gates, the resulting $S_R$ curves show the same trends as the data in the main text (Fig. 2).

\subsection{Learnability protocol and data for BSFP gates} \label{sec:learning}

\begin{figure}
    \centering
    \includegraphics[width=0.7\linewidth]{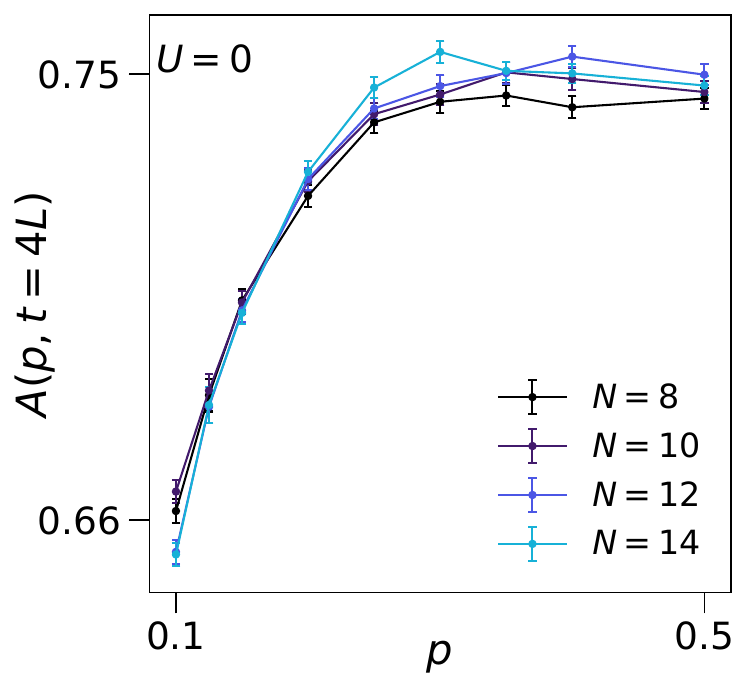}
    
    \caption{\textbf{Learnability numerics for BSFP monitored dynamics.} Accuracy $A(p)$ of an exact classical decoder for a monitored circuit composed only of BSFP gates.  In line with the main text numerics, the data is consistent with a phase at high measurement rates in which the ``decodability time'' is $O(L)$.}
    \label{fig:learn0}
\end{figure}

In the main text, we proposed an experimental protocol to detect a MIPT by computing the entanglement entropy $S_R$ of an ancilla qubit globally coupled to an $L$-mode system. In the near term, post-selection is experimentally feasible because current circuit-QED implementations are limited to system sizes of at most $L \sim 10$ modes~\cite{huang2025fastsidebandcontrolweakly}; however, post-selection quickly becomes untenable with even modest increases in $L$. Recent works have sought to bypass this post-selection bottleneck by viewing the MIPT as a ``learnability'' transition accessible to an exact classical decoder~\cite{Barratt_2022, Agrawal_2024, singh2025, kim2025}. In this approach, the order parameter is the decoder’s accuracy in predicting which initial state generated a given measurement record~\cite{Barratt_2022}. Here we outline such a learnability protocol and present corresponding numerical results. This protocol also has the advantage of not requiring an ancilla qubit, thereby reducing potential sources of experimental error.

In this protocol, we choose two Haar-random initial states $\ket{\psi_0}$ and $\ket{\psi_1}$ such that $\braket{\psi_0|\psi_1}=0$. We then run a standard monitored circuit starting from $\ket{\psi_\alpha}$, where $\alpha\in\{0,1\}$ is chosen at random. After the circuit is executed, the measurement record $\vec m$---including the locations and outcomes of all measurements---along with the complete gate list and a full description of the states $\ket{\psi_0}$ and $\ket{\psi_1}$, is passed to a classical decoder. The decoder's task is to predict $\alpha$, \emph{i.e.}, which initial state generated the record $\vec m$.

To make this prediction, the decoder simulates the full monitored evolution starting from $\ket{\psi_0}$ and $\ket{\psi_1}$ to compute the Born probabilities $P(\vec m|\alpha)$ of obtaining the record $\vec m$ given the initial state $\ket{\psi_\alpha}$. It then predicts the label corresponding to the larger probability. Averaged over $N$ measurement records, the decoder accuracy is
\begin{equation}
A=\frac{1}{N}\sum_{\vec m}\max\{P(\vec m|0),\,P(\vec m|1)\}.
\end{equation}
The decoder accuracy undergoes a transition from an ``encoded'' phase to a ``decodable'' phase~\cite{Barratt_2022}.

While this protocol avoids the need for experimental post-selection, it is not truly scalable because the decoder must simulate the full evolution of the state, which remains exponentially costly in $L$. The main advantage of the learnability approach is therefore that the exponential difficulty is shifted from an experimental task to a numerical one.

For monitored dynamics with on-site Hubbard gates ($U=2$) and parity measurements, we numerically find a transition in the decoder accuracy consistent with $z=1$ near $p_c \approx 0.25$, as discussed in the main text. In addition, for purely BSFP dynamics ($U=0$), the decoder accuracy is consistent with the main-text trends: at high measurement rates, the system appears to enter a phase in which the ``decodability time'' (analogous to the purification time) scales linearly with system size (Fig.~\ref{fig:learn0}).

\subsection{Number measurements} \label{sec:den}

\begin{figure*}
    \centering
    \includegraphics[width=0.4\linewidth]{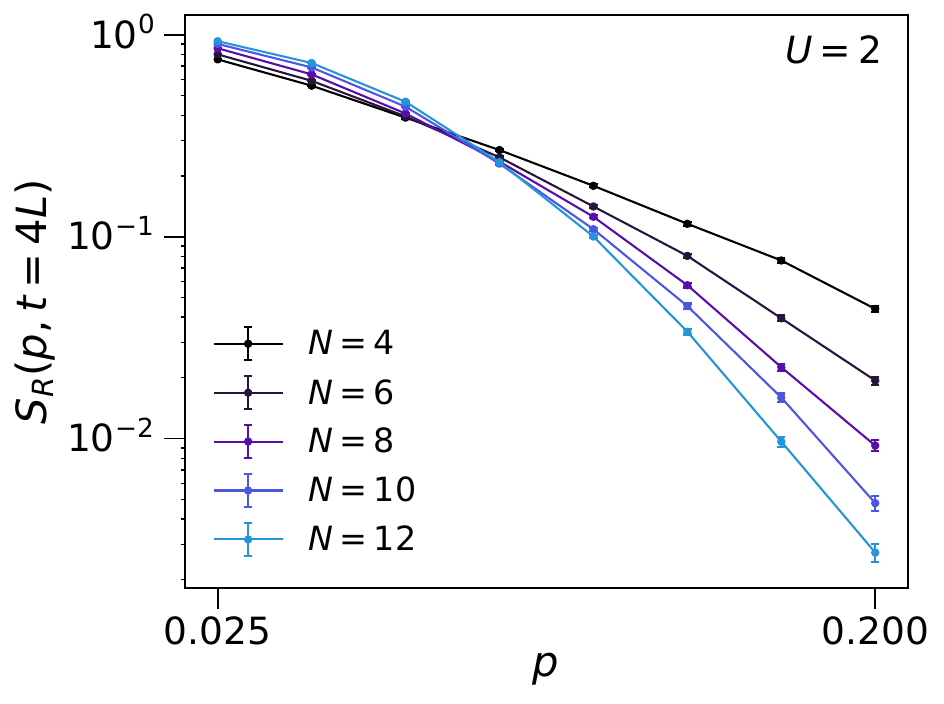}
    \hspace{0.01\linewidth}
    \includegraphics[width=0.4\linewidth]{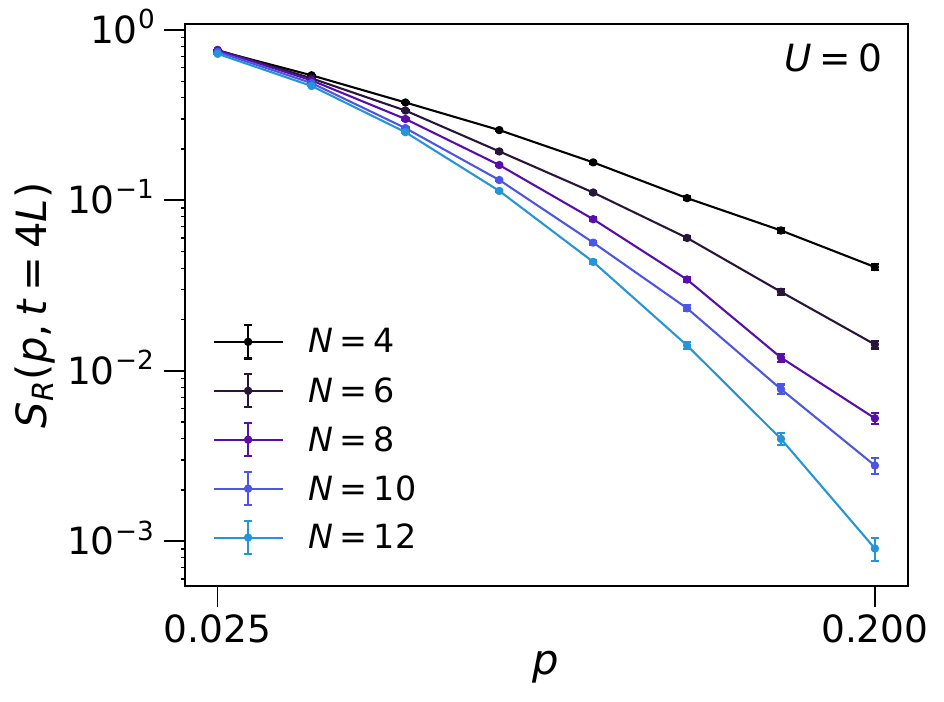}
    \caption{\textbf{Ancilla entanglement entropy for monitored dynamics with number measurements.}  Data is shown for monitored dynamics with ($U=2$, left) and without ($U=0$, right) on-site Hubbard gates with density measurements for a circuit of depth $4L$.  The circuit with on-site Hubbard gates produces a phase transition at critical measurement rate $p_c \approx 0.1$; in contrast, there is no transition in the $U=0$ circuit and the ancilla qubit purifies.}
    \label{fig:den}
\end{figure*}

The main text presented results for parity measurements; here, we present additional results for measurements of photon number.  As was the case for parity measurements, the behavior of the ancilla entanglement entropy curves $S_R(p,t)$ depend on whether the monitored circuit is composed of only BSFP gates or not (Fig.~\ref{fig:den}).  For a circuit of only BSFP gates ($U=0$), the data in Fig.~\ref{fig:den} indicates that there is no MIPT with $z=1$ and that the ancilla purifies for any measurement rate $p>0$.  In contrast, for circuits with on-site Hubbard gates ($U=2$), there appears to be a transition with $z=1$ at the critical measurement rate $p_c \approx 0.1$.

It is interesting to compare the impact of turning on $U>0$ for parity and number measurements.  Despite the fact that neither measurement type features a transition when $U=0$, the ancilla is in a different phase depending on the measurement type-- the ancilla always purifies for density measurements but appears to be in either a purified or critical phase for parity measurements.  Additionally, while there is a transition that appears when $U \neq 0$ for both measurement types, the mechanism through which this occurs appears to be qualitatively different.  Since performing a number measurement locally collapses the state onto a single Fock basis state, the standard picture of MIPTs as a competition between unitary dynamics that hide information and measurements that reveal information is accurate for number measurements.  This intuition is consistent with the fact that for number measurements, turning on on-site Hubbard gates allows the unitary dynamics to hide information more efficiently, which results in the emergence of a mixed phase.  This qualitative picture is entirely different than that of the parity measurement, where increasing the scrambling power of the gates helps the purification process (see Sec. \ref{sec:scrambling}) and produces a phase with a $\mathcal O(1)$ purification time.

Performing measurements onto the mod-$n$ subspace for a generic choice of $n$ is natural to implement on circuit-QED platforms, and understanding the impact of tuning $n$ on the MIPT remains an interesting question.  In practice, it is difficult to experimentally or numerically realize system sizes large enough to meaningfully study cases apart from the parity ($n=2$) and density ($n=Q+1$) cases, and so future investigations of this subject will likely be largely theoretical.

\subsection{Bipartite entanglement entropy} \label{subsec:bipartite}

\begin{figure*}
    \centering
    \includegraphics[width=0.4\linewidth]{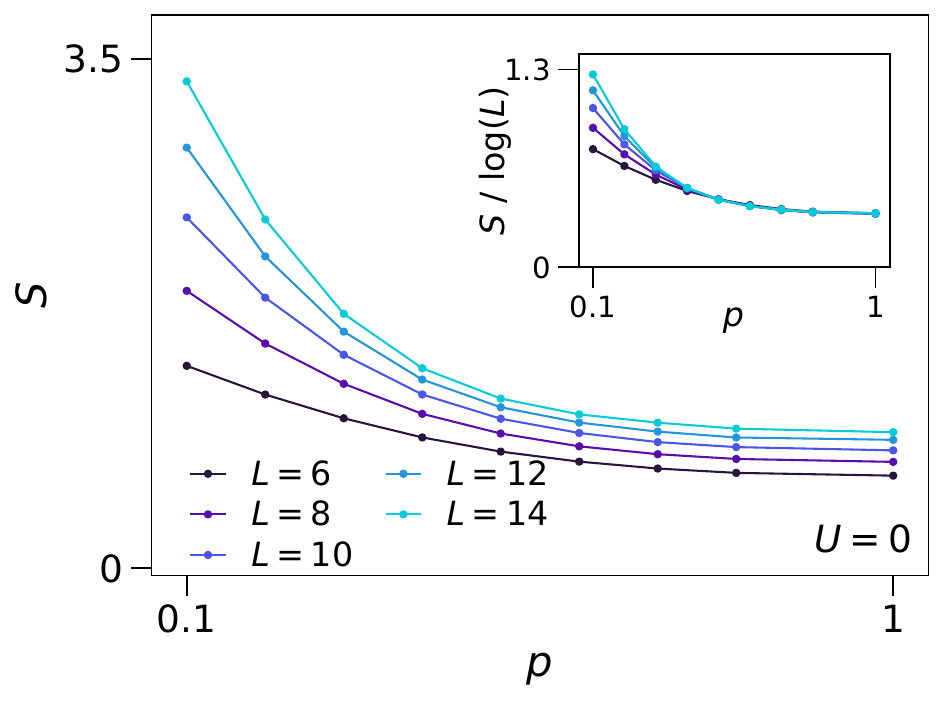}
    \hspace{0.01\linewidth}
    \includegraphics[width=0.4\linewidth]{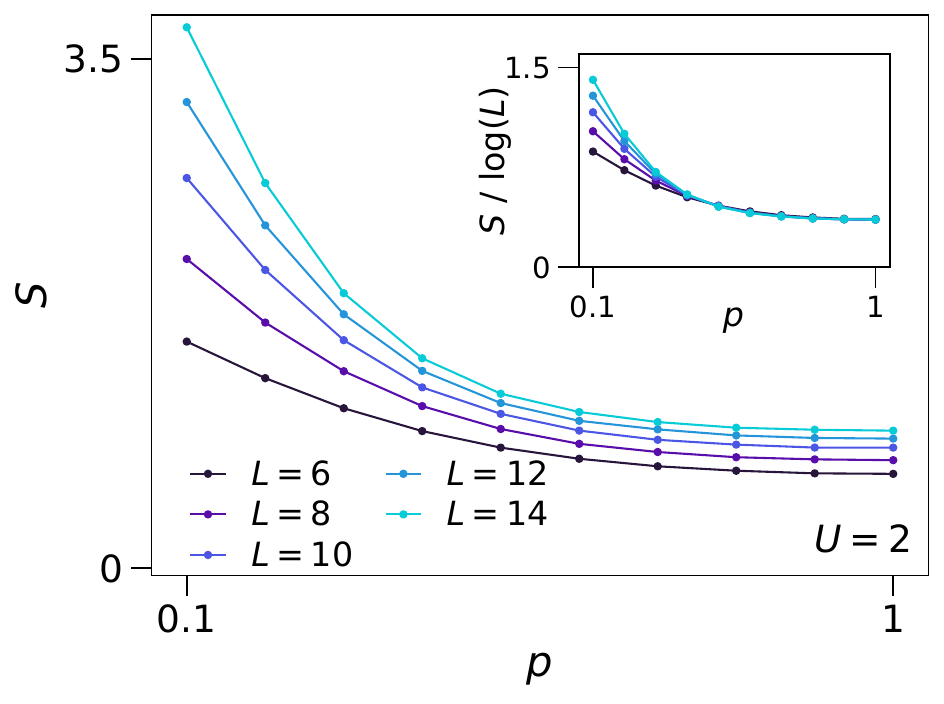}
    \caption{\textbf{Bipartite entanglement entropy} Bipartite entanglement entropy $S$ for systems with (left) and without (right) on-site Hubbard gates shown for system sizes $L=4-14$.  Both cases appear to feature $S$ that scales logarithmically with system size at large $L$ for numerically-accessible system sizes (insets).}
    \label{fig:bipartite}
\end{figure*}

Here we briefly present results for the bipartite entanglement entropy $S = -\mathrm{Tr}[ \rho_A \ln \rho_A]$ with parity measurements for monitored dynamics with and without on-site Hubbard gates.  In contrast to the ancilla entanglement entropy data presented in the main text, the bipartite entanglement entropy $S$ looks similar for both $U=0$ and $U>0$ (\ref{fig:bipartite}).  Interestingly, the data for both values of $U$ appear consistent with a phase at high measurement rates in which $S \sim \log L$, although it is hard to accurately verify this scaling with only numerical access to small system sizes.

Considering the striking differences in the ancilla entanglement entropy data, it is interesting that the bipartite entanglement entropy looks similar for both cases.  The data is possibly consistent with a decoupling between the ancilla entanglement entropy $S_R$ for the $U=2$ , which purifies at high measurement rates, and the half-cut entanglement entropy, which appears to have a critical-like $\sim \log L$ scaling in the same regime.  Another possibility that we raise here is that this is a matter of protocol-- following the procedure in \cite{zabalo2020critical}, we calculate $S$ immediately after applying gates but before measuring.  Given that the parity measurements project onto an entire subspace, rather than individual Fock basis states, it is possible that choosing this protocol obscures the effect of the measurements.

\subsection{Alternative values of the dynamical exponent} \label{sec:alternative}

\begin{figure*}
    \centering
    \includegraphics[width=0.4\linewidth]{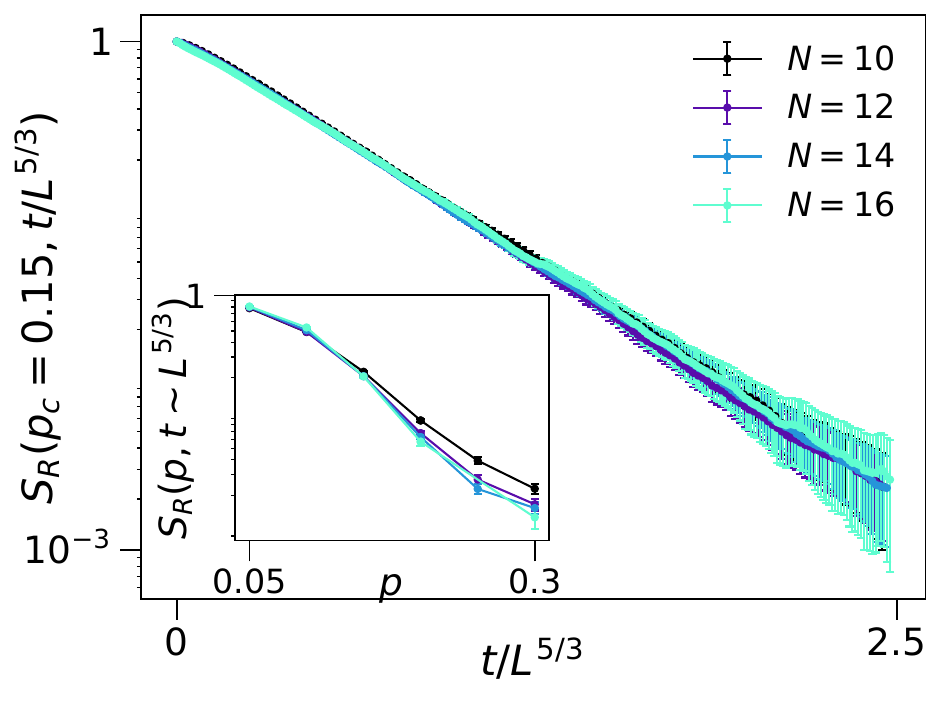}
    \includegraphics[width=0.4\linewidth]{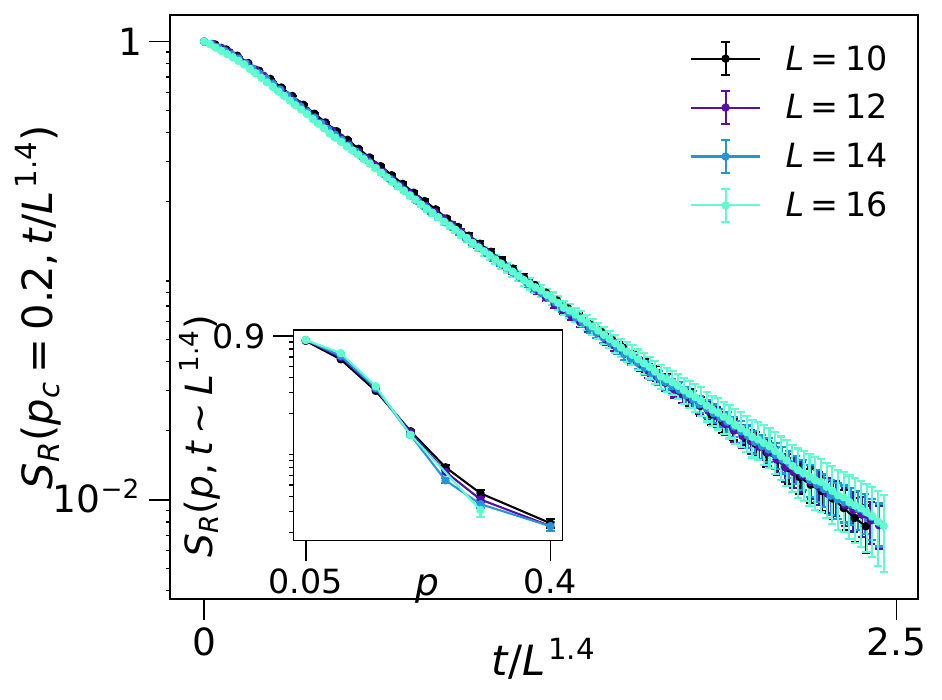}
    \caption{\textbf{Alternative values of $z$.}  Alternative values of $(z,p_c)$ that yield an apparent collapse in the ancilla entanglement entropy $S_R(p_c,t/L^z)$ for BSFP gates with parity measurements.  Both $p_c=0.15, \ z=5/3$ (left) and $p_c=0.2, \ z=1.4$ (right) produce reasonable collapses-- without an independent prediction for either the dynamical exponent or the critical measurement rate, determining if there is a pair of $p_c,\ z$ from numerics alone is challenging.}
    \label{fig:zneq1}
\end{figure*}

In the main text, we presented numerical results based on the assumption that any MIPT would have the scaling exponent $z=1$.  The basis for this assumption is the theoretical argument for $z=1$ in the $d \rightarrow \infty$ limit with no symmetries \cite{Skinner_2019, Jian_2020} as well as numerical evidence showing $z=1$ in the qubit limit \cite{zabalo2020critical}.  The BSFP numerical data was consistent with the emergence of a novel critical phase at high measurement rates with a purification time that scales $\sim L$.  In principle, however, there is no reason to rule out the existence of a MIPT from a phase with an exponentially-long purification time to one with an $O(1)$ purification time with $z \neq 1$.  Furthermore, since there is no way to independently pinpoint the location of the transition with access only to the ancilla entanglement entropy, and since numerics are limited to small system sizes, there are multiple possible combinations of $(z,p_c)$ that are consistent with the numerics, two of which are illustrated in Fig. \ref{fig:zneq1} for $U=0$ (though the same holds true for the $U=2$ case as well).

Without an independent prediction of either $p_c$ or $z$, it is difficult to determine which, if any, combination of $(p_c \ z)$ corresponds to a true transition.  The approach taken in the main text is to assume $z=1$ and look for a transition consistent with this scaling exponent.  Regardless of  whether the system exhibits an anomalous transition with $z>1$ or if the system is in a phase with an $O(L)$ purification time at high measurement rates, the numerics demonstrate that the ancilla purifies faster when on-site Hubbard gates are used in addition to BSFP gates.

\subsection{Qualitative similarity of $z=1$ MIPT away from BSFP gates}\label{sec:qualitative}

\begin{figure}[h!]
    \centering
    \includegraphics[width=0.49\linewidth]{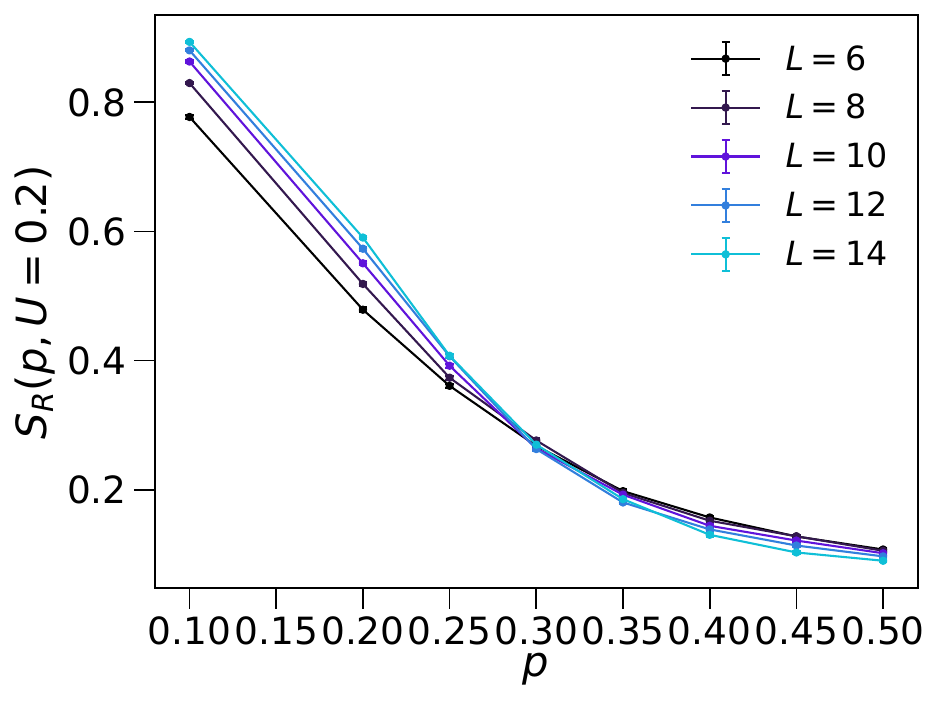}
    \includegraphics[width=0.49\linewidth]{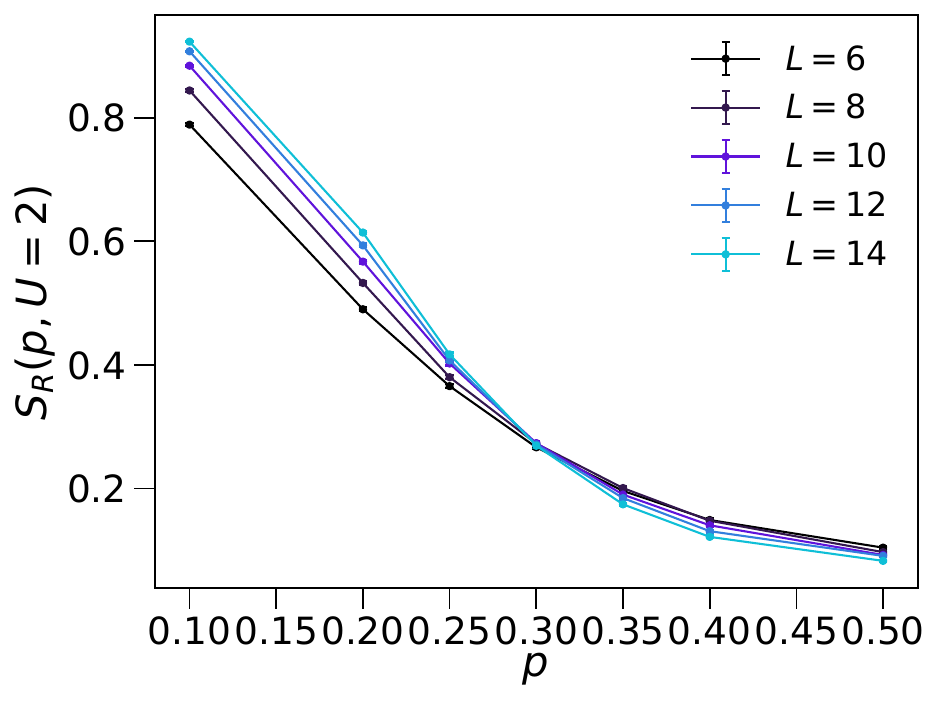}
    \caption{\textbf{Stability of crossing for different $U$.} Data is shown for circuits of depth $t_f=4L$ with parity measurements for circuits of $U=0.2$ (left) and $U=2$ (right).  Tuning $U \neq 0$ does not appear to qualitatively affect the nature of the transition.}
    \label{fig:varyU}
\end{figure}

\begin{figure}[h!]
    \centering
    \includegraphics[width=0.49\linewidth]{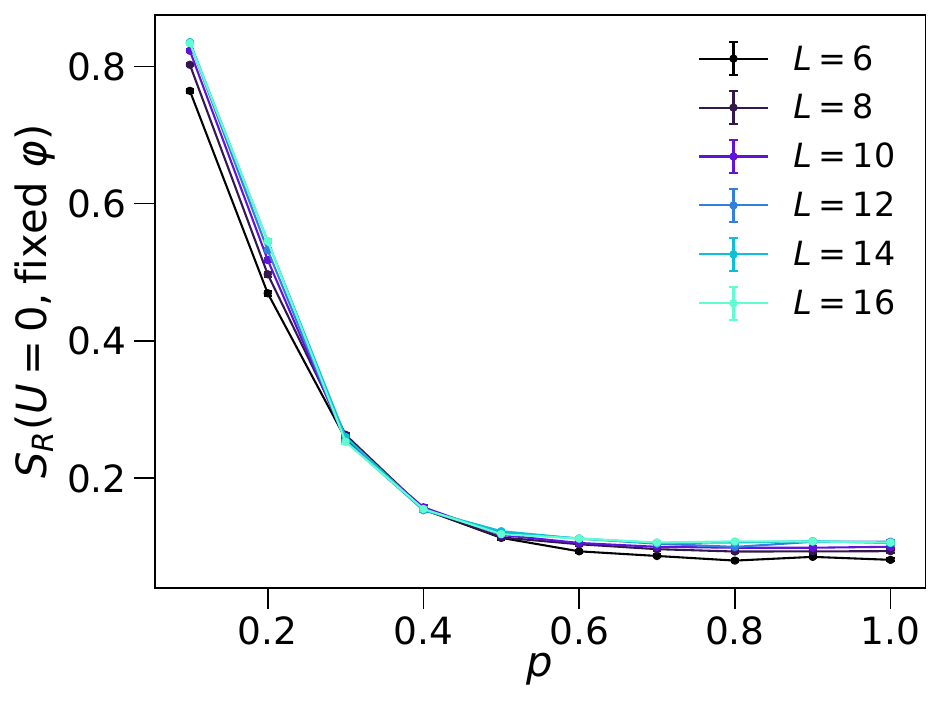}
    \includegraphics[width=0.49\linewidth]{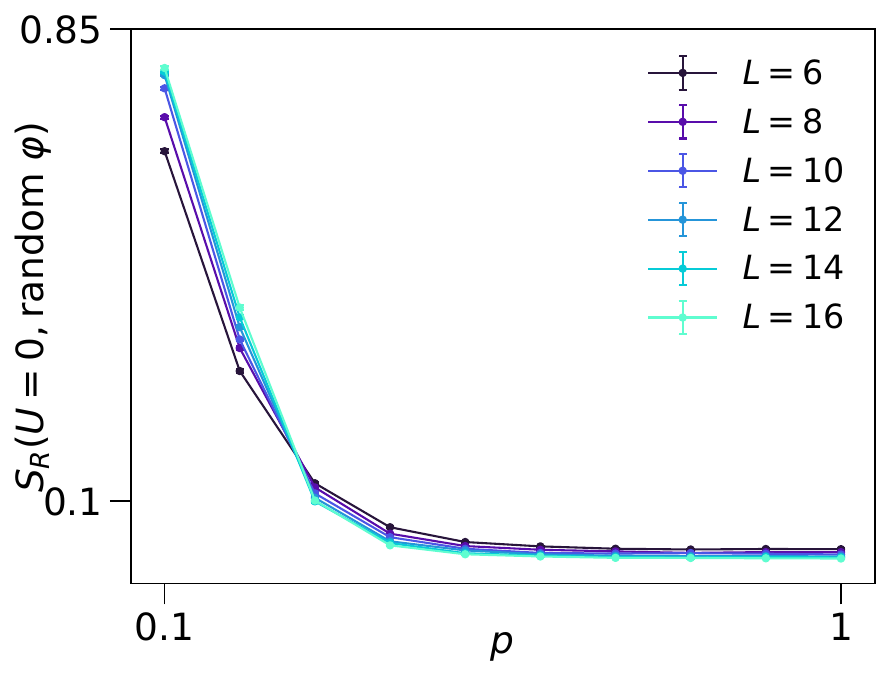}
    \caption{\textbf{Beam-splitter gates with fixed vs random phases.} Comparison of ancilla entanglement entropy $S_R(p, t \sim L)$ for BSFP (left) and BSRP (right) gates.  Choosing random phases produces a crossing in the $S_R$ curves.}
    \label{fig:bsrp}
\end{figure}

In the main text, we claim that there is a sharp distinction between monitored dynamics generated with purely BSFP gates and more generic circuits; here, we present numerical results validating this claim for two cases.  In particular, the presence of a crossing in the $S_R$ curves for the monitored dynamics generated by a combination of BSFP and on-site Hubbard gates remains when varying $U > 0$ (Fig. \ref{fig:varyU}); for a very small $0 < U \ll 1$, we would expect the crossing to emerge for sufficiently large circuit depths.  The precise mechanism through which this takes place remains an interesting open question.  Additionally, choosing beam-splitter unitaries with random phases (BSRP gates) restores a crossing, in comparison to the BSFP gates discussed in the main text (Fig. \ref{fig:bsrp}).

\section{Scrambling and purification with parity measurements} \label{sec:scrambling}

\begin{figure}
    \centering
    \includegraphics[width=\linewidth]{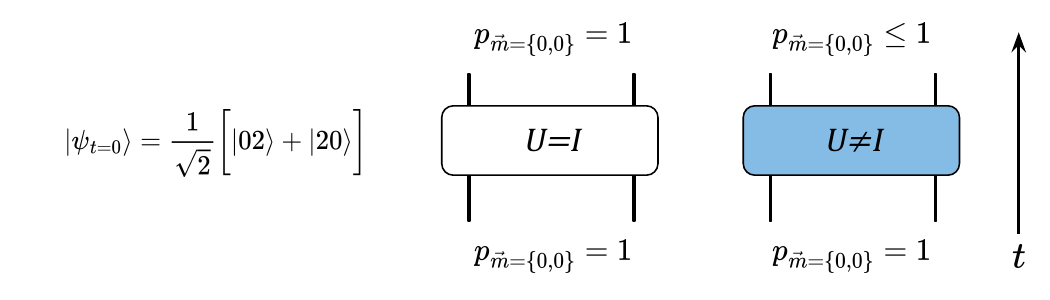}
    \caption{\textbf{Scrambling with parity measurements.} This figure illustrates a scenario in which increasing the scrambling power of the unitary dynamics can \emph{decrease} a system’s entanglement entropy. We consider a two-mode system with a conserved total photon number $Q=2$, initialized in a superposition within the even-parity subspace, and evolve it for one timestep under a unitary $U$. At $t=1$, we perform parity measurements on both modes. Evolving with the identity deterministically yields the measurement record $\vec{m}_{t=1}=\{0,0\}$, whereas evolving with a generic random unitary can yield $\vec{m}_{t=1}=\{1,1\}$, thereby reducing the entanglement entropy between the two modes.}
    \label{fig:scramble}
\end{figure}

As discussed in the main text, performing local measurements that project the state onto a subspace of the on-site Hilbert space, as opposed to onto a particular Fock basis state, leads to an effect in which scrambling helps the state purification process.  This effect originates from the fact that scrambling local degrees of freedom allows an observer to measure the same site multiple times and obtain different measurement outcomes, which results in a decrease in entanglement entropy.  Below, we will provide a concrete example of this effect taking place.

Consider a two-mode system with $Q=2$ photons.  The global two-mode state is spanned by the states $\{ \ket{20}, \ket{02}, \ket{11} \}$.  Consider time-evolving the initial state $\ket{\psi_{t=0}} = \frac{1}{\sqrt 2} \big[ \ket{20} + \ket{02}\big]$ by some unitary $U$ for one timestep before parity measurements are performed on both modes at $t=1$ (Fig.~\ref{fig:scramble}). For a generic random gate $U$, it is possible to either measure the outcome $\vec m_{t=1} =\{0,0\}$ or $\vec m_{t=1} =\{1,1\}$; if the outcome $\vec m_{t=1} = \{1,1\}$ is measured, the entanglement entropy between the two modes will decrease to zero.  However, choosing a less scrambling gate (e.g. the identity, or a constrained unitary that only allows for photons to hop in pairs) deterministically leads to the outcome $\vec m_{t=1} = \{0,0\}$ after the application of the gate.  In this case, the entanglement between the two modes will remain the same.  In contrast, for a situation in which number measurements are performed on both sites at both $t=0$ and $t=1$, the gate chosen does not impact the entanglement entropy between the two modes since the number measurements definitively collapse the state onto a single Fock basis state.

While somewhat contrived, the above example clearly illustrates that for measurements which project onto a subspace of the full Hilbert space and dynamics that mix between those sets of subspaces, increasing the scrambling power of the gate can lead to a decrease in the entanglement entropy in the system.  This intuition is consistent with the numerical results showing that BSFP and on-site Hubbard gates purify the ancilla faster than BSFP gates alone when using parity measurements.  Note that the argument above does not answer the interesting question of exactly how scrambling a gate has to be in order to produce a purified phase-- it only serves to illustrate the concept that increasing scrambling can favor purification in the presence of projective measurements onto subspaces.

\section{Analysis of the SNAIL coupler} \label{sec:AnalysisoftheSNAILcoupler}
In this section we derive the Hamiltonian of a Superconducting NonlineAr Inductive eLement (SNAIL) tunable coupler which we use to perform beam-splitter operations on our cavity modes.

\begin{figure}
    \centering
    \includegraphics[width=\linewidth]{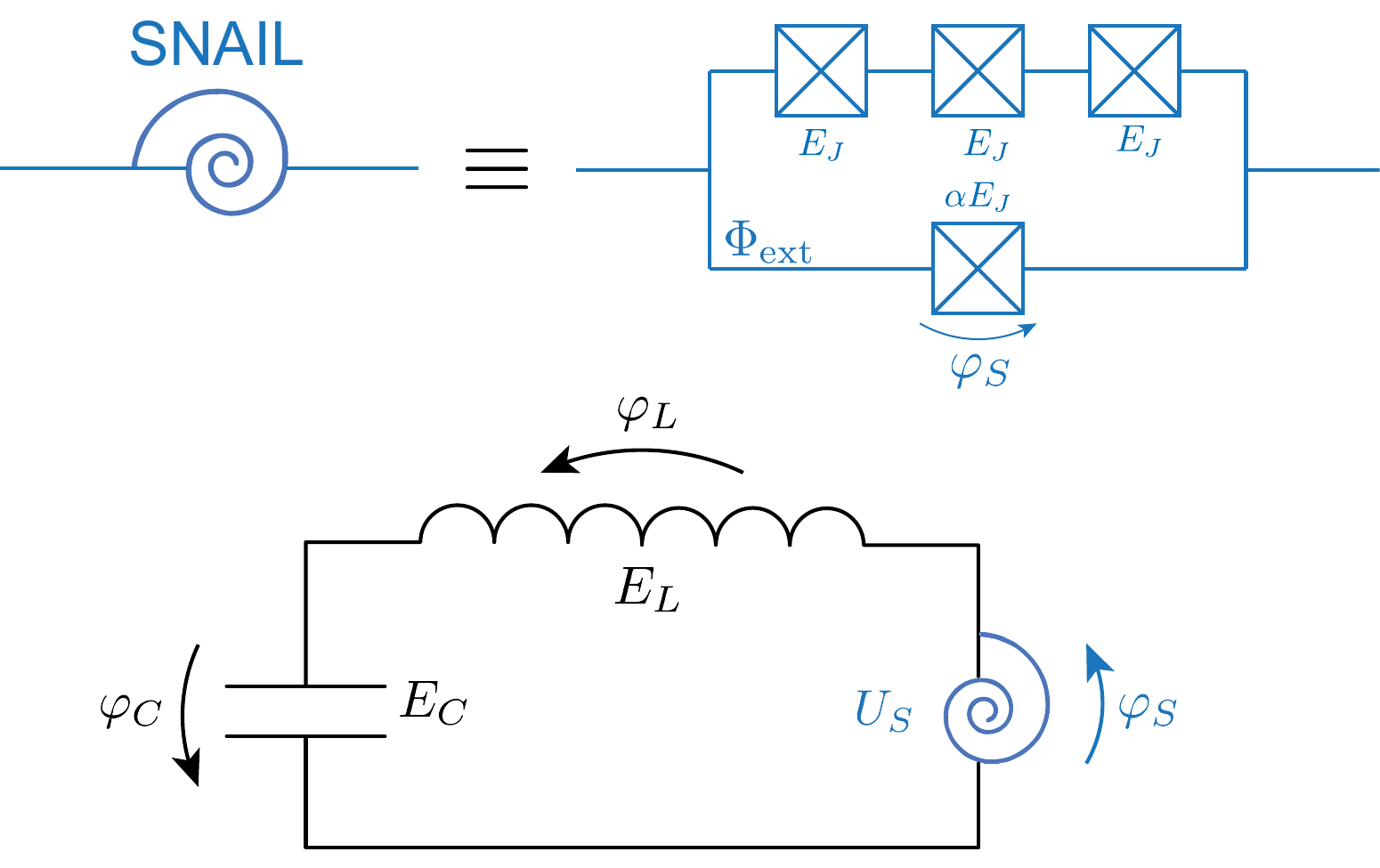}
    \caption{\textbf{Schematic of the SNAIL superconducting circuit used to activate beams-splitters between modes.} The SNAIL consists of three Josephson junction with junction energy $E_J$ in parallel with a junction with smaller energy $\alpha E_J$. The loop is threaded with an external flux $\Phi_\text{ext}$. We connect the SNAIL to a capacitor with charging energy $E_C$ and an inductor with inductive energy $E_L$ to make a SNAILmon circuit.}
    \label{fig:SNAIL_schematic}
\end{figure}

A SNAIL consists of three identical Josephson junctions with Josephson energy $E_J$ in parallel with a smaller junction of energy $\alpha E_J$, forming a superconducting loop threaded by an external magnetic flux $\Phi_{\mathrm{ext}}$, as shown in Fig.~\ref{fig:SNAIL_schematic}. The superconducting phase across the smaller junction, $\varphi_s$, serves as a quantum degree of freedom, while the external flux tunes the circuit's effective nonlinearity and resonance frequency.

The potential energy of the SNAIL is 
\begin{equation}\label{SNAIL potential}
    U_S(\varphi_S) = -E_J[\alpha\cos{\varphi_S} + 3\cos{\frac{\varphi_s - \varphi_{\rm ext}}{3}}],
\end{equation}
where $\varphi_{\mathrm{ext}}=2\pi\frac{\Phi_\mathrm{ext}}{\Phi_0}$ and $\Phi_0$ is the flux quantum. The first term in the potential arises from the smaller junction while the second term corresponds to the three larger junctions. 

To form a tunable coupler, the SNAIL is connected in series with a capacitor of capacitance $C$ (charging energy $E_C=\frac{e^2}{2C}$), where $e$ is the electron charge, and a linear inductor of inductance $L$ (inductive energy $E_L=\frac{\varphi_0^2}{L}$ where $\varphi_0=\Phi_0/2\pi$). Denoting the phases across the capacitor, inductor, and SNAIL by $\varphi_C, \varphi_L$, and $\varphi_S$, respectively, Kirchhoff's loop rule gives 
\begin{equation}
    \varphi_c = \varphi_L + \varphi_S.
\end{equation}

The Lagrangian of the circuit is 
\begin{equation}
    \mathcal{L} = \frac{1}{2} C \varphi_0^2 \Dot{\varphi}_C^2 -
            \frac{1}{2} E_L \varphi_L^2 -
            U_S(\varphi_S).
\end{equation}
The absence of a kinetic term for $\varphi_S$ yields the current conservation equation,
\begin{equation}
E_L(\varphi_c - \varphi_S) = \frac{\partial U_S}{\partial \varphi_S},
\end{equation}
which implicitly defines $\varphi_S$ as a nonlinear function of $\varphi_C$.  

We treat the nonlinearity perturbatively by first linearizing the SNAIL's potential about its minimum, $U_S \approx \frac{1}{2} E_J c_2 \varphi_S^2$, where $c_2$ is the second order Taylor coefficient. Plugging this expression into the current conservation equation and performing some algebra gives
\begin{equation}
\varphi_s = p\varphi_c, \qquad
p = \frac{x_J}{c_2 + x_J},
\end{equation}
where $x_J=\frac{E_L}{E_J}$, and $p$ is the participation ratio of the SNAIL which quantifies the fraction of the total inductive energy contributed by the SNAIL. The remaining $1-p$ contribution comes from the linear inductor.

Using this linear approximation, we incorporate the SNAIL's nonlinearity by defining an effective nonlinear potential as a function of $\varphi_C$,
\begin{equation}
U^{\mathrm{NL}}(\varphi_c) =
\frac{1}{2} E_L (\varphi_c - \varphi_s)^2 + U_S^{\mathrm{NL}}(\varphi_s),
\end{equation}
where $U_S^{\mathrm{NL}}=U_S(\varphi_S)-\frac{1}{2}E_Jc_2\varphi_S^2$ is the nonlinear part of the SNAIL's potential. Expanding $U^{\mathrm{NL}}$ in powers of $\varphi_C$ gives effective Taylor coefficients $\tilde{c}_n$ that are related to those of the bare SNAIL potential,
\begin{equation}
    \tilde{c}_n = -x_J \frac{d^{n-1}\varphi_S}{d \varphi_C^{n-1}},
\end{equation}
providing a systematic way to dress the SNAIL nonlinearity by the linear inductor.

The Hamiltonian of the SNAILmon tunable coupler circuit is then
\begin{equation}
    H_{\mathrm{TC}} = 4E_CN^2 + E_J\sum_{n=2}^{\infty}\frac{\tilde{c}_n}{n!}\varphi^n,
\end{equation}
where $N=\frac{Q}{2e}$ is the dimensionless charge on the capacitor and we choose to drop the subscript from $\varphi_C$. We promote $\varphi$ and $N$ to operators and impose canonical quantization using $[\hat{\varphi},\hat{N}]=i$ and 
\begin{align}
   \hat {\varphi} &= \varphi_{\mathrm{zpf}}(\hat{c}^\dag + \hat{c}) \\
   \hat{N} &= i N_{\mathrm{zpf}}(\hat{c}^\dag - \hat{c}),
\end{align}
where $\varphi_{\mathrm{zpf}}$ and $N_{\mathrm{zpf}}$ are the zero-point fluctuations of the phase and charge, respectively, and $\hat{c}^\dag,\hat{c}$ are the creation and annihilation operators of the SNAILmon tunable coupler. Then the Hamiltonian takes the form
\begin{equation} 
    \hat{H}_{\mathrm{TC}} = \omega_c \hat{c}^\dag \hat{c} + \sum_{n=3}^{\infty} g_n (\hat{c}^\dag + \hat{c})^n,
    \label{eq:SNAILmon Hamiltonian}
\end{equation}
where $\omega_C=\sqrt{8 \tilde{c}_2 E_C E_J}$ is the resonant oscillator frequency of the tunable coupler circuit and the coefficients $g_n=\varphi^n_{\mathrm{zpf}}\frac{\tilde{c}_n}{n!}$.

\subsection{SNAIL-mediated beam-splitter interactions} \label{sec:Beam-splitter Hamiltonian}
The SNAILmon tunable coupler circuit is engineered to have a large third-order nonlinearity, $\frac{\tilde{c}_3}{\tilde{c}_n} \gg 1$, which leads strong three-wave mixing enabling fast high fidelity beam-splitter operations between cavity modes coupled to it \cite{chapman_high--off-ratio_2023,lu2023high, frattini2017}.

We consider a pair of cavity modes coupled to an externally driven SNAILmon tunable coupler. The system Hamiltonian is,
\begin{align}
    \hat{H}_{\mathrm{sys}} &= \hat{H}_{\mathrm{TC}} + \hat{H}_{\mathrm{cavity}} + \hat{H}_{\mathrm{coupling}} + \hat{H}_{\mathrm{drive}},\\
    \hat{H}_{\mathrm{cavity}} &= \omega_a \hat{a}^\dag \hat{a} + \omega_b \hat{b}^\dag \hat{b},\\
    \hat{H}_{\mathrm{coupling}} &= i \sum_{k\in\{a,b\}} g_{k} (\hat{k}^\dag+ \hat{k})(\hat{c}^\dag - \hat{c}),\\
    \hat{H}_{\mathrm{drive}} &= (\epsilon e^{-i(\omega_d t - \varphi)} - \epsilon^* e^{i(\omega_d t + \varphi)})(\hat{c}^\dag - \hat{c}).
\end{align}
$\hat{H}_{\mathrm{TC}}$ is given in Eq.~\ref{eq:SNAILmon Hamiltonian} (up to third order) with $\hat{c}^\dag$ / $\hat{c}$ being the tunable coupler's creation/annihilation operator. $\hat{H}_{\mathrm{cavity}}$ is the Hamiltonian of the cavity oscillator mode Alice(Bob), with frequency $\omega_{a(b)}$ and creation and annihilation operators $\hat{a}^\dag(\hat{b}^\dag)$ and $\hat{a}(\hat{b})$. $\hat{H}_{\mathrm{coupling}}$ describes the capacitive coupling of the tunable coupler to the cavity mode Alice(Bob) with coupling strength $g_{a(b)}$. Finally, $\hat{H}_{\mathrm{drive}}$ describes an external microwave tone of amplitude $\epsilon$ and frequency $\omega_d$ driving the charge on the SNAILmon capacitor, and phase $\varphi$. 

To see the beam-splitter operation in the interaction picture, we first diagonalize $\hat{H}_{\mathrm{coupling}}$ via a Schrieffer-Wolff transformation
\begin{equation}
    \hat{U}_S = e^{i\sum_{k \in a,b} \frac{g_k}{\Delta_k}(\hat{k}\hat{c}^\dag + \hat{k}^\dag\hat{c})},
\end{equation}
where $\Delta_k = \omega_c -\omega_k$. This transformation hybridizes the tunable coupler mode with Alice and Bob, $\hat{c} \rightarrow \hat{c} - i \sum_{k \in a,b} \frac{g_k}{\Delta_k} \hat{k}$. Next, we eliminate the oscillator terms and external drive by moving to a displaced, rotating frame using
\begin{align}
    \hat{U}_D &= e^{\tilde{\zeta}(t) \hat{c}^\dag - \tilde{\zeta}^*(t) \hat{c}},\\
    \hat{U}_R &= e^{-i(\omega_a \hat{a}^\dag \hat{a}
                +\omega_b \hat{b}^\dag \hat{b}
                +\omega_c \hat{c}^\dag \hat{c})t},
\end{align}
where $ \tilde{\zeta}(t) = \zeta e^{-i \omega_d t}, \zeta = \frac{\epsilon}{-\Delta_d} e^{i\varphi},$ and $\Delta_d = \omega_d - \omega_c$. After these transformations the dominant contribution arises from the cubic nonlinearity. Setting the drive frequency to the difference of the Alice and Bob modes, $\omega_d = \omega_a - \omega_b$, and taking a rotating-wave approximation yields the beam-splitter Hamiltonian,
\begin{equation}
    \hat{H}_{\mathrm{BS}} = g_{\mathrm{b s}}(e^{i\varphi}\hat{a}\hat{b}^\dag + \mathrm{h.c.}),
\end{equation}
where the beam-splitter rate $g_{\mathrm{b s}}=6 g_3 \frac{g_a}{\Delta_a} \frac{g_b}{\Delta_b} \frac{\epsilon}{\Delta_d}$. 

\section{Transmon-mediated Parity Measurement} \label{sec:Parity Measurement}
We describe how to perform a projective measurement of the photon number parity of a cavity mode using a transmon ancilla. Later, we generalize this protocol to enable full photon number measurements. 
The cavity mode is dispersively coupled to the transmon via
\begin{equation}\label{eq:Dispersive Hamiltonian}
    \hat{H}_\mathrm{disp} = \frac{\chi}{2} \hat{\sigma}_z \hat{n},
\end{equation}
where $\chi$ is the dispersive shift, $\hat{\sigma}_z$ is the Pauli-Z operator acting on the two lowest transmon states $\{\ket{g}, \ket{e}\}$, and $\hat{n}$ is the number operator of the cavity mode. This dispersive interaction causes the transmon transition frequency to shift depending on the number of photons in the cavity. As a result, the cavity's photon number information can be encoded in the phase of the transmon and subsequently readout using a Ramsey sequence. 

Take the initial state of the transmon, cavity system to be $\ket{\psi}=\ket{g,n}$, where $n$ is the photon number in the cavity mode. A resonant $\frac{\pi}{2}$-pulse on the transmon prepares it in a superposition, bringing the state to $\ket{\psi}=\frac{1}{\sqrt{2}}(\ket{g,n} + \ket{e,n}).$ The system is left to evolve freely under its dispersive interaction for a time $T$. During this interval, the transmon's Bloch vector precesses around the equator at a rate $\chi n$. This evolution imprints a photon-number-dependent relative phase between the transmon basis states,
\begin{equation}
    \ket{\psi(T)} =  \frac{1}{\sqrt{2}}(e^{i \frac{\chi}{2}  n T}\ket{g,n} + e^{-i \frac{\chi}{2}  n T}\ket{e,n}).
\end{equation}
A second $\frac{\pi}{2}$-pulse is then applied to the transmon after which the probability of the transmon being in $\ket{g}$ or $\ket{e}$ is given by
\begin{align}
    P_g = \sin^2 \left(\frac{\chi}{2} n T \right), \\
    P_e = \cos^2 \left(\frac{\chi}{2} n T \right).
\end{align}
If we set the idle time between $\frac{\pi}{2}-$pulses to $T=\frac{\pi}{\chi}$, the argument in the above expressions becomes $\frac{n\pi}{2}$. In this case reading out the transmon will deterministically project it to $\ket{e}$ or $\ket{g}$ for even or odd photon numbers, respectively. Thus, this Ramsey sequence performs a projective measurement of the cavity's photon number parity.

\section{Photon counting through repeated measurements of binary digits}\label{sec:Repeated Measurement Protocol For Photon Counting}
\label{sec: parity measurement}
\begin{figure*}\label{fig:Measurement protocol}
    \centering
    \includegraphics[width=\linewidth]{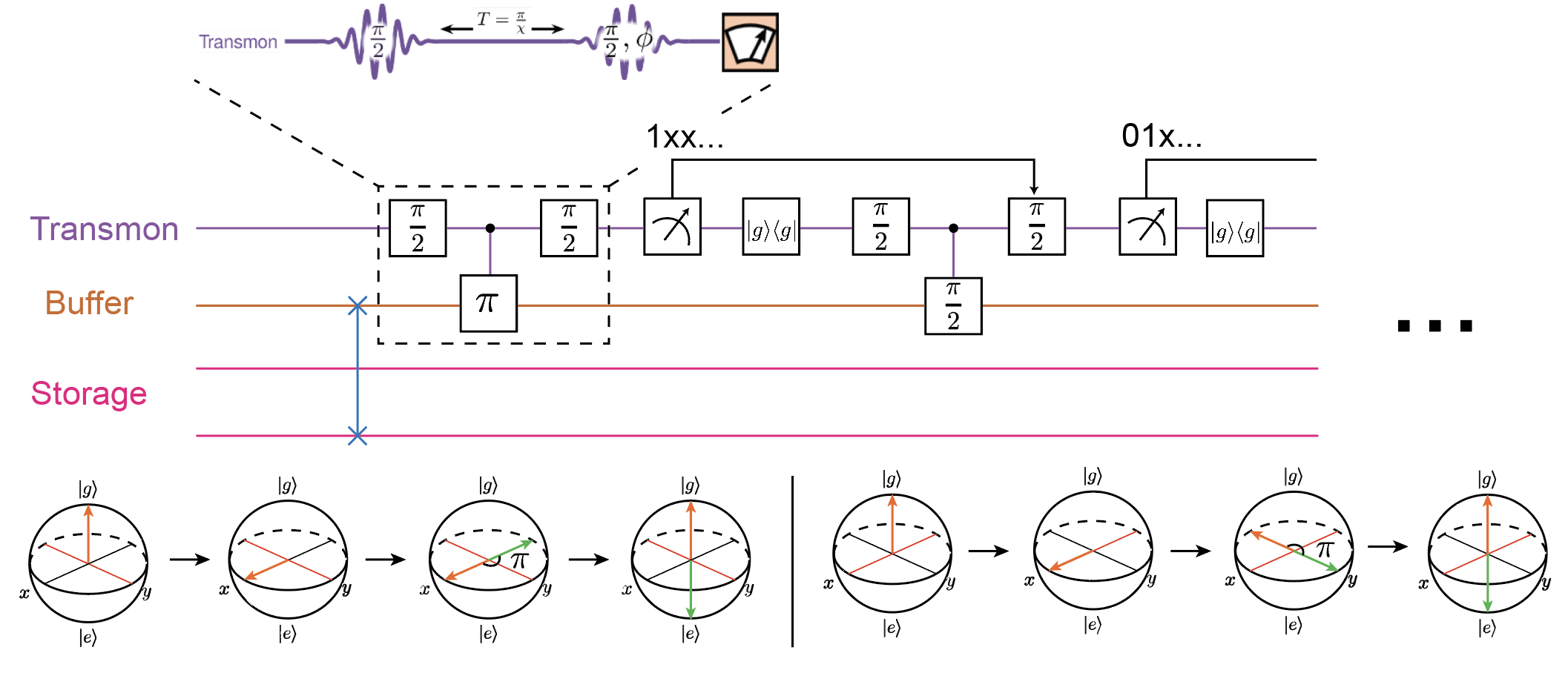}
    \caption{\textbf{Protocol for measuring the photon number of an oscillator mode using a transmon.} To measure the photon number of a storage mode, the mode is swapped into the buffer cavity using the SNAILmon (blue line). The photon number is then determined bitwise in the binary representation using consecutive Ramsey sequences (dashed box). In each sequence, the wait time between the $\pi/2$ pulses is adjusted and the phase of the final pulse is set using feedforward from previous bit measurements, and the transmon is reset to its ground state before the next sequence. The Bloch spheres show the evolution of the transmon’s Bloch vector during measurements of the 0th and 1st bits, conditioned on whether the bit is 0 (green) or 1 (orange).}
    \label{fig:Photon_count}
\end{figure*}
In the previous section, we showed a Ramsey sequence which performs a parity measurement of the cavity's photon population, which corresponds to measuring the $0^{\mathrm{th}}$ bit in the binary expansion of $n$. We now generalize this protocol to measure up to an arbitrary bit of $n$, thereby enabling a fully projective photon number measurement \cite{Wang_2020}. 

To access higher order bits, we modify the idle time $T$ between the $\pi/2$-pulses  and by introducing a phase shift $\phi$ to the axis about which we drive the final $\pi/2$-pulse. With this modification, the probabilities to measure the transmon in $\ket{g}$ or $\ket{e}$ become
\begin{align}
    P_g = \sin^2 \left(\frac{\chi}{2} n T + \phi \right), \\
    P_e = \cos^2 \left(\frac{\chi}{2} n T + \phi \right).
\end{align}
As was the case for the parity measurement, the goal is to choose the idle time $T$ and phase $\phi$ such that the argument in the above expressions is an integer multiple of $\pi$, causing the measurement outcome of the transmon to be conditioned on the bit we wish to measure.

To see how this can be achieved, we consider the binary expansion of $n$,
\begin{equation}
    n = \sum_{i=0}^{K} b_i 2^i, \qquad b_i \in \{0,1\}.
\end{equation}
Suppose we wish to measure the $m$-th bit $b_m$. Dividing $n$ by $2^m$ separates the contribution of this bit from others: bits $>m$ make an even contribution to $\frac{n}{2^m}$ while all bits < $m$ make a total fractional contribution strictly less than one. Consequently, the parity of $\lfloor\frac{n}{2^m}\rfloor = \frac{n}{2^m} - \sum_{j=0}^{m-1} b_j \frac{2^j}{2^m}$ is determined by the value of $b_m$. 

In the transmon-cavity system, dividing by $2^m$ is implemented by setting the idle time $T=\frac{\pi}{2^m\chi}.$ The fractional contribution from the preceding bits is canceled by setting $\phi=-\frac{\tilde{n}\pi}{2^m}$, where $\tilde{n}$ denotes the photon number revealed by measuring all bits before the $m$-th. With these parameters, the measurement probabilities of the transmon state become,
\begin{align}
    P_g = \sin^2 \left(\frac{n - \Tilde{n}}{2^m} \pi \right), \\
    P_e = \cos^2 \left(\frac{n - \Tilde{n}}{2^m} \pi \right),
\end{align}
which deterministically map $\ket{e}$ to $b_m=0$ and $\ket{g}$ to $b_m=1$. 

Importantly, measuring the $m$-th bit requires prior measurements of all preceding bits since $\phi$ depends on $\tilde{n}$, necessitating real-time feedback. By performing these Ramsey sequences sequentially to measure all bits, we build up to a full photon number measurement. This repeated measurement protocol, shown in Fig.~\ref{fig:Photon_count}, is enabled by the long lifetimes of the cavity modes, which ensures photon loss remains minimal over the duration of the sequence.

\section{Preparing an Entangled State of the Reference Mode and the Multimode Cavity Register} \label{sec:Preparation Sequence For Entangled State of the Ancilla and the Multimode Cavity Register}
To probe the bosonic MIPT through ancilla purification, we prepare an initial state in which a dedicated ancillary storage mode is maximally entangled with two orthogonal half-filled configurations of the system storage modes: $\frac{1}{\sqrt{2}}(\ket{0101...}\ket{0} + \ket{1010...}\ket{1})$. This section details a gate sequence using universal sideband control on the RAQM architecture to construct this entangled "checkerboard" state ~\cite{huang2025fastsidebandcontrolweakly}.

We label the Hilbert space as $\ket{s_1s_2s_3..., \mathrm{ancilla}, \mathrm{transmon};ij}$, where $s_k$'s denote storage modes, the ancilla is a dedicated storage mode, the transmon is indexed by its three lowest energy states, $\ket{g}, \ket{e},\ket{f}$, and $i,j$ label the buffer cavity modes. Starting from $\ket{000...,0;g;0,0}$, we sequentially apply gates labeled as $\Theta^I_G$, where $\Theta$ is the pulse-angle, $G$ specifies the state manifold supported, and $I$ indicates the subsystems targeted: $q$ for transmon and $i,j$ for the buffer modes. For example, $\pi^i_{f0g1}$ denotes a $f0-g1$ sideband on buffer mode $i$, while $\frac{\pi}{2}^q_{ef}$ denotes a $\frac{\pi}{2}-$pulse between the $\ket{e}$ and $\ket{f}$ states of the transmon.

The first set of gates prepares an entangled superposition of single photon excitation states of the buffer modes to the $\ket{g}, \ket{e}$ states of the transmon,
\begin{equation}\label{eq:Shelving_protocol}
  \pi^j_{f0g1} \cdot \pi^q_{ge} \cdot \pi^q_{ef} \cdot \pi^i_{f0g1} \cdot \frac{\pi}{2}^q_{ef} \cdot \pi^q_{ge}
\end{equation}
Restricting attention to the transmon buffer subspace, $\ket{\mathrm{transmon; ij}}$, the state evolves as
\begin{align}\label{eq:State_evol_shelving}
    \ket{g;00} &\xrightarrow{\pi^q_{ge}} \ket{e;00} \notag \\
    &\xrightarrow{\frac{\pi}{2}^q_{ef}} \frac{1}{\sqrt{2}}(\ket{e;00} + \ket{f;00})  \notag\\ &\xrightarrow{\pi^i_{f0g1}} \frac{1}{\sqrt{2}}(\ket{e;00} + \ket{g;10})  \notag\\ 
    &\xrightarrow{\pi^q_{ef}} \frac{1}{\sqrt{2}}(\ket{f;00} + \ket{g;10})  \notag\\ 
    &\xrightarrow{\pi^q_{ge}} \frac{1}{\sqrt{2}}(\ket{f;00} + \ket{e;10})  \notag \\
    &\xrightarrow{\pi^j_{f0g1}} \frac{1}{\sqrt{2}}(\ket{g;01} + \ket{e;10}). 
\end{align}
Using the tunable coupler, we swap buffer mode $i$ into storage mode $s_1$ and buffer mode $j$ into $s_2$, resulting in the system state $\frac{1}{\sqrt2}(\ket{1000...000,0;e;00} +\ket{010...000,0;g;00})$.

To populate the next pair of storage modes, we perform a $\pi^q_{ef}$-pulse to bring the state to $\frac{1}{\sqrt{2}}(\ket{g;00} + \ket{f;00}$, then repeat the gate sequence in Eq.~\ref{eq:Shelving_protocol}. This brings our transmon-buffer cavity subsystem to the same final state in Eq.~\ref{eq:State_evol_shelving}. The repeated $\pi^q_{ef}$ and $\pi^q_{ge}$ pulses serve to "shelve" the $\ket{g}$ branch of the transmon's superposition in the $\ket{e}$ manifold, making it invisible to the proceeding $f0g1$ sideband, rendering it insensitive to off-resonant driving during the proceeding $f0g1$ sideband and preserving the coherence of the superposition state~\cite{huang2025fastsidebandcontrolweakly}. We again swap buffer mode $i$ into $s_3$ and buffer mode $j$ into $s_4$, resulting in the system state 
$\frac{1}{\sqrt2}\left(\ket{0101...000,0;g;00} + \ket{1010...000,0;e;00}\right)$.

Repeating this second set of gates populates the storage cavity in an alternating pattern, $\frac{1}{\sqrt{2}}\left(\ket{010101...10,0;g;00} +\ket{1010101...01,0;e;00}\right)$. To end the protocol we perform $\pi_{ef}-$pulse followed by a $\ket{f0}\!-\!\ket{g1}$ sideband to transfer the excitation in the $\ket{e}$ branch into buffer mode $i$, then swap the excitation into the ancilla mode resulting in the desired checkerboard state: $\frac{1}{\sqrt2}\left(\ket{010101...10,0;g;00} +\ket{1010101...01,1;g;00}\right)$.

We take the gate time for a $\pi_{ge}$-pulse to be 20 ns, $\pi_{ef}$-pulse to be 30 ns, and a $\pi_{f0g1}$ sideband to be 100 ns. The total time for state preparation is therefore
\begin{equation}
    T_\mathrm{state\,prep}=1.9\,\mu\text{s} + (L/2) \times 1.3\mu s,
\end{equation}
where $L$ is the system size.

\section{Hubbard-like interactions via Selective number dependent arbitrary phase gates} \label{sec:Hubbard-like interactions via Selective number dependent arbitrary phase gates}
In this section we show how to achieve a Hubbard-like onsite interaction via selective number dependent arbitrary phase (SNAP) gates enabled by a transmon dispersively coupled to a cavity mode.

The dispersive interaction discussed in Sec.~\ref{sec: parity measurement} shifts the transmon's transition frequency depending on the cavity's photon population, allowing for independent control over the transmon within each photon number subspace. By driving the transmon at the photon number resolved frequencies and steering its Bloch vector along a closed path, an arbitrary phase can be imprinted on each Fock state,
\begin{equation}
    \hat{S} = \sum_n e^{i\theta_n} \ket{n}\bra{n}.
\end{equation}
The SNAP gate is performed in two stages. First, a selective multitone drive applies photon-number-dependent phases phase shifts to the state while exciting the transmon. Second, a fast, nonselective $\pi-$pulse returns the transmon to its ground state. 

Optimal selectivity favors longer pulse times, though longer pulse times increase incoherent errors, while faster times increase coherent errors. Recent work has shown that SNAP gates can instead be executed rapidly, on timescales $\mathcal{O}(\frac{1}{\chi})$, where coherent errors can be systematically suppressed through pulse optimization \cite{landgraf2024fastquantumcontrolcavities}. The multitone drive takes the form
\begin{equation}
    \Omega(t)
= \sum_n \lambda_n \, e^{i\left(\omega_n t + \alpha_n - \Delta \omega_n T / 2\right)},
\end{equation}
where $\lambda_n$, $\omega_n$ and $\alpha_n$ are the photon number ($n$) dependent drive amplitudes, frequencies, and phases, respectively, $\Delta \omega_n=\omega_n - \chi n$, and $T$ is the pulse duration. Starting from an initial guess, the drive parameters are iteratively updated according to 
 \begin{equation}
\Delta \lambda_n = -\frac{\epsilon_n^{(L)}}{2T}, 
\qquad
\Delta \alpha_n = -\Delta \theta_n, 
\qquad
\Delta \omega_n = \frac{\pi \epsilon_n^{(T)}}{2T}.
\end{equation}
Here $\Delta\theta_n$ is the phase error relative to the target $\theta_n$, while $\epsilon^{(L)}$ and $\epsilon^{(T)}$ quantify the longitudinal and transverse errors of the transmon's final Bloch vector from the desired state. These errors are extracted by projecting the Bloch vector onto a tangent plane to the $\ket{e}$ state on the Bloch sphere. Iterating the parameters according to the above rules perturbatively suppresses coherent errors while maintaining a fixed short gate time. We refer the reader to ~\cite{landgraf2024fastquantumcontrolcavities} for full details.

Choosing the SNAP phases as
\begin{equation}
    \theta_n=Un^2,
\end{equation}
 we realize the non-Gaussian Hubbard gates of our model.

\section{Error Budget for the experiment} \label{sec:Error Budget for the experiment}

\begin{table*}[t]
\caption{Hamiltonian parameters used in Lindblad simulations.}
\label{tab:SM_hamiltonian_parameters}
\begin{ruledtabular}
\begin{tabular}{llll}
Subsystem & Parameter & Definition & Value \\ 
\hline

\multirow{4}{*}{Cavity--tunable coupler} 

& $g_{1}$ & Coupling strength between cavity mode 1 and coupler & 0.533 GHz \\
& $g_{2}$ & Coupling strength between cavity mode 2 and coupler & 0.592 GHz \\

& $\Delta_{1}$ & Detuning between cavity mode 1 and coupler & 2.261 GHz \\
& $\Delta_{2}$ & Detuning between cavity mode 2 and coupler & 2.590 GHz \\
& $g_{\mathrm{bs}}$ & Effective beam-splitter coupling rate & 0.500 MHz \\

\hline

\multirow{1}{*}{Cavity--transmon (SNAP)} 

& $\chi$ & Dispersive shift between cavity mode and transmon & -5.352 MHz \\

\multirow{1}{*}{Cavity--transmon (Parity)} 
& $\chi$ & Dispersive shift between cavity mode and transmon & -2.0 MHz \\

\end{tabular}
\end{ruledtabular}
\end{table*}

\begin{table*}[t]
\caption{Coherence time scales used in Lindblad simulations.}
\label{tab:SM_noise_parameters}
\begin{ruledtabular}
\begin{tabular}{llll}
Subsystem & Parameter & Definition & Value \\ 
\hline

\multirow{2}{*}{Cavity} 
& $T_1$ & Intrinsic relaxation time & 1.5(10) ms \\
& $n_a$ & Thermal occupation & 0.1\% \\

\hline

\multirow{9}{*}{Cavity--tunable coupler} 
& $T_1^C$ & Coupler energy relaxation time & $50$ $\mu s$ \\
& $T_\phi^C$ & Coupler pure dephasing time & $5$ $\mu s$  \\
& $n_C$ & Coupler thermal occupation & 2\% \\
& $T_{\mathrm{Th}}$ & Drive-induced thermal dephasing time & 2.5 ms \\
& $\tilde{T}_{1}^{\,a_1}$ & Effective relaxation time of cavity mode 1 & $562(824)$ $\mu s$ \\
& $\tilde{T}_{\phi}^{\,a_1}$ & Effective dephasing time of cavity mode 1 &  $7$ $\mu s$\\

& $\tilde{T}_{1}^{\,a_2}$ & Effective relaxation time of cavity mode 2 & $584(874)$ $\mu s$ \\
& $\tilde{T}_{\phi}^{\,a_2}$ & Effective dephasing time of cavity mode 2 & $92$ $\mu s$ \\

\hline

\multirow{3}{*}{Cavity--transmon qubit} 
& $T_1^q$ & Energy relaxation time & $200$ $\mu s$ \\
& $n_q$ & Thermal occupation & 0.5\% \\
& $T_\phi^q$ & Pure dephasing time & $100$ $\mu s$ \\

\end{tabular}
\end{ruledtabular}
\end{table*}

To model the impact of accumulation of errors, we simulate the full circuit by composing a noisy quantum error channel, which includes relevant dissipation and decoherence, for its elementary operations. Each error channel is obtained by numerically simulating Lindbladian evolution on the minimal target subsystem of the operation. We use the QuTiP python package~\cite{qutip5, johansson2013qutip, johansson2012qutip} to numerically time evolve the Lindblad master equation,
\begin{equation}
\label{eq:Lindbladian master equation}
\begin{split}
\dot{\rho}(t) =\;& -\frac{i}{\hbar} [\hat{H}(t), \rho(t)] \\
&+ \sum_n \frac{1}{2}
\left(
2 \hat{C}_n \rho(t) \hat{C}_n^\dagger
- \rho(t) \hat{C}_n^\dagger \hat{C}_n
- \hat{C}_n^\dagger \hat{C}_n \rho(t)
\right).
\end{split}
\end{equation}
where $\rho(t)$ is the system's density matrix at time $t$, $\hat{H}(t)$ is the time dependent Hamiltonian of the system, and $\hat{C}_n = \sqrt{\gamma_n}\hat{A}_n$ are collapse operators that describe coupling to the environment with strength $\gamma_n$.

The quantum error channel is a superoperator $U$ that maps an initial density matrix to a final density matrix under the corresponding noisy operation. A library of such superoperators for each operation is collected to compose a full end-to-end simulation. We consider four primary error channels corresponding to the main operations in our model. All relevant circuit and cavity parameters and coherence time scales used in simulation are included in Tables~\ref{tab:SM_hamiltonian_parameters},~\ref{tab:SM_noise_parameters}.

\subsection{Decay error} \label{subsec:Decay error}
As a baseline, we simulate the photon loss and thermal excitation of a single idle cavity mode, modeled as a truncated harmonic oscillator evolving freely for a time $T$. The decay and heating processes are modeled by $\sqrt{\kappa_a(1+n_a)} \hat{a}$ and $\sqrt{\kappa_an_a} \hat{a}^\dagger$, respectively, where $\kappa_a=1/T_1$ is the cavity's decay rate and $n_a$ is the average thermal occupation. Throughout, we take $T_1=1.5(10)$ ms and $n_a=0.001$.

\subsection{Beam-splitter gate error} \label{Beam-splitter gate error}
To model noisy BSFP operations, we simulate two cavity modes evolving under the beam-splitter Hamiltonian, with $\varphi=0$. 

Though we do not explicitly model the SNAILmon tunable coupler in our Lindblad simulations, we incorporate the inherited decay and decoherence of cavity modes from the coupler. The decay rate inherited by the cavity is 
\begin{equation}
    \tilde{\kappa}_a = \kappa_a + \left(\frac{g}{\Delta} \right)^2\kappa_C,
\end{equation}
where $g$ is the coupling strength between the cavity mode and the tunable coupler, $\Delta$ is the detuning between them, and $\kappa_C$ is the decay rate of the coupler. 
In the white noise regime, the inherited dephasing rate by the cavity mode is 
\begin{equation}
    \gamma_a \approx \left(\frac{g}{\Delta} \right)^4 \gamma_C,
\end{equation}
where $\gamma_C$ is the pure dephasing rate of the coupler. However, if the noise spectrum is pink and low frequency noise dominants, the inherited dephasing rate is instead,
\begin{equation}
    \gamma_a \approx \left(\frac{g}{\Delta} \right)^2 \gamma_C.
\end{equation}
As a moderate choice, we assume pink noise, consistent with experimental observations~\cite{chapman_high--off-ratio_2023}. In the presence of an external drive, the dephasing of the tunable coupler manifests as heating, inducing additional dephasing. This thermal dephasing rate is given by
\begin{equation}
    \gamma_{\mathrm{Th}} \approx n_C \kappa_C,
\end{equation}
where $n_C$ is the average thermal population of the coupler. We take the total inherited dephasing rate by cavity mode to be,
\begin{equation}
    \tilde{\gamma}_a = \left(\frac{g}{\Delta} \right)^2 \gamma_C + \gamma_{\mathrm{Th}},
\end{equation}
with the collapse operator taken to be $\hat{a}^\dag \hat{a}$.
We extract a noisy beam-splitter propagator with an average gate fidelity of $0.997(0)$.

Randomizing the phase $\varphi$ amounts to changing the phase of the external charge drive on the SNAIL and thus does not separate a spererate error channel. To obtain a noisy BSRP gate with phase $\alpha$, we simply apply the following unitary operation on the noisy BSFP gate,
\begin{equation}
    \hat{V}_\mathrm{BSRP}(\varphi=\alpha)=e^{i\alpha \hat{a}^\dag\hat{a}} \hat{V}_\mathrm{BSFP}(\varphi=0) e^{-i\alpha \hat{a}^\dag\hat{a}},
\end{equation}
 where $\hat{a}$ is the destruction operator for one of the target modes.

\subsection{Parity measurement error} \label{subsec:Parity measurement error}
Parity measurement errors are modeled by simulating a single cavity mode dispersively coupled to a transmon qubit. Intrinsic photon loss and thermal excitation for the cavity are included as before, and transmon decay, thermal excitation, and dephasing are modeled using the collapse operators $\sqrt{\kappa_q(1+n_q)} \hat{\sigma}_-,\sqrt{\kappa_q n_q} \hat{\sigma}_+$, and $\sqrt{\gamma}\hat{\sigma}_z$, respectively. We realize a parity measurement fidelity of $0.997(0)$. In addition to the intrinsic noise of the parity measurement, we model readout error by introducing a misassignment probability $\epsilon=0.004$, corresponding to the fraction of measurements in which the outcome is flipped and the state collapses onto the incorrect parity.

\subsection{SNAP gate error} \label{subsec:SNAP gate error}
The noisy SNAP gates are modeled using the cavity transmon system as in the parity measurements and the same noise parameters. Averaged over random cavity superposition states, we obtain a noisy SNAP gate fidelity of $0.997(0)$.

\subsection{Full circuit simulation} \label{subsec:Full circuit simulation}
Having obtained propagators for cavity dissipation, beam-splitter gates, SNAP (on-site Hubbard) gates, and parity measurements, we now simulate full monitored circuit realizations. This allows us to quantify how different noisy gates affect our ability to identify the transition. These simulations neglect correlated many-body dephasing errors and only include local decay and dephasing errors as described above.

The full circuit consists of four cavity modes, a dedicated ancilla mode, and a transmon qubit. The cavity modes are modeled as harmonic oscillators with a truncated local Hilbert space dimension of 3 so that at half-filling each oscillator can be occupied by a maximum of two photons. The ancilla and transmon are modeled as two-level systems. The total Hilbert space is therefore indexed as $\mathrm{mode}_1\, \otimes \, \mathrm{mode}_2\, \otimes \, \mathrm{mode}_3\, \otimes \, \mathrm{mode}_4\, \otimes \, \mathrm{transmon}\, \otimes \, \mathrm{ancilla}.$
The propagators obtained from each subsystem simulation are embedded into this larger Hilbert space. Since the noisy gates are modeled as superoperators, we require simulations on the doubled Hilbert space of dimension 104,976. 

The system is initialized in the globally entangled checkerboard state described in Sec.5, $\frac{1}{\sqrt{2}} \left( \ket{1010,g,1} + \ket{0101,g,0} \right)$. For any given layer of our circuit, the operations are applied in the following order:

\textbf{Beam-splitter gates}
\begin{itemize}
    \item Apply beam-splitter propagators for a random durations ($\theta$) and phases $\varphi$ (if BSRP) on modes (1,2) and (3,4), alternating with (1,2) to implement the brick-layer geometry with open boundary conditions.
    \item For every application of a beam-splitter gate, apply a decay propagator to the ancilla qubit as well as spectator modes to account for the idle time while performing the operation.
\end{itemize}

\textbf{On-site Hubbard gates}

For every cavity mode:
\begin{itemize}
    \item Apply a 500 ns decay propagator to all modes and the ancilla to simulate idle time during the swap of the target mode to the buffer cavity.
    \item Apply the SNAP gate propagator to the target mode and transmon. 
    \item Apply decay propagators to all spectator modes and the ancilla to simulate idle time during the SNAP operation.
    \item Reset the transmon ancilla to the ground state and apply a decay propagator to all modes and the ancilla qubit to simulate idle time during the reset operation.
    \item Apply another 500 ns decay propagator to all modes and ancilla to simulate idle time for the swap of the target mode back to the storage cavity.
\end{itemize}

\textbf{Parity Measurements}

For each mode selected for measurements:
\begin{itemize}
    \item Apply a 500 ns decay propagator to all modes and the ancilla qubit to simulate idle time during the swap the target mode to the buffer cavity.
    \item Apply the parity measurement propagator to the target mode and transmon.
    \item Apply decay propagators to all spectator modes and the ancilla to account for idle time during the parity measurement operation.
    \item Apply decay propagators to all modes, the transmon, and the ancilla to account for the idle time during readout. 
    \item Collapse the transmon into the ground or excited state according to the appropriate Born probabilities obtained from its reduced density matrix. Apply a 4\% chance to flip the outcome and collapse onto the opposite parity, corresponding to a readout error.
    \item Apply another 500 ns decay propagator too all modes and ancilla to simulate the idle time for the swap of the target mode back to the storage cavity.
    \item Repeat the step above for every mode that has to be measured.
\end{itemize}
Two clarifying remarks: (1) Although we account for the idle time during swaps between storage and buffer cavities, we do not explicitly simulate the buffer mode, as doing so would significantly increase computational cost. (2) Transmon dissipation and decoherence are included in the SNAP gates and parity measurements. During other operations the transmon remains in its ground state and acts as a spectator, so additional decay modeling is unnecessary.

 The circuit begins with $S$ scrambling layers consisting of BSFP or BSRP gates followed by on-site Hubbard gates but no measurements, which prepares a sufficiently random state (see Sec. \ref{sec:stateprep}). This is followed by $M$ monitored layers where random parity measurements are also included at a rate $p$ and on-site Hubbard gates are optional depending on the model.
Including the overhead from swapping between the storage and buffer cavity and transmon reset, the effective experimental duration of a SNAP operation is $T_{\mathrm{SNAP}} \sim 1.32~\mu\mathrm{s}$, while a parity measurement takes $T_{\mathrm{parity}}\sim1.47~\mu\mathrm{s}$. The total experimental wall time is therefore
\begin{equation}
T_{\mathrm{wall}} = L\,\tilde{S}\,T_{\mathrm{SNAP}} + L M p\,T_{\mathrm{parity}} + L\,\frac{\tau_\mathrm{bs}}{2} (S+M),
\end{equation}
where $\tilde{S}=S$ for BSFP and BSRP monitored circuits and $\tilde{S}=S+M$ for monitored circuits with on-site Hubbard gates, and $\tau_\mathbf{bs}=0.25~\mu\mathrm{s}$ is the average duration of a random BSFP gate. For the end-to-end simulations reported here ($L=4$ and $S=M=8$), at $p=1$, this yields an average circuit duration of $\sim104~\mu\mathrm{s}$ (BSFP) and $152~\mu\mathrm{s}$ (BSFP and on-site Hubbard). BSRP circuits share the same time scale as BSFP circuits since randomizing $\varphi$ doesn't cost any additional time.

\begin{figure*}
    \centering
    \includegraphics[width=0.5\linewidth]{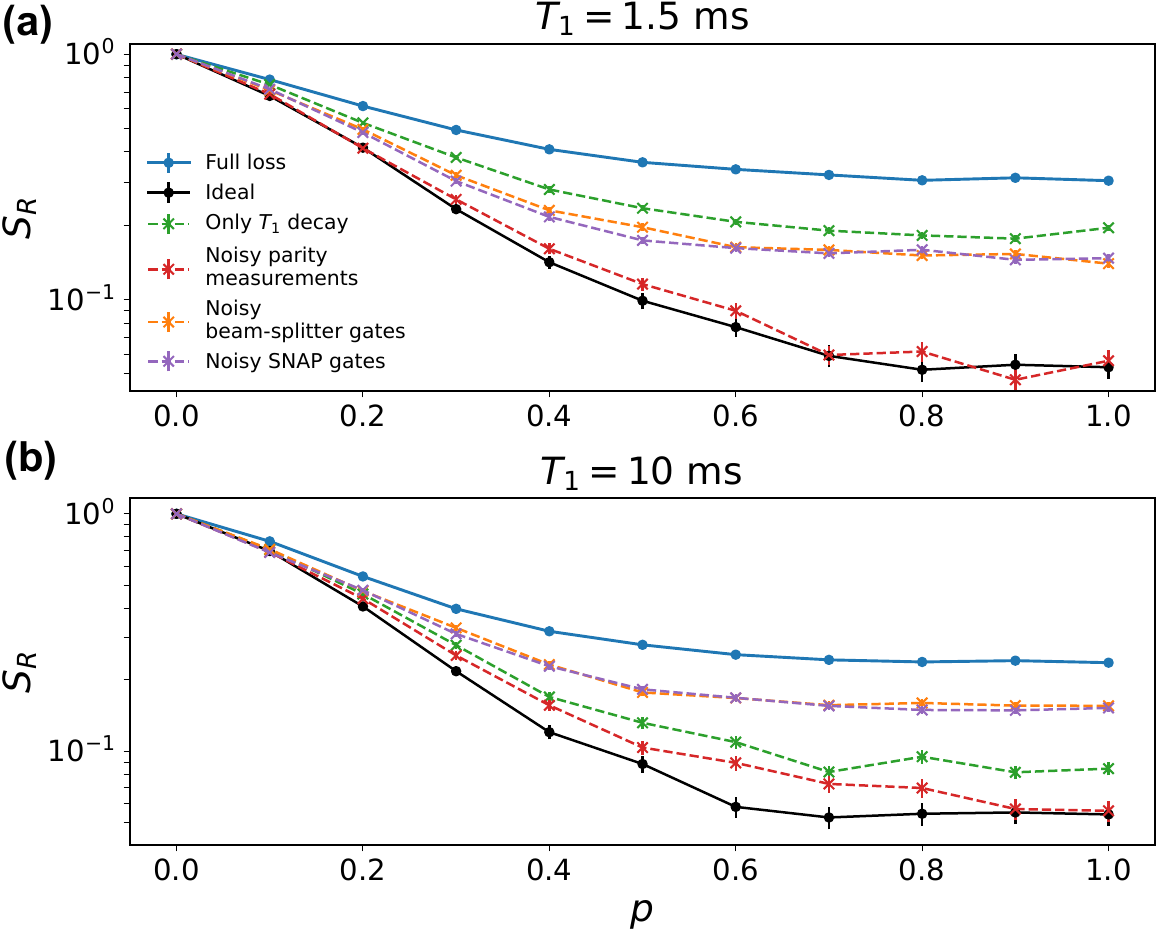}
    \caption{(a) Averaged entropy curves for circuits with BSFP and on-site Hubbard gates with different noise sources active given $T_1=1.5$ ms: beam-splitter (orange) noise, parity measurement noise (red), SNAP gate noise (in scrambling) (purple), $T_1$ decay only (green), all sources of noise active (blue), and the ideal noiseless case (black). (b) Averaged entropy curves for same noise scenarios but with $T_1=10$ ms.}
    \label{fig:Noisy gate analysis int}
\end{figure*}

\begin{figure*}
    \centering
    \includegraphics[width=1.0\linewidth]{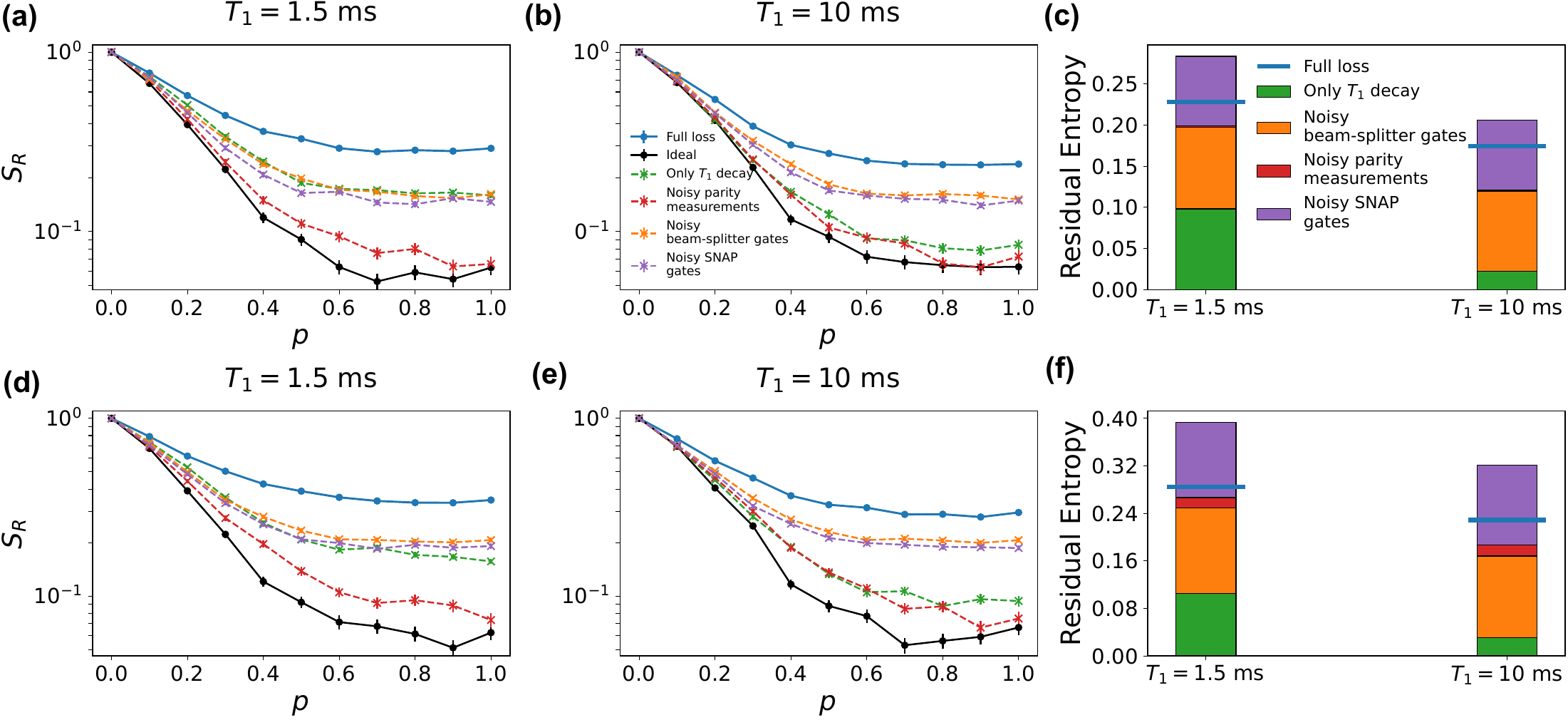}
    \caption{\textbf{Analysis of individual noisy gates}(Top row) BSFP circuits with no on-site Hubbard gates (a) Averaged entropy curves with different noise sources active given $T_1=1.5$ ms: beam-splitter (orange) noise, parity measurement noise (red), SNAP gate noise (in scrambling) (purple), $T_1$ decay only (green), all sources of noise active (blue), and the ideal noiseless case (black). (b) Averaged entropy curves for same noise scenarios but with $T_1=10$ ms. (c) Bar graph showing the excess residual entropy generated by different noise sources relative to the case where all noise sources are active. (Bottom row) BSRP circuits (d-f) Same plots as top row but for BSRP circuits.}
    \label{fig:Noisy gate analysis BSFP+BSRP}
\end{figure*}

In Fig.\ref{fig:Noisy gate analysis int}, we show the breakdown of the noisy gates on BSFP circuits with on-site Hubbard gates included in the monitored layers, as discussed in the main text. The simulations consider several individual noise channels including: noisy BSFP gates, noisy parity measurements, noisy SNAP gates (during scrambling), $T_1$ decay, and the case where all noise sources are activate. For each noise channel, we report the excess residual entropy relative to the ideal dynamics. Consistent with the analysis discussed in the main text, noisy parity measurements show a small residual entropy of 0.001(9) compared to noisy BSFP gates, noisy SNAP gates, and $T_1$ decay which are the primary sources of residual entropy. For a cavity lifetime of $T_1=1.5$ ms, when individually activated, $T_1$ decay gives a residual entropy of 0.140(8), noisy BSFP gates give 0.099(7), and noisy SNAP gates give 0.090(8). Increasing the lifetime to $T_1=10$ ms suppresses the contribution from the $T_1$ decay, bringing its residual entropy down to 0.031(7), after which the BSFP and SNAP gate imperfections become the dominant error sources. When all noise channels are activated simultaneously, the residual entropy is 0.251(8) for $T_1=1.5$ ms and 0.182(8) for $T_1=10$ ms.

In the top row of Fig.\ref{fig:Noisy gate analysis BSFP+BSRP}, we show the averaged entropy curves (over 1000 random samples) for BSFP circuits with $L=4$ modes and 16 layers (8 scrambling and 8 monitored). BSFP circuits follow similar qualitative lossy behavior to BSFP and on-site Hubbard gate model. The residual from noisy parity measurements is small at 0.001(9) compared to the rest of the noisy gates. The residual entropy from $T_1$ decay decreases from 0.098(8) to 0.022(7) when the lifetime is increased from $T_1=1.5$ ms to 10 ms. These residuals are smaller than those seen in BSFP and on-site Hubbard circuits due to the shorter wall times introduced from fewer SNAP operations. The residual error from noisy BSFP gates is similar to the BSFP with on-site Hubbard gate case at 0.096(7), and noisy SNAP gates (in scrambling) give 0.084(8). When all noise channels are active, the total residual entropy is 0.228(8) for $T_1=1.5$ ms and 0.174(8) for 10 ms. 

In the bottom row of Fig.\ref{fig:Noisy gate analysis BSFP+BSRP}, we show the breakdown of the noisy gates on BSRP circuits where $\varphi$ is randomized in the beam-splitter gates. The residual entropies from $T_1$ decay are consistent with the  BSFP results with no on-site Hubbard gates since the wall times for both circuits are similar. The residuals from the noisy beam-splitter gates, SNAP gates (in scrambling), and parity measurements are slightly higher than their BSFP counterparts at 0.145(9), 0.126(9), and 0.017(9) respectively. When all noise channels are active, the total residual entropy is 0.284(9) at $T_1=1.5$ ms and 0.229(8) at $T_1=10$ ms.

\bibliography{references}